\documentclass[structabstract]{aa}
\usepackage{amsmath,amsfonts,amssymb}
\usepackage{xspace}
\usepackage[dvipsnames,usenames]{color}
\usepackage{graphicx}
\graphicspath{{./}{./fig/}}
\usepackage{txfonts}
\usepackage{color}

\makeatletter
\let\NAT@parse\undefined
\makeatother

\usepackage{natbib}

\usepackage{multirow}
\bibpunct{(}{)}{;}{a}{}{,} 

\definecolor{LinkColor}{rgb}{0.0,0.2,0.5}
\usepackage[pdfpagemode=UseNone,pdfstartview=FitB]{hyperref}
\hypersetup{
  breaklinks=true,
  colorlinks=true,
  linkcolor=LinkColor,
  citecolor=LinkColor,
  filecolor=LinkColor,
  menucolor=LinkColor,
  pagecolor=LinkColor,
  urlcolor=LinkColor,
}

\makeatletter
\newcommand{\MathFunc}[1]{\mathop{\operator@font #1}\nolimits}
\newcommand{\MathFuncWithLimits}[1]{\mathop{\operator@font #1}\limits}
\makeatother

\newcommand{\Tag}[1]{\mathsf{#1}}        
\newcommand{\BS}[1]{{\boldsymbol{#1}}}   
\newcommand{\V}[1]{\boldsymbol{#1}}      
\newcommand{\M}[1]{\mathbf{#1}}          
\newcommand{\FT}[1]{\hat{#1}}            
\newcommand{\Var}{\MathFunc{Var}}        

\newcommand{\mathd}{\mathrm{d}}
\newcommand{\mathe}{\mathrm{e}}
\newcommand{\mathi}{\mathrm{i}}

\newcommand{\norm}[1]{\Vert #1\Vert}
\newcommand{\Norm}[1]{\left\Vert #1\right\Vert}
\newcommand{\abs}[1]{\vert #1\vert}
\newcommand{\Abs}[1]{\left\vert #1\right\vert}
\newcommand{\avg}[1]{\langle #1\rangle}

\newcommand{\Expected}[1]{\mathrm{E}\left\{#1\right\}}
\newcommand{\argmin}{\MathFuncWithLimits{arg\,min}}

\newcommand{\bydef}{\stackrel{\mathrm{def}}{=}}

\newcommand{\cf}{\emph{cf.}\xspace}
\newcommand{\eg}{\emph{e.g.}\xspace}
\newcommand{\ie}{\emph{i.e.}\xspace}

\newcommand{\etc}{\emph{etc}.\xspace}
\newcommand{\Fig}[1]{Fig.~\ref{#1}}
\newcommand{\Tab}[1]{Table~\ref{#1}}

\newcommand{\micron}{\ensuremath{\muup\mathrm{m}}} 

\newcommand{\ComplexVis}{V}
\newcommand{\VisPhase}{\varphi}
\newcommand{\VisAmp}{\varrho}
\newcommand{\Freq}{\nu}               
\newcommand{\VFreq}{\V{\Freq}}
\newcommand{\Dirn}{\theta}            
\newcommand{\VDirn}{\V{\Dirn}}
\newcommand{\Time}{t}                 

\newcommand{\uv}{\ensuremath{(u,v)}\xspace}
\newcommand{\para}{{/\!/}}

\newcommand{\Param}{x}
\newcommand{\VParam}{\V{\Param}}
\newcommand{\Err}{r}         
\newcommand{\ErrBar}{\sigma} 
\newcommand{\Image}{I}
\newcommand{\BasisFunc}{b}
\newcommand{\Baseline}{B}
\newcommand{\VBaseline}{\V{\Baseline}}
\newcommand{\MaxBaseline}{\Baseline_\Tag{max}}
\newcommand{\Wavelength}{\lambda}

\newcommand{\DataTag}{\Tag{data}}
\newcommand{\PriorTag}{\Tag{prior}}

\newcommand{\Fcost}{f}
\newcommand{\Fdata}{\Fcost_\DataTag}
\newcommand{\Fprior}{\Fcost_\PriorTag}
\newcommand{\Weight}{w}

\newcommand{\Mira}{MiRA\xspace}

\newcommand{\Ndata}{N_\Tag{data}}
\newcommand{\Nedf}{N_\Tag{edf}}

\definecolor{FireBrick}{rgb}{0.70,0.13,0.13}

\definecolor{SeaGreen}{rgb}{0.13,0.70,0.13}

\begin{document}
\title{Image reconstruction in optical interferometry: Benchmarking the regularization}

   \author{S. Renard
          \inst{1}
          \and E. Thi\'ebaut\inst{2}
          \and F. Malbet\inst{1}
          }

   \offprints{E. Thi\'ebaut}

   \institute{IPAG,
              CNRS/UMR 5571, Universit\'e J. Fourier, BP-53, 38041 Grenoble Cedex, France \\
              \email{Stephanie.Renard,Fabien.Malbet@obs.ujf-grenoble.fr}
         \and
             Centre de Recherche Astrophysique de Lyon, CNRS/UMR5574, 69561 St-Genis-Laval, France \\
             \email{thiebaut@obs.univ-lyon1.fr}
             }

   \date{Received; accepted }


   \abstract %
   {With the advent of visible and infrared long-baseline interferometers with more than two telescopes, both the size and the completeness of interferometric data sets have significantly increased, allowing images based on models with no a priori assumptions to be reconstructed with an aperture synthesis technique.} %
   {Our main objective is to analyze the multiple parameters of the image reconstruction process with particular attention to the regularization term and the study of their behavior in different situations (types of astrophysical objects, telescope array configurations, level of noise, \etc). The secondary goal is to derive practical rules for the users.} %
   {Using the Multi-aperture image Reconstruction Algorithm (\Mira), we performed multiple systematic tests, analyzing 11 regularization terms commonly used. The tests are made on different astrophysical objects, different \uv plane coverages and several signal-to-noise ratios to determine the minimal configuration needed to reconstruct an image. We establish a methodology and we introduce the \emph{mean-square errors} (MSE) to discuss the results.} %
   {From the $\sim$24000 simulations performed for the benchmarking of image reconstruction with \Mira, we are able to classify the different regularizations in the context of the observations. We find typical values of the regularization weight. A minimal \uv coverage is required to reconstruct an acceptable image, whereas no limits are found for the studied values of the signal-to-noise ratio. We also show that super-resolution can be achieved with increasing performance with the \uv coverage filling.}  %
   {Using image reconstruction with a sufficient \uv coverage is shown to be reliable. The choice of the main parameters of the reconstruction is tightly constrained. We recommend that efforts to develop interferometric infrastructures should first concentrate on the number of telescopes to combine, and secondly on improving the accuracy and sensitivity of the arrays.} %


\maketitle



\section{Introduction}

Many astrophysical studies require milli-arcsecond (hereafter mas) resolution images at optical  wavelengths (visible and infrared), for example, the understanding of the interplay between accretion and ejection in the inner part of the disks of young stellar objects, the expansion mechanisms in novae just a few hours or days after the explosion, and the nature of dust in active galactic nuclei. Information at such a high resolution at the optical wavelengths requires diffraction-limited images with pupil sizes of the order of tens to hundreds of meters that can only be achieved by interferometrically combining light from separate apertures. The \emph{Very Large Telescope Interferometer} (VLTI) and the \emph{Center for High Angular Resolution Array} (CHARA) are facilities that provide interferometric measurements that can be used to reconstruct images of stellar surfaces \citep[e.g.,][]{Monnier2007, Hautbois2009, Zhao2009}, binaries \citep[e.g.,][]{Zhao2008,Kraus2009}, circumstellar shells around evolved stars \citep[e.g.,][]{LeBouquin2009}, and the close environment of young stars \citep[][]{Renard2010,Kraus2010}.

Owing to the sparse \uv coverage, the image reconstruction process is ill posed as there are more unknowns, e.g. the pixels of the image, than measurements. The data alone are insufficient to reconstruct an unambiguous image and some additional constraints, so-called the regularizations, are needed to converge to a unique and stable solution. Compared to radio interferometry, the data are much sparser in optical interferometry; hence, we expect the image reconstruction problem to be much more sensitive to the choice and the tuning of the regularization.  Since the general study of regularization by \citet{Titterington1985}, many different methods have been proposed in the literature to adjust the regularization level. Since image reconstruction for optical interferometry is still in its infancy, it is fundamental to analyze the different types of regularization to find those that are the most  suitable for the different astrophysical problems and to be able to tune the weight of the regularization. In this context, we carried out systematic tests using the image reconstruction algorithm devoted to optical interferometry data developed by \citet{Thiebaut2008}, called the Multi-aperture image Reconstruction Algorithm (\Mira). The analysis of these tests allow us to extract some general conclusions and establish practical rules for the users.

The mathematical principles of the image reconstruction technique is presented in Sect.~\ref{sec:principles}. The parameters of the simulated data are presented together with the characteristics of the images and the strategy in Sect.~\ref{sec:simus}. The results of the simulations are presented and discussed in Sect.~\ref{sec:discussion}, with an analysis of the role of the different terms and parameters in the image reconstruction. Finally, our conclusions are summarized in Sect.~\ref{sec:conclusions}.


\section{Principles of image reconstruction from optical interferometric data}
\label{sec:principles}

We do not intend to provide a formal and precise description of image reconstruction in optical  interferometry, but instead sufficient details to ensure that the paper is self-contained. Readers interested by a more detailed description of the method should refer to \citet{Thiebaut_Giovannelli-2010-interferometry}.


\subsection{Data from optical interferometric observations}
\label{sec:OIdata}

The principle of interferometry is to recombine coherently the beams from two or more independent telescopes and measure the so-called complex visibilities of the fringe patterns produced by the interferences. According to the van Cittert-Zernicke theorem, for an ideal interferometer, the complex visibility $\ComplexVis_{j_1,j_2}(\Time)$ of the fringes produced by the interferences of the telescopes $j_1$ and $j_2$ at time $\Time$ is proportional to the Fourier transform of the object brightness distribution $\FT{\Image}(\VFreq_{j_1,j_2}(\Time))$ at spatial frequency $\VFreq_{j_1,j_2}(\Time) = \VBaseline_{j_1,j_2}^\perp(\Time)/\Wavelength$, where $\Wavelength$ is the wavelength, and the so-called \emph{baseline} $\VBaseline_{j_1,j_2}^\perp(\Time)$ represents the separation between the two telescopes projected on a plane perpendicular to the line of sight \citep{Lawson2000, Malbet2007}.  Since the number of measurements is finite, to simplify the equations we introduce some notation for the $m$-th measured complex visibility and the corresponding spatial frequency, given by
\begin{align}
  \ComplexVis_{m} &\bydef  \ComplexVis_{\!j_{1,m},j_{2,m}}(\Time_m) \, , \\
  \VFreq_m &\bydef \VBaseline_{\!j_{1,m},j_{2,m}}^\perp(\Time_m)/\Wavelength  \, ,
\end{align}
where $j_{1,m}$ and  $j_{2,m}$ are the interfering telescopes and $\Time_m$ is the time of observation.


\subsection{Description of the image model}
\label{sec:ImgModel}

The final product of the image reconstruction is an image that can be treated as a grid of square  pixels. In this context, the object brightness distribution as a function of the position $\VDirn$ can be approximated using the parametrization
\begin{equation}
  \label{eq:image-model}
  \Image(\VDirn) = \sum_{n=1}^{N} \Param_n\,\BasisFunc_n(\VDirn) \, ,
\end{equation}
where $\VParam=\{\Param_n\}_{n=1}^{N}$ are the image parameters, \eg the pixel values of the  image, and $\{\BasisFunc_n(\VDirn)\}_{n=1}^{N}$ is the chosen basis of functions, \eg the response function of each pixel.  The image reconstruction then consists of estimating the $N$ parameters $\VParam$ that most closely fit the interferometric data.  In this paper, we chose $x_n$ to be proportional to the value of the $n$-th pixel of a sampled image and $\BasisFunc_n(\VDirn)=\BasisFunc(\VDirn-\VDirn_n)$, where $\BasisFunc(\VDirn)$ is the pixel shape and $\VDirn_n$ the position of the $n$-th pixel; thus
\begin{equation}
  \label{eq:sampled-image-model}
  x_n \bydef \alpha \, \Image(\VDirn_n) \, ,
\end{equation}
where $\alpha>0$ is a scaling factor such that $\VParam$ is normalized (this is required by the interferometric data format, \cf \citet{Pauls2005}). With this model, the exact Fourier transform of the brightness distribution is given by
\begin{equation}
  \label{eq:TF-image-model}
  \FT{\Image}(\VFreq)
  = \sum_n x_n\,\FT{\BasisFunc}_n(\VFreq)
  = \FT{\BasisFunc}(\VFreq) \, \sum_n x_n\,\mathe^{-\mathi\,2\,\pi\,\VDirn_n\cdot\VFreq} \, ,
\end{equation}
where $\FT{\BasisFunc}_n(\VFreq)$ and $\FT{\BasisFunc}(\VFreq)$ are the Fourier transforms of the basis functions. In our case, they correspond to the Fourier transform of the pixel response function, \ie the pixel shape. Hence, the model of the $m$-th complex visibility is given by
\begin{equation}
  \label{eq:discrete-TF-image-model}
  \FT{\Image}_m
  = \FT{\Image}(\VFreq_m)
  = \sum_n A_{m,n}\,x_n
  = (\M{A}\cdot\VParam)_m \, ,
\end{equation}
where $\M{A}$ is a matrix with the complex coefficients
\begin{equation}
  \label{eq:model-coefs}
  A_{m,n} = \FT{\BasisFunc}_n(\VFreq_m)
  =  \FT{\BasisFunc}(\VFreq_m)\,\mathe^{-\mathi\,2\,\pi\,\VDirn_n\cdot\VFreq_m} \, .
\end{equation}
This matrix multiplication performs a linear transformation that contains the Fourier transform, the pixel shape, and the sparse sampling of the \uv plane.

The main problem in optical interferometry is the small number of telescopes (currently up to four or six), which leads to a sparse sampling of the spatial frequencies, the so-called \uv plane (see \Fig{fig:pluv}).  Owing to random effects caused by the atmospheric turbulence, the visibility phase cannot be calibrated and the power spectrum and the closure phase are used.  This results in a partial loss of the Fourier phase information \citep{Thiebaut_Giovannelli-2010-interferometry}.


\subsection{Inverting the problem of interferometric imaging}
\label{sec:InvApproach}

Because of the sparse \uv coverage and the possible lack of other information such as the phase,  the reconstruction of an image obtained by the interferometric data alone is an ill-posed inverse problem. It needs additional \emph{a priori} constraints to be recasted into a problem that has a unique and stable solution.  A general prescription is to express the solution as one that minimizes a penalty function $\Fcost$ under some strict constraints \citep{Thiebaut2005,Thiebaut_Giovannelli-2010-interferometry}
\begin{equation}
  \VParam^{+} = \argmin_{\VParam} \Fcost(\VParam)
  \quad\text{s.t.}\quad
  \VParam \geqslant 0 \text{ and } \sum\nolimits_n \Param_n = 1 \, ,
  \label{eq:optim-prob}
\end{equation}
with
\begin{equation}
  \Fcost(\VParam) = \Fdata(\VParam) + \mu\, \Fprior(\VParam) \, ,
  \label{eq:penalty}
\end{equation}
where the so-called \emph{likelihood} term $\Fdata(\VParam)$ measures the discrepancy between the model and the available data, while the so-called \emph{regularization} term $\Fprior(\VParam)$ measures the discrepancy with the \emph{prior} information.  In other words, minimizing the likelihood term $\Fdata(\VParam)$ enforces the fit with the actual data, while minimizing the regularization term $\Fprior(\VParam)$ enforces the agreement with the priors. The so-called hyperparameter $\mu>0$ is used to adjust the relative weight of the constraints set by the measurements and the ones set by the priors.  In Eq.~(\ref{eq:optim-prob}), the positivity ($\VParam \geqslant 0$) and the normalization ($\sum\nolimits_n \Param_n = 1$) of the brightness distribution are also included by default.

For the image reconstruction, we use \Mira \citep[][]{Thiebaut2008} to find solutions of Eq.~(\ref{eq:optim-prob}).  \Mira can deal with the various kinds of data provided by an optical interferometer and implements a number of different regularizations \citep{Thiebaut2008,Thiebaut_Giovannelli-2010-interferometry}.


\subsubsection{The likelihood term $\Fdata$ and the data model}
\label{sec:fdata}
\label{sec:model}

We focus on the choice of the priors and the tuning of their parameters.  It is not within the scope of the paper to deal with global optimization issues and the search for a global minimum of the penalty function in Eq.~(\ref{eq:penalty}).  We therefore assume that the available measurements consist of complex visibilities, \ie amplitude and phase, in order to have a \emph{convex} likelihood term $\Fdata(\VParam)$.  If the regularization term $\Fprior(\VParam)$ is also convex, the global penalty function $\Fcost(\VParam)$ will be convex, which ensures that the solution of Eq.~(\ref{eq:optim-prob}) is unique.  Current optical interferometers only provide phase closures and power-spectrum data (\ie the phase of the bispectrum and the squared amplitude of the complex visibilities), this means that our assumption will give somewhat optimistic results because some Fourier phase information is missing and because the likelihood term $\Fdata(\VParam)$ is non-convex when dealing with real data.  However, as the number of simultaneous interfering telescopes increases, the number of missing phases becomes much less important and they can be reliably derived using \emph{self-calibration} \citep{Pearson_Readhead-1984-self_calibration_review} to achieve a situation similar to the case studied in our simulations. Moreover, new interferometers will make use of a phase reference source to directly measure the phase of the complex visibilities \citep{Delplancke_at_al-2003-Prima}.

However, the OI-FITS standard \citep{Pauls2005} imposes the use of complex visibilities in their polar representation with independent error bars. We therefore simulate each measured complex visibility as
\begin{equation}
  \VisAmp_m = \left|\ComplexVis_m\right| + \delta\VisAmp_m \, ,
  \quad \mbox{and} \quad
  \VisPhase_m = \arg \ComplexVis_m + \delta\VisPhase_m \, ,
\end{equation}%
where $\VisAmp_m$ and $\VisPhase_m$ are the measured amplitude and phase of the $m$-th measure, $\ComplexVis_m$ the corresponding complex visibility computed from the true object brightness distribution, and $\delta\VisAmp_m$ and $\delta\VisPhase_m$ are additive noise terms. In our simulations, the noise terms have independent Gaussian statistics such that
\begin{equation}
  \Var(\delta\VisAmp_m) = \avg{\VisAmp_m}^2 \, \Var(\delta\VisPhase_m) \, ,
\end{equation}
where $\avg{\VisAmp_m} = |\ComplexVis_m|$ is the expected value of the amplitude that can be computed from the complex visibility of the true image.  This particular choice follows approximately the model of \citet{Goodman-statistical_optics}.

In this context, to define the likelihood term we use the \emph{local approximation} \citep{Meimon2005b}
\begin{equation}
  \Fdata(\VParam) = \sum_m \left\{
    \left(\frac{\Err_{\para,m}(\VParam)}{\ErrBar_{\para,m}}\right)^2 +
    \left(\frac{\Err_{\perp,m}(\VParam)}{\ErrBar_{\perp,m}}\right)^2
  \right\}  \, ,
\end{equation}
where $\Err_{\para,m}(\VParam)$ and $\Err_{\perp,m}(\VParam)$ are the two components of the complex residuals, $\Err_{m}(\VParam)= \VisAmp_m\,\mathe^{\mathi\,\VisPhase_m} - (\M{A}\cdot\VParam)_m$, respectively, along and orthogonally to the measured complex visibility. Given the error bars $\ErrBar_{\VisAmp,m}$ and $\ErrBar_{\VisPhase,m}$ of the amplitude and the phase of the complex visibility, the standard deviations in the components of the complex residuals are \citep{Pauls2005}
\begin{equation}
  \ErrBar_{\para,m} = \ErrBar_{\VisAmp,m}
  \quad \mbox{and} \quad
  \ErrBar_{\perp,m} = \VisAmp_m\,\ErrBar_{\VisPhase,m} \, .
\end{equation}


\subsubsection{The regularization term $\Fprior$}
\label{sec:fprior}

In our simulations, we test 11 different regularization terms that are commonly used in image reconstruction methods and are implemented in \Mira (see Appendix~\ref{app:fprior} for detailed expressions).
\begin{list}{XXXXXXX}{\setlength{\itemsep}{1em}}
\item[1.] \textbf{Quadratic smoothness}, which most closely describes a smooth image and helps us to avoid unmeasured high frequencies.
\item[2-3.] \textbf{Compactness}, which describes compactness in the image plane and hence smoothness in the Fourier plane \citep{LeBesnerais2008}. Two different cases were studied in the simulations, with penalties of the second and third orders with respect to the distance of the center of the field of view.
\item[4.] \textbf{Total variation} (hereafter TV), which minimizes the total gradient of the image and helps us to describe uniform areas in the sought image with steep but localized changes  \citep{Strong_Chan-2003-total_variation}.
\item[5.] \textbf{$\ell_1$ smoothness}, which is useful for an extended object with sharp edges since it is linear for strong gradients.
\item[6-8.] \textbf{$\ell_p$-norm} with $p=1.5$, $p=2$ and $p=3$. For $p>1$, the $\ell_p$-norm regularization tends to produce a smooth image as it reduces the variance in the pixels.
\item[9-11.] \textbf{Maximum entropy methods} (hereafter MEM) aims to identify the least informative image consistent with the data \citep{Gull1984,Narayan1986}.  We try three types of entropies \emph{MEM-sqrt}, \emph{MEM-log}, and \emph{MEM-prior}, respectively. The first two tend to reproduce an image with a flux spread across a minimum number of pixels. The last one is minimum when the image is as close as possible to a prior image. This prior image is  the Gaussian that most closely reproduces the amplitude visibility data.
\end{list}

The different regularization terms are expected to behave as follows. The positivity and the normalization imposed in all the reconstructions are an $\ell_1$ norm and lead to the sparsity of the solution, \ie to a minimum number of bright pixels to explain the data. As most astrophysical images are smooth and compact, we expect that the regularizations that can describe these images will behave well, \ie smoothness (quadratic or $\ell_1$), compactness, and TV. The $\ell_p$-norm regularization with $p=2$ (and by extension for $p>1$ as the regularization has the same behavior) has the tendency to force to zero the spatial frequencies that have not been measured (according to the Parseval theorem). Since the regularization has to interpolate correctly between the data, which is closer to the reality, it is not expected to give good results.  Finally, the MEM-prior is expected to yield more reliable results than the other type of entropy because our choice of the \emph{a priori} image can closely describe a compact object.


\section{Description of the simulations}
\label{sec:simus}

We now describe all the various simulated data that we compute for different objects, \uv coverages, and signal-to-noise ratios (SNR), as well as the parameters used for the image reconstructions.


\subsection{Simulated data}
\label{sec:data}

Our simulated data sets are saved as OI-Fits files \citep{Pauls2005} and depend on several setting of the object type, the \uv coverage, and the SNR. The 90 data files are available on the JMMC website\footnote{\url{http://apps.jmmc.fr/oidata/shared/srenard/}}.

\subsubsection{Astrophysical objects}
\label{sec:obj}

\begin{figure*}
  \centering
  \includegraphics[width=17cm]{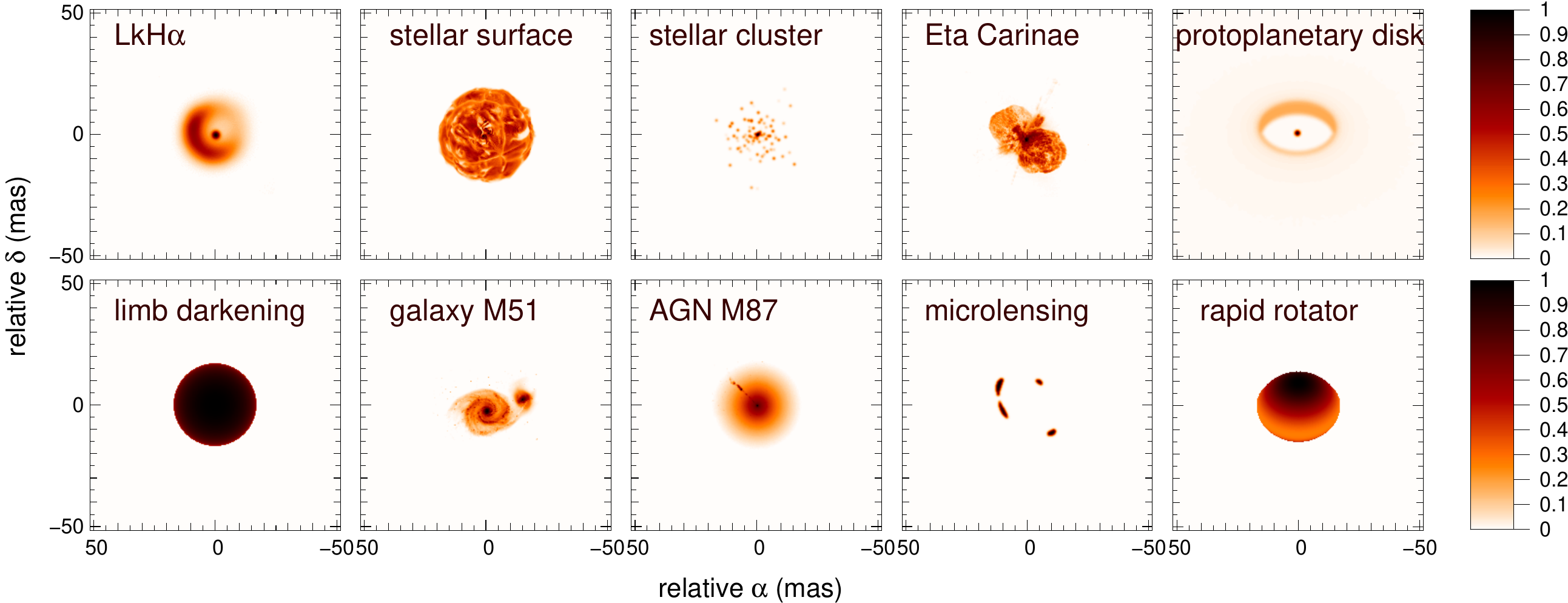}
  \caption{Astrophysical objects used in the simulations.}
  \label{fig:Img}
\end{figure*}

For our simulations, we consider ten astrophysical objects (see \Fig{fig:Img}) that differ in term of their morphology and the typical length scales of their structures.

\begin{enumerate}
\item \textbf{LkH$\BS{\alpha}$:} the model of LkH$\alpha$ describes a compact object with a peak of intensity and a smooth envelope. The model comes from the 2004 International Imaging Beauty Contest organized by P. Lawson for the IAU \citep{Lawson2004}.
\item \textbf{A stellar surface:} this is a model of the supergiant $\alpha$ Ori that presents some convective cells, producing small scales on a smooth background. This model was produced by A. Chiavassa for the 2009 Workshop on Interferometry Imaging (WII09) organized by J.-P. Berger and F. Malbet (Berger et al., in prep.).
\item \textbf{A stellar cluster:} the model consists of a hundred point sources. The position and the brightness of the sources follow a normal law.
\item \textbf{Eta Carinae:} the image of Eta Carinae presents many different scales and structures, such as the extended gas and the stars. This image was retrieved from the Hubble Space Telescope's website\footnote{\url{http://hubblesite.org/}}. Some treatments have been applied to the image, \ie a mean of the three different color channels to produce a grayscale image and a cut of the low intensities to produce a null background.
\item \textbf{A protoplanetary disk:} the model represents an Herbig\,Ae/Be star with a point source (the star) and an extended structure (the disk). This model was computed by J.-P. Berger for the phase A science case of the VLTI-Spectro Imager instrument \citep{Filho2008}.
\item \textbf{A limb-darkened star:} we used the power-law model of \citet{Hestroffer1997} with an exponent $\alpha = 0.3$. The image has a very smooth core with steep edges.
\item \textbf{The galaxy system M51:} this image of M51 consists of as many different scales and structures as Eta Carinae (gas, stars, spiral arms). This image was retrieved from the Hubble Space Telescope's website and was processed in a way similar to the image of Eta Carinae.
\item \textbf{The AGN M87:} this AGN has a jet that consists of a narrow structure surrounded by a smooth background due to the gas. The image was retrieved from the Hubble Space Telescope's website and the same treatments as the Eta Carinae's image have been applied.
\item \textbf{A gravitational microlensing image:} Gravitational microlensing is an astronomical phenomenon due to the gravitational lens effect. When a distant star or quasar becomes sufficiently aligned with a massive compact foreground object, the bending of light due to its gravitational field leads to two distorted unresolved images resulting in an observable magnification. The image shows four very compact structures. This model was developed by J. Surdej for the phase A science case of the VLTI-Spectro Imager instrument \citep{Filho2008}.
\item \textbf{A rapid rotator:} the rapid rotation of a star affects the stellar shape and the local emitted flux. We used the model of \citet{Domiciano2002}, with parameters $D=0.78$ and $T_p=35000$\,K. The resulting light distribution was projected onto a plane with an inclination of 45\degr.
\end{enumerate}

Since the goal of our tests is to determine the influence of the object's type on the image reconstruction, all the considered objects have a similar angular size of $\sim34$\,mas, which is consistent with the typical size of the field of view of an optical interferometer such as Amber/VLTI. To generalize our results, we expect the most important figure for a given object and instrument to be the number of \emph{resolved elements}, which is equal to the ratio of the angular size of the object support to the effective resolution of the imaging system.  Estimated by this ratio, the complexity of the objects that we have considered is in the range of $200-600$ resolved elements.

\subsubsection{\uv coverage}
\label{sec:uv-coverage}

\begin{figure}
  \resizebox{\hsize}{!}{
    \includegraphics{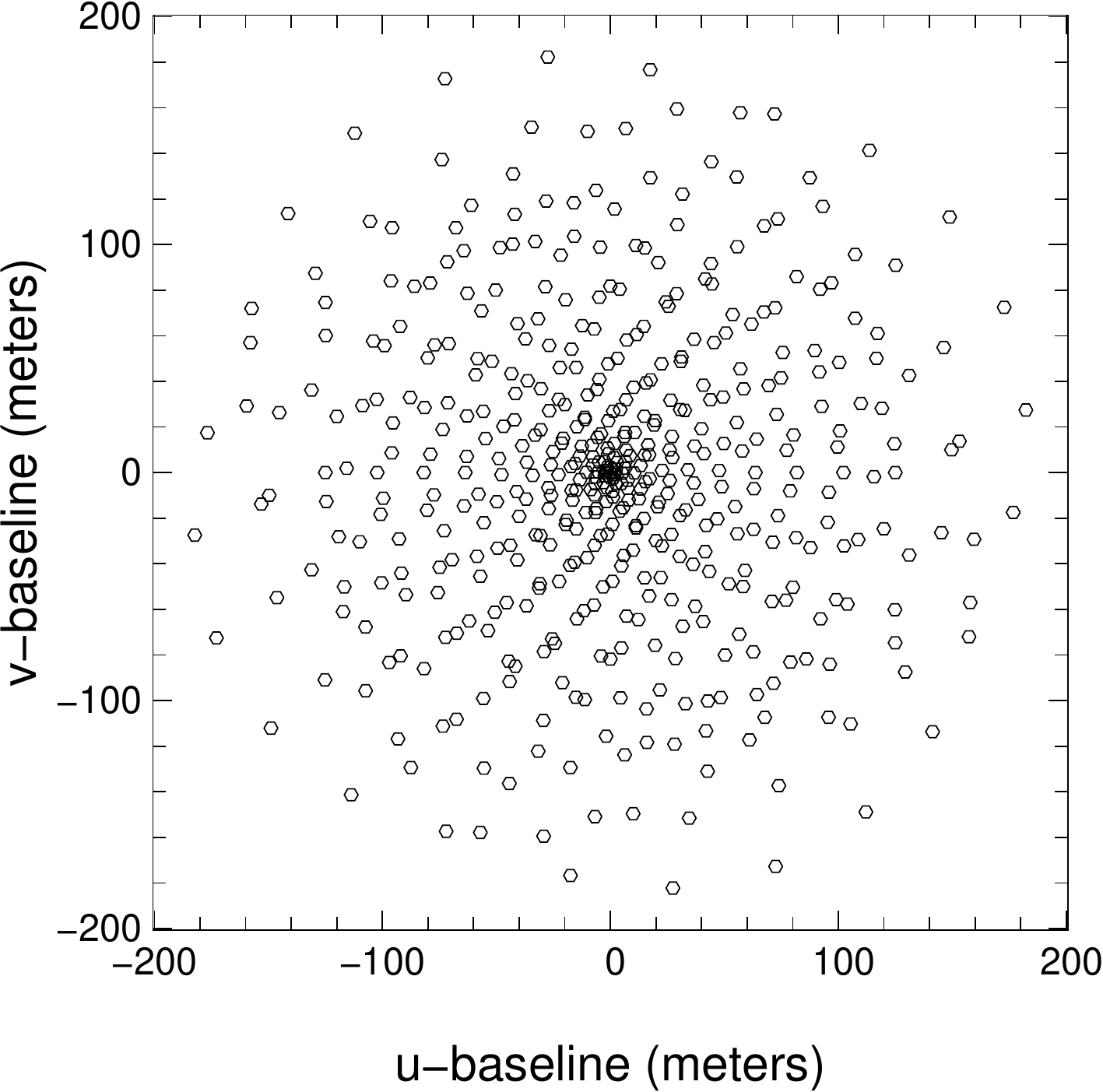}
    \includegraphics{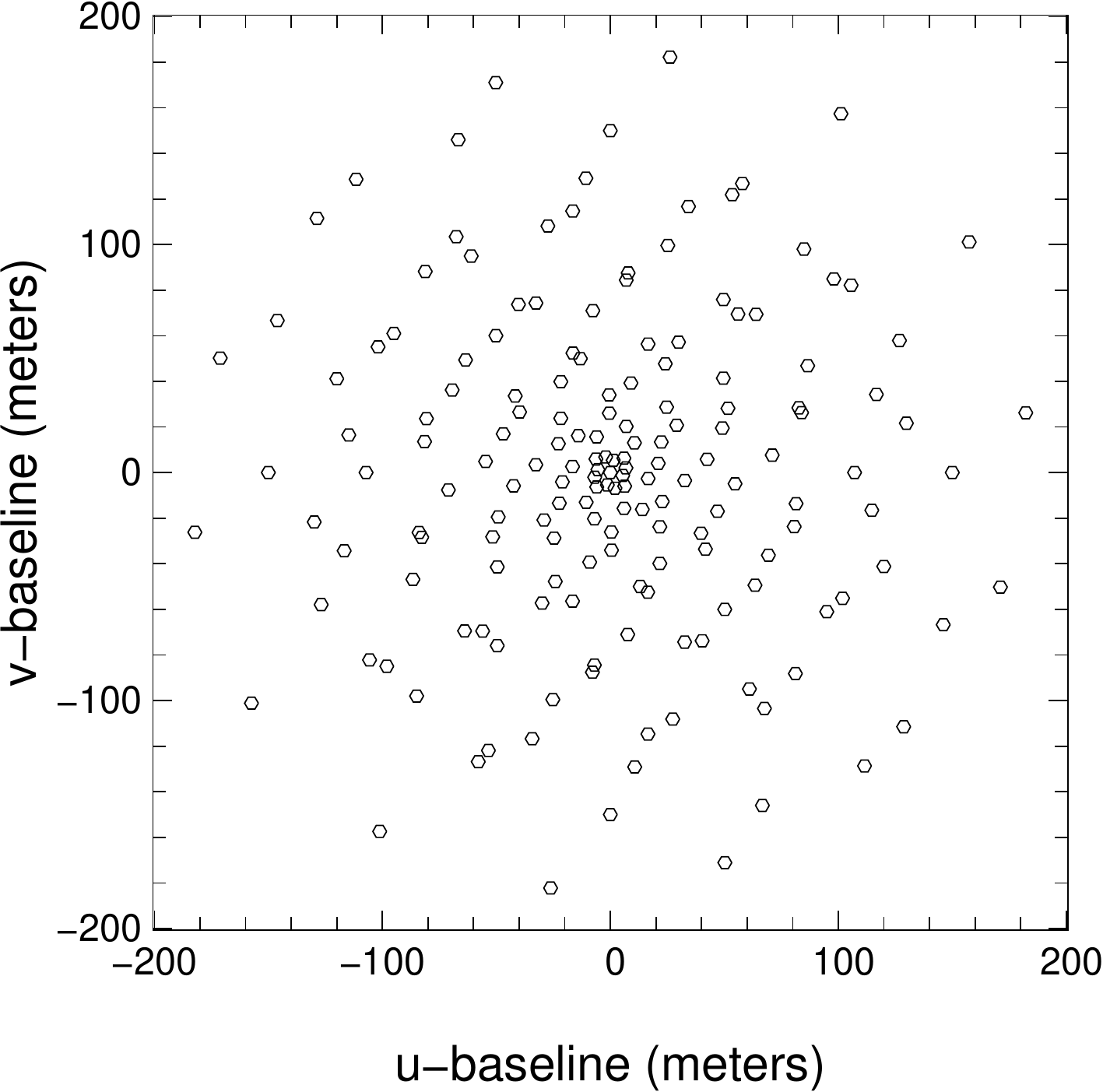}
    \includegraphics{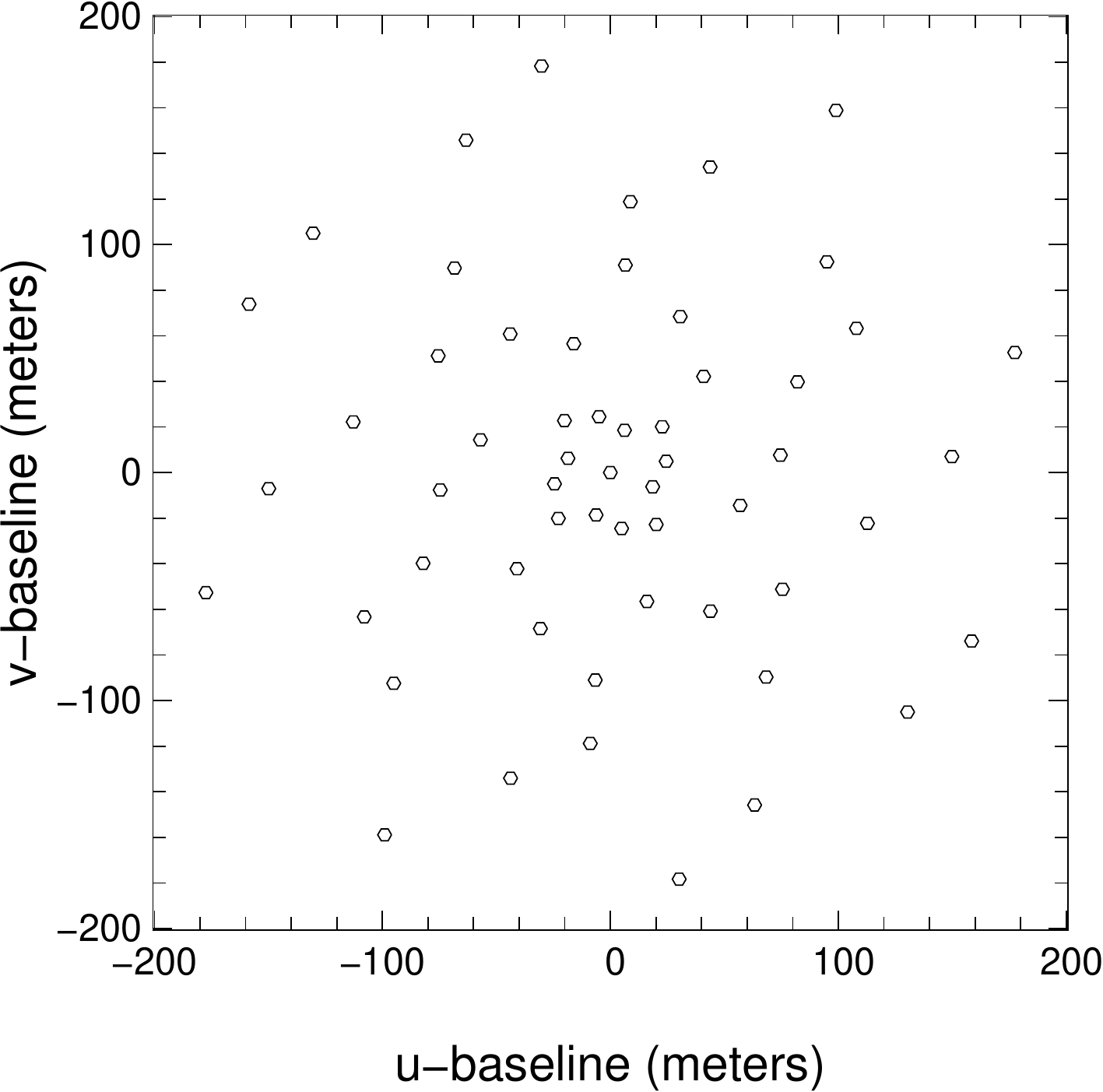}}
  \caption{\uv coverage. \emph{From left to right:} rich (245 sampled frequencies), medium (88 sampled frequencies), and poor (31 sampled frequencies).}
  \label{fig:pluv}
\end{figure}

To study the influence of the instrumental configuration and analyze the capability of the regularization to interpolate the available data and thus fill the voids in the \uv plane, we consider several sparse \uv coverages (see \Fig{fig:pluv}). To remain general, our different \uv coverages were constructed to be uniformly spread. Each \uv coverage consists of concentric rings modulated by a sinusoid along the ring and with phases of the modulations chosen to maximize the distance between the points of two adjacent rings.  This also avoids symmetries in the \uv coverage. The concentration of the rings is more important on small baselines than on the largest ones to insure a good sampling of low spatial frequencies. In this paper, we consider three \uv coverages which differ in terms of the number of samples, called hereafter \emph{rich} (245 sampled frequencies), \emph{medium} (88 sampled frequencies), and \emph{poor} (31 sampled frequencies) coverages. The chosen \uv configurations could be considered as very good by the standards of existing data, but the goal of the paper is not to cover all possible cases but to show the general trend.

\subsubsection{Signal-to-noise ratio}
\label{sec:snr}

To investigate the effects of varying the SNR, we use a SNR factor $\gamma$ in the standard deviations given by
\begin{equation}
  \ErrBar_{\VisAmp,m}  = \gamma \, \avg{\VisAmp_m}
  \quad \mbox{and} \quad
  \ErrBar_{\VisPhase,m} = \gamma \, .
\end{equation}
Therefore with these settings, the error bars become
\begin{equation}
  \ErrBar_{\para,m} = \ErrBar_{\perp,m} = \gamma \, \avg{\VisAmp_m} \, .
\end{equation}

To analyze the influence of the SNR on the reconstructed images, three values of $\gamma$ are tested: \emph{high} SNR (1\%) , \emph{intermediate} SNR (5\%), and \emph{low} SNR (10\%). We limit our study to these SNR values to maintain a small amount of results.  Therefore, there is no systematic attempt to search for the limit of SNR and we realize that our  worst case (10\%) might be considered moderate noise in real data.


\subsection{Parameters of the synthesized image}
\label{sec:img-params}

The resolution and the field of view (hereafter FOV) of the reconstructed image have to be chosen with care. On the one hand, the pixel size $\Delta\theta$ must be small enough to properly account for the highest spatial frequencies available in the data. Using Shannon's rule, we find that $\Delta\theta \leqslant \Wavelength/(2\,\MaxBaseline)$, where $\MaxBaseline$ is the maximum length of the observed baselines and $\Wavelength$ the wavelength. In our reconstructions, the pixel size is one third of the limit set by Shannon's rule. This oversampling allows us to check whether image reconstruction can achieve some level of super-resolution.  For instance, this yields $\Delta\theta=0.4$ mas at $\Wavelength=2.2\,\micron$ with $\MaxBaseline = 190$\,m. On the other hand, the FOV has to be large enough to avoid field aliasing and therefore we take an image three times larger than the object itself. As all our objects have a size of 34\,mas and thus approximately fit into $85\times85$ pixels, the reconstructed  images have $256\times256$ pixels.


\subsection{Reconstruction strategy}
\label{sec:strategy}

Given the data (determined by the object, the $(u,v)$ coverage, and the noise realization) and the regularization, we conduct a sequence of image reconstructions for different values of the regularization weight $\mu$. Since the problems solved are convex (as explained in Sect.~\ref{sec:model}), their solutions do not depend on the starting point. To reduce the calculation time, we therefore try to use a starting point that is as close as possible to the final solution. We begin the sequences of reconstructions with the highest value of $\mu$ and use the true image as the starting point for this first reconstruction. We then reduce $\mu$ and use the image previously obtained as the starting point. This latter step is repeated until we reach the lowest value of $\mu$. Each reconstruction is an iterative process that is stopped when the norm of the gradient of the penalty function $\Fcost(\VParam)$ is below a preset threshold.
This threshold is derived from the true image according to
\begin{equation}
  \Norm{\nabla \Fcost(\VParam^\Tag{rec})} \le \eta \,
  \frac{\Abs{\Fcost_\mu\left(\VParam^\Tag{ref}\right)}}{\Norm{\VParam^\Tag{ref}}}
  = 10^{-5}\,\Abs{\Fcost_\mu\left(\VParam^\Tag{ref}\right)} \, ,
\end{equation}
where $\eta>0$ is a small value and $\Fcost_\mu(\VParam_\Tag{ref})$ the penalty computed for the true image and for each value of $\mu$. The $\ell_2$ norm of the true image is normalized and we assume that $\eta=10^{-5}$.


\subsection{Image quality criterion:  the mean-squared error}
\label{sec:quality-criterion}

To assess the quality of the reconstructed images, we consider the mean-squared error (hereafter MSE) of the reconstructed images.  In our simulations, we use normalized input images with the same pixel size as for the reconstructed image. Hence, to compare the reconstructed image $\VParam^\Tag{rec}$ and the true image $\VParam^\Tag{ref}$, we can simply use the MSE defined as
\begin{equation}
  \Tag{MSE} = \frac{1}{N}\,\sum_n \left(\Param^\Tag{rec}_n
  - \Param^\Tag{ref}_n\right)^2  = \frac{1}{N}\,\Norm{\VParam^\Tag{rec}
  - \VParam^\Tag{ref}}^2 \, ,
\end{equation}
where $N$ is the total number of pixels. We note that, since we use complex visibilities and our priors, apart from the FOV one, are shift-invariant, the reconstructed image is correctly centered and there is no need to compensate for registration errors.


\section{Results and discussion}
\label{sec:discussion}

We performed a total of $\sim$24000 simulations corresponding to the reconstruction of all cases described in Sect.~\ref{sec:simus} with different values of the hyperparameter $\mu$. We present here the results of these simulations and analyze the consequences for the image reconstruction process in order to draw general conclusions.

We begin by discussing the optimal value of the hyperparameter and determining whether the MSE is a good quality criterion. We then discuss the effects of the  following parameters:
\begin{itemize}
\item \textrm{the regularization:} what are the good and bad regularizations?
\item \textrm{the limits:} what are the minimal \uv coverage and SNR value?
\item \textrm{the hyperparameter $\mu$:} what is the optimal value? With which parameters does it vary?
\item \textrm{the likelihood term:} how does it vary? Can it be used to tune the regularization term instead of the hyperparameter $\mu$?
\item \textrm{the effective resolution:} what degree of super-resolution can be achieved? How does it vary?
\end{itemize}


\subsection{Optimal regularization weight $\mu^+$}
\label{sec:optim-regul}

\begin{figure*}[t]
  \centering
  \begin{minipage}[c]{.3\linewidth}
    \centering
    \includegraphics[width=5.1cm]{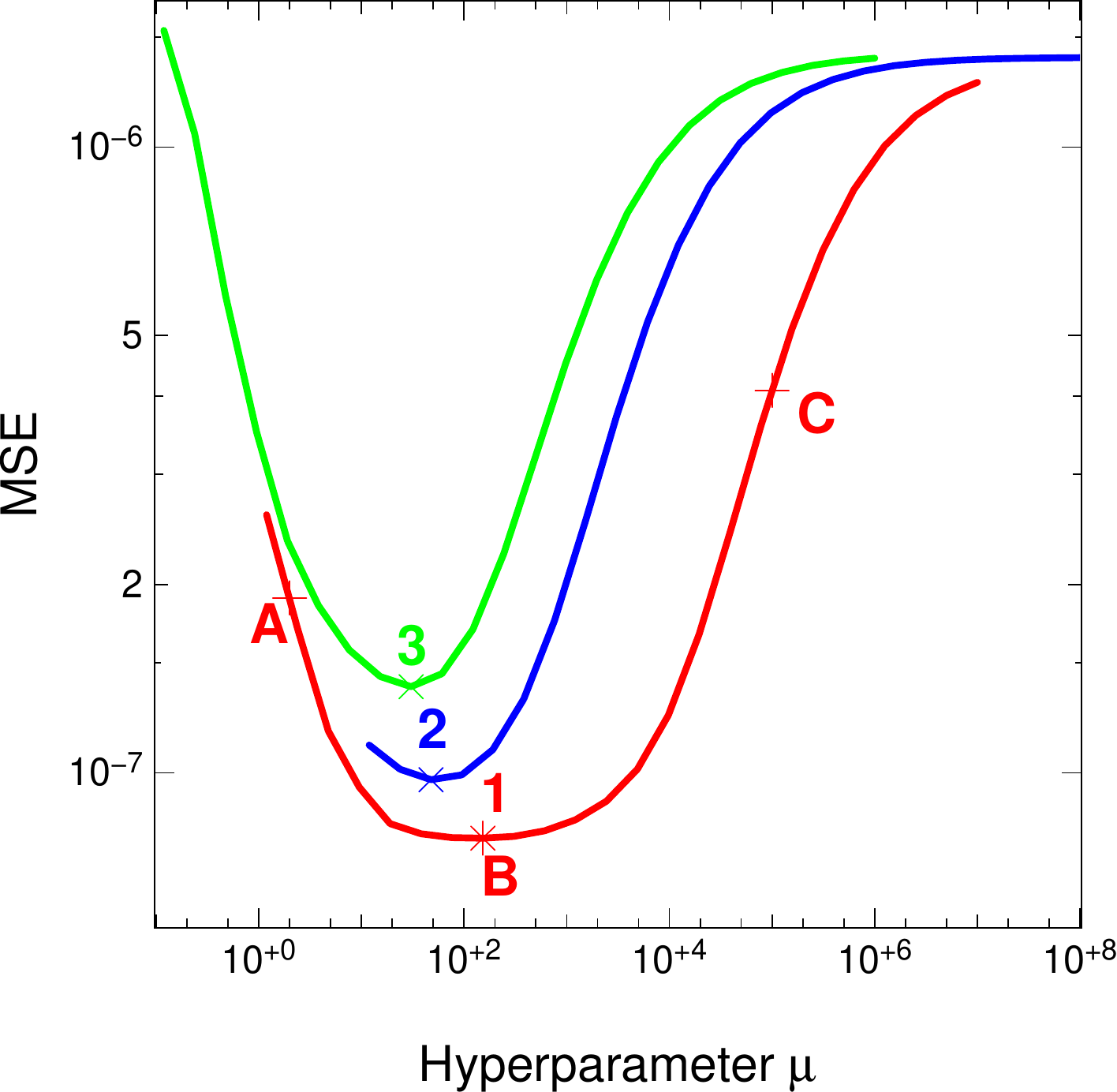}
  \end{minipage}
  \hfill
  \begin{minipage}[c]{.6\linewidth}
    \centering
    \begin{tabular}{ccc}
      \includegraphics[width=3.4cm]{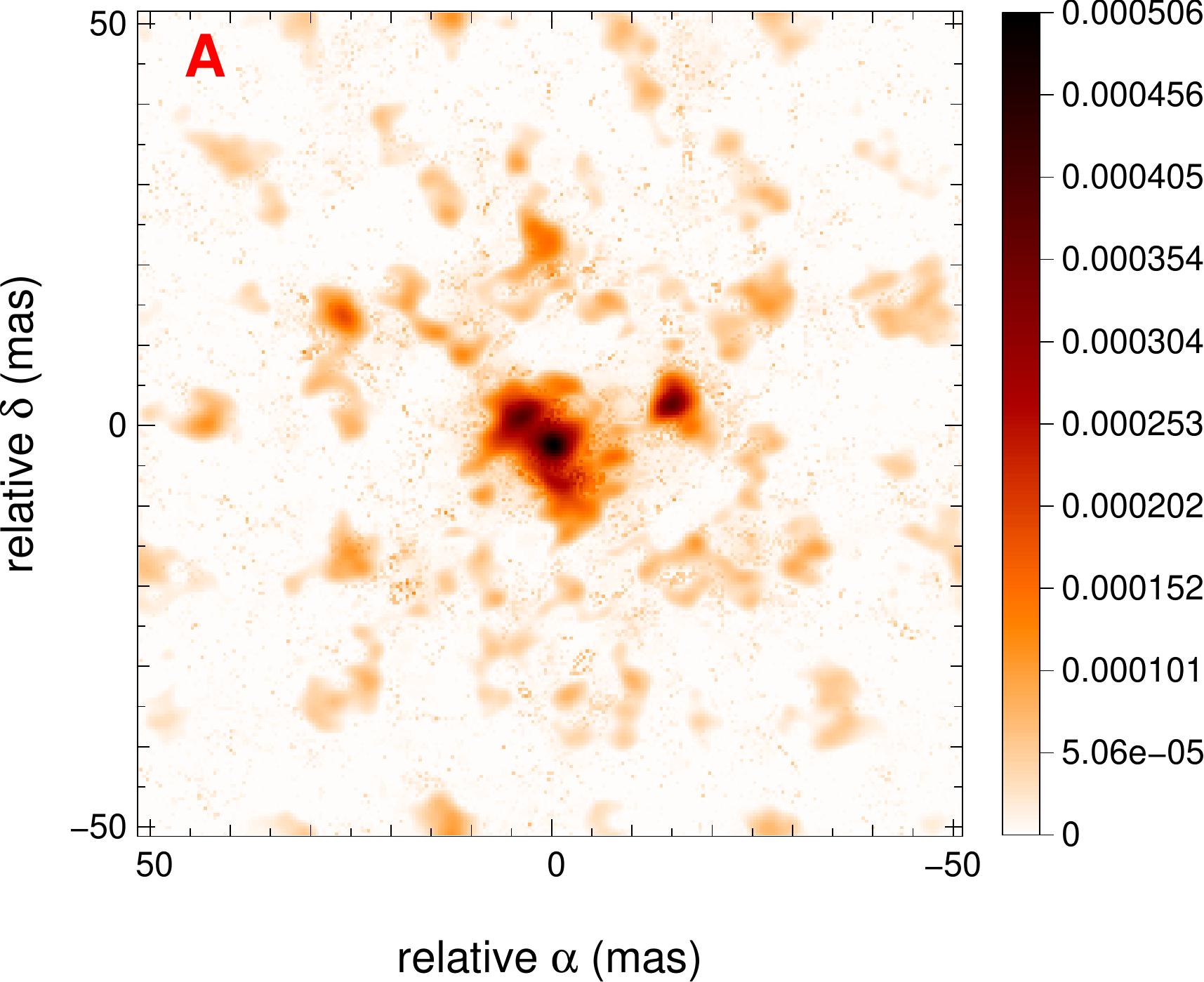}&
      \includegraphics[width=3.4cm]{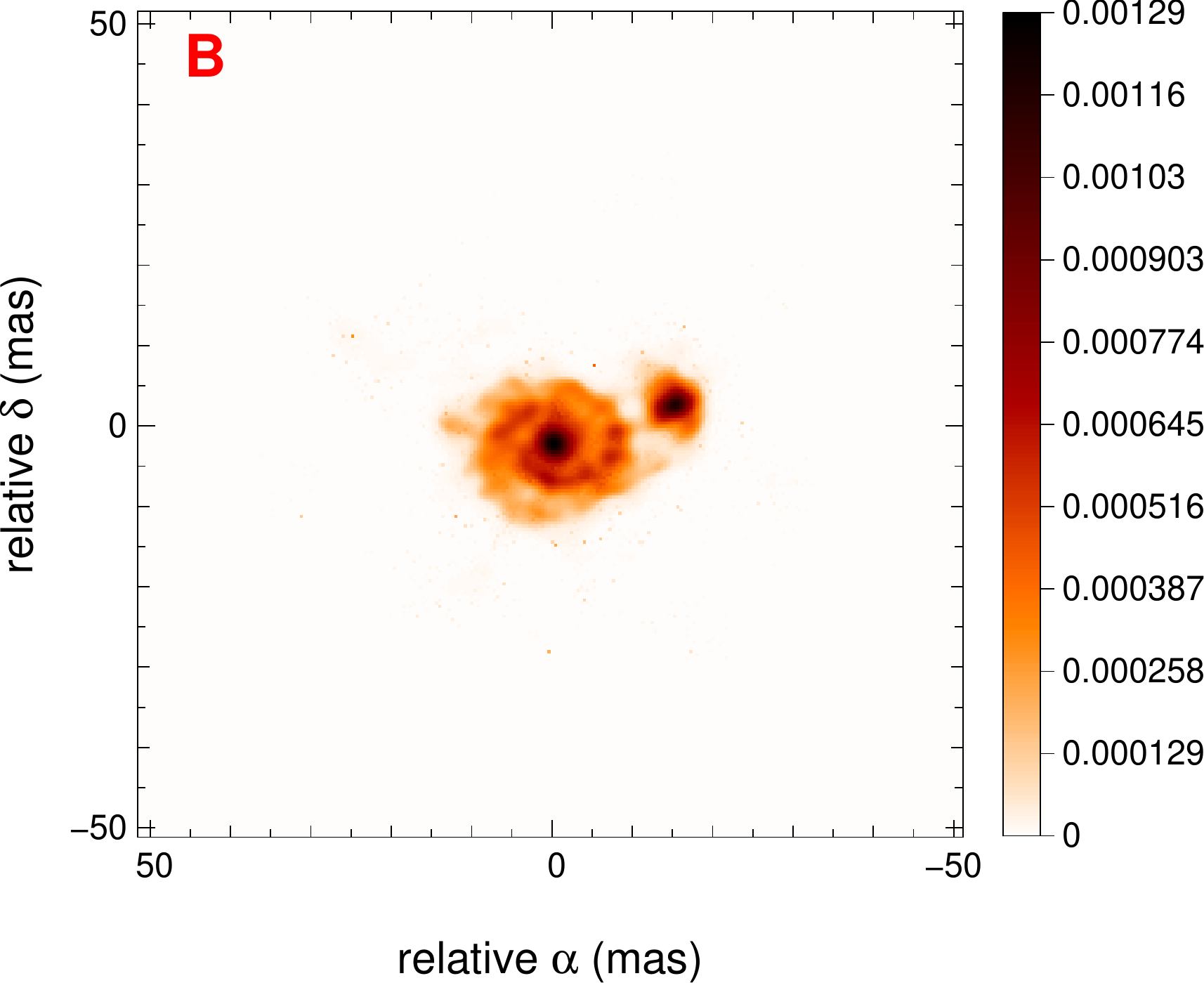}&
      \includegraphics[width=3.4cm]{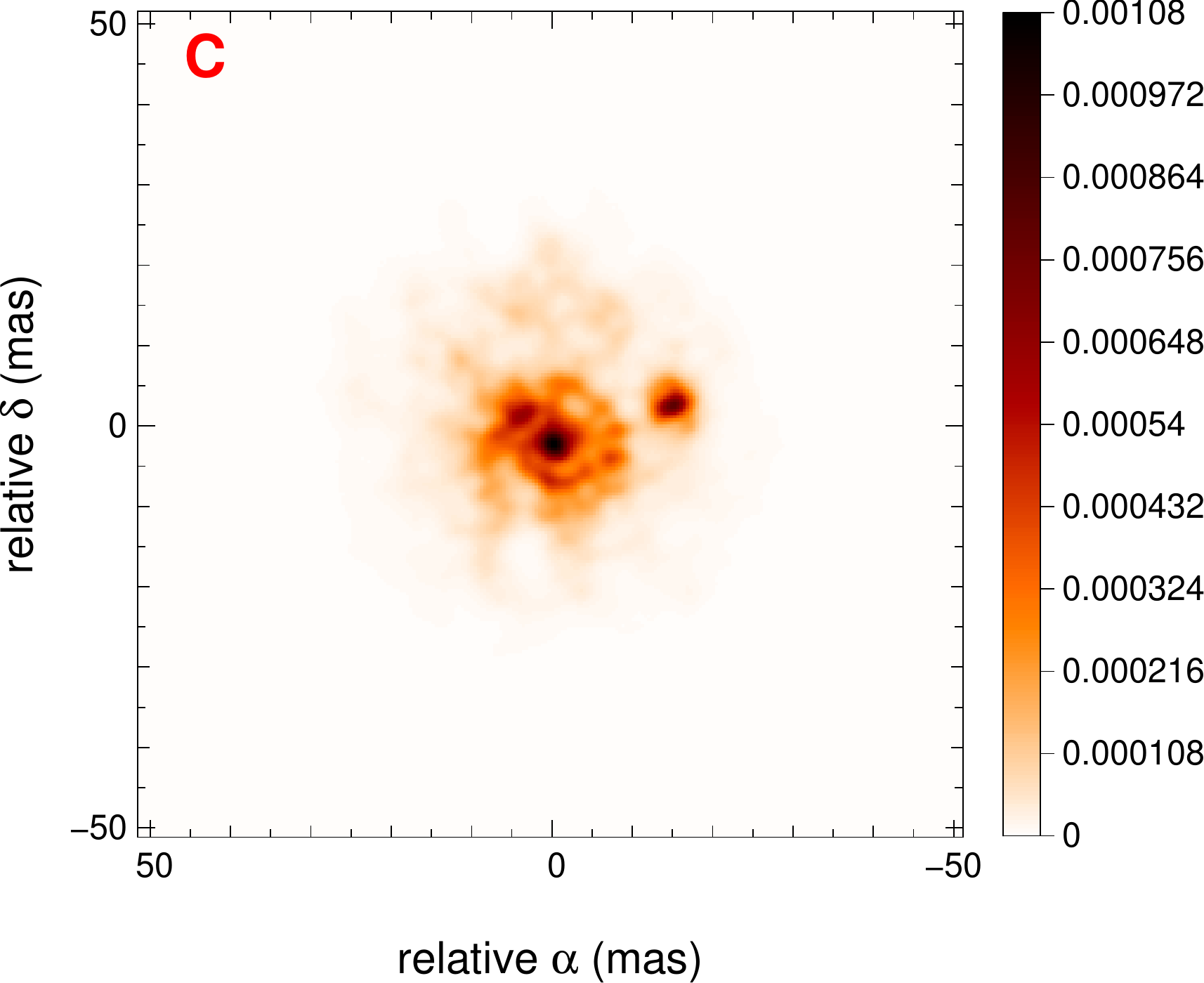}\\
      \includegraphics[width=3.4cm]{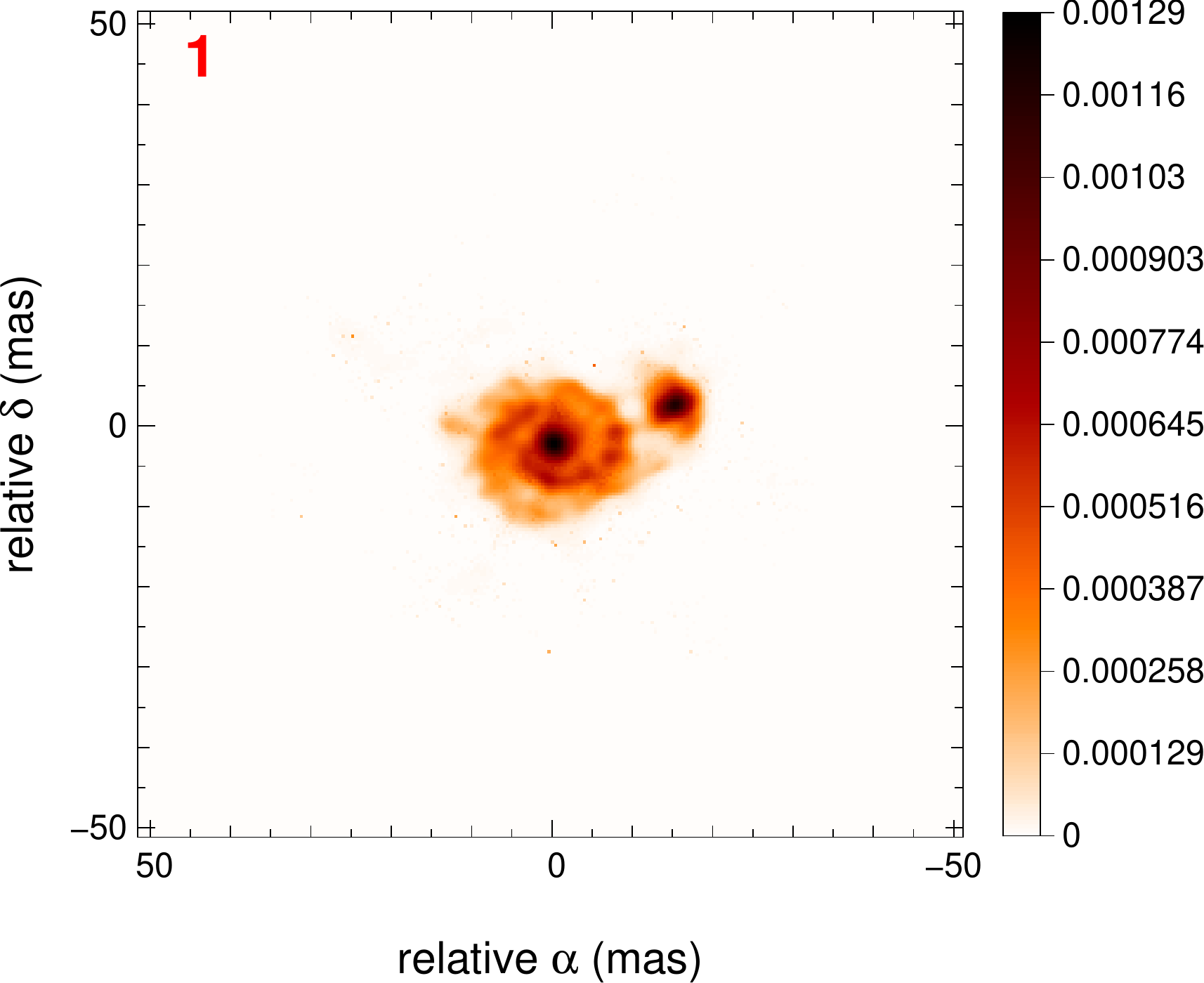}&
      \includegraphics[width=3.4cm]{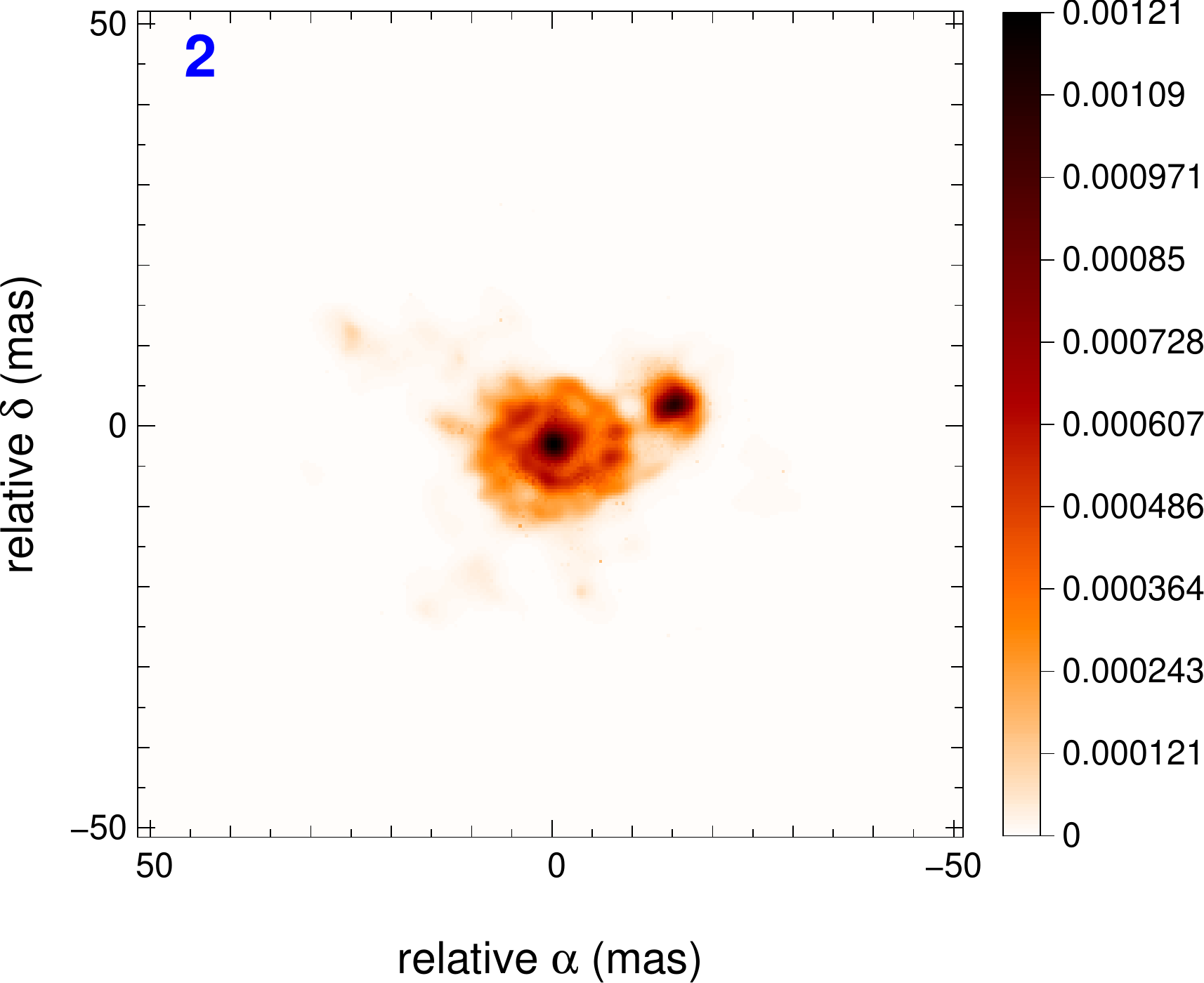}&
      \includegraphics[width=3.4cm]{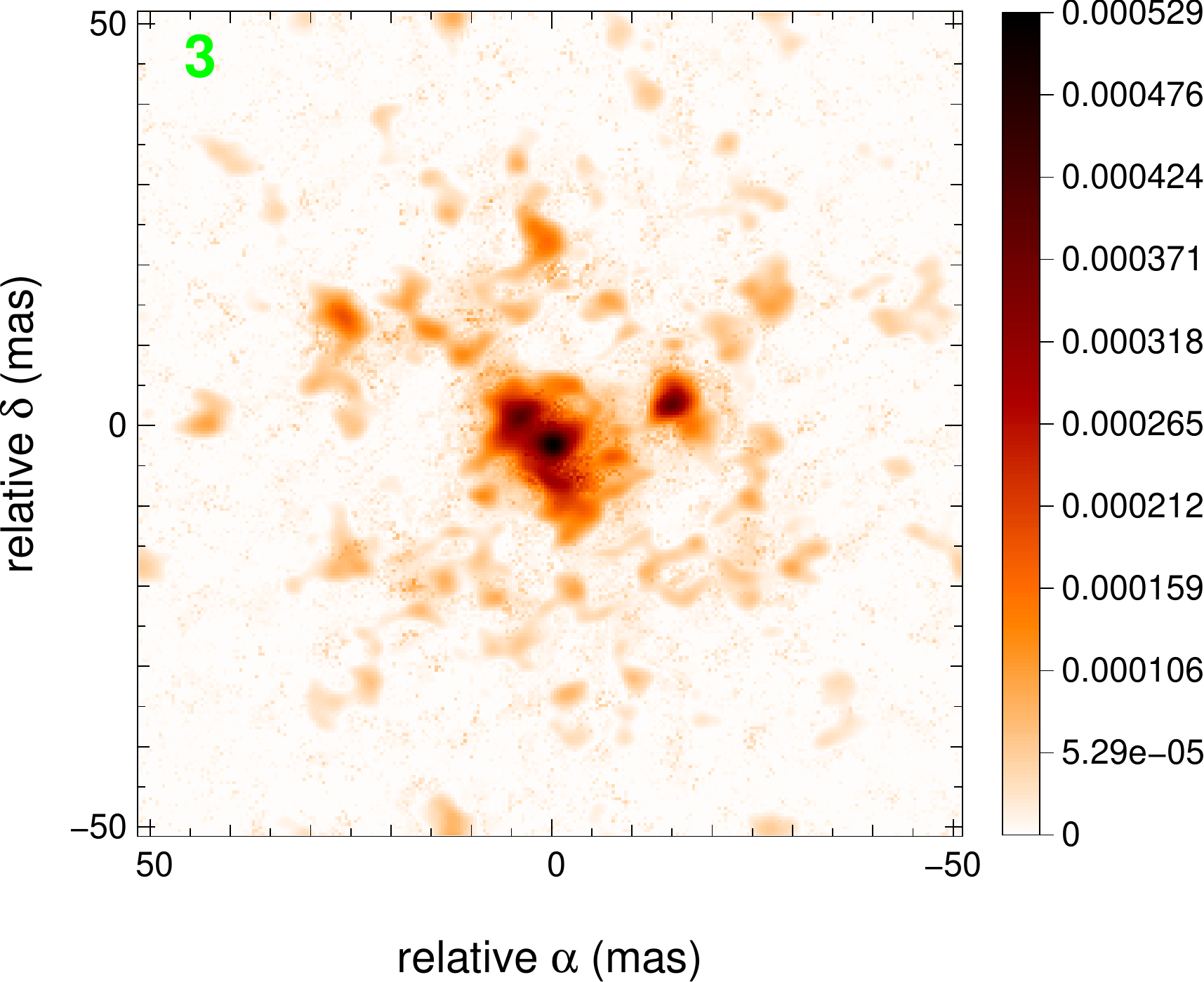}
    \end{tabular}
  \end{minipage}
  \caption{\emph{Left panel} shows a plot of the MSE as a function of the hyperparameter $\mu$. The different colors correspond to different levels of SNR (\emph{red} high, \emph{blue} intermediate, \emph{green} poor). For each curve, the optimal value $\mu^+$ is labeled by a number (\emph{1}, \emph{2}, and \emph{3}). The corresponding images are shown in the \emph{bottom row of the right panel}. The \emph{top row of the right panel} shows three reconstructed images with different values of $\mu$, labeled by a letter on the red curve of the left part (\emph{A} an under-regularized image, \emph{B} the best image, and \emph{C} an over-regularized image). This example is made for the galaxy object and the medium \uv coverage. The regularization is the MEM-prior one.}
  \label{fig:MSEexpl}
\end{figure*}

We first investigate whether there is an optimal regularization weight $\mu$ for a given situation. Therefore, for each object, configuration, SNR level, and regularization, we reconstruct an image for different values of the hyperparameter $\mu$.

The top row of the right panel of Fig.~\ref{fig:MSEexpl} shows the effects of different values of $\mu$, whereas the left panel displays the obtained MSE (see Sect.~\ref{sec:quality-criterion}) for each values of $\mu$. For too small a value of $\mu$, the under-regularized image (labeled with an \emph{A}) has plenty of artifacts. In contrast, for too large a $\mu$, the over-regularized image (labeled with a \emph{C}) is blurred, and many fine features are lost. These two extreme situations correspond to high values of the MSE (A and C), but there is a situation where the MSE reaches a minimum and the image appears to have far fewer artifacts (B).  Visual comparison of the A and C images with the one obtained for B confirms that the MSE is a good criterion to correctly set the regularization weight.

We conclude that there is indeed an optimal value of $\mu$, the one that gives the smallest MSE
\begin{equation}
  \label{eq:best-mu}
  \mu^+ = \argmin_\mu \Norm{\VParam^\Tag{rec}_\mu - \VParam^\Tag{ref}}^2 \, ,
\end{equation}
where $\VParam^\Tag{rec}_\mu$ is the image reconstructed with a regularization weight set to $\mu$. As the minimum of the curve is quite flat, the optimal value of $\mu$ is not precisely defined but may vary by a factor as large as either two or three with a negligible influence on the result.

This procedure cannot be used in practice because the true image is obviously unknown.  Nevertheless, this procedure allows us to define the most accurate image that can be reconstructed given the data and the type of regularization. \textbf{In the following analysis, the reconstructed images are always obtained with $\mu=\mu^+$}.


\subsection{Dependence of $\mu^+$ on the MSE quality criterion}
\label{sec:MSE_obj}

\begin{figure*}
  \centering
  \begin{tabular}{cc}
    \includegraphics[width=8.5cm]{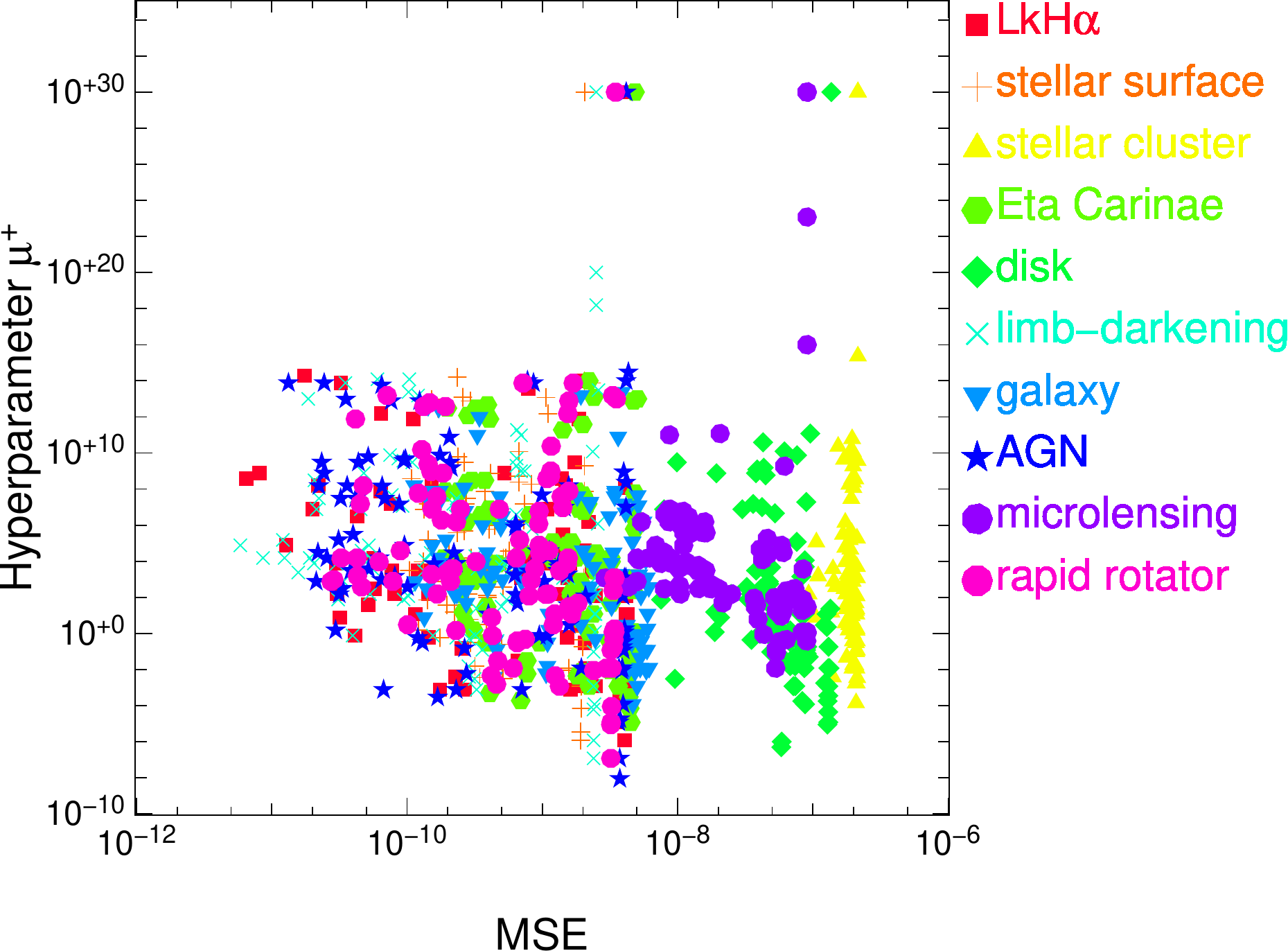}&
    \includegraphics[width=8.5cm]{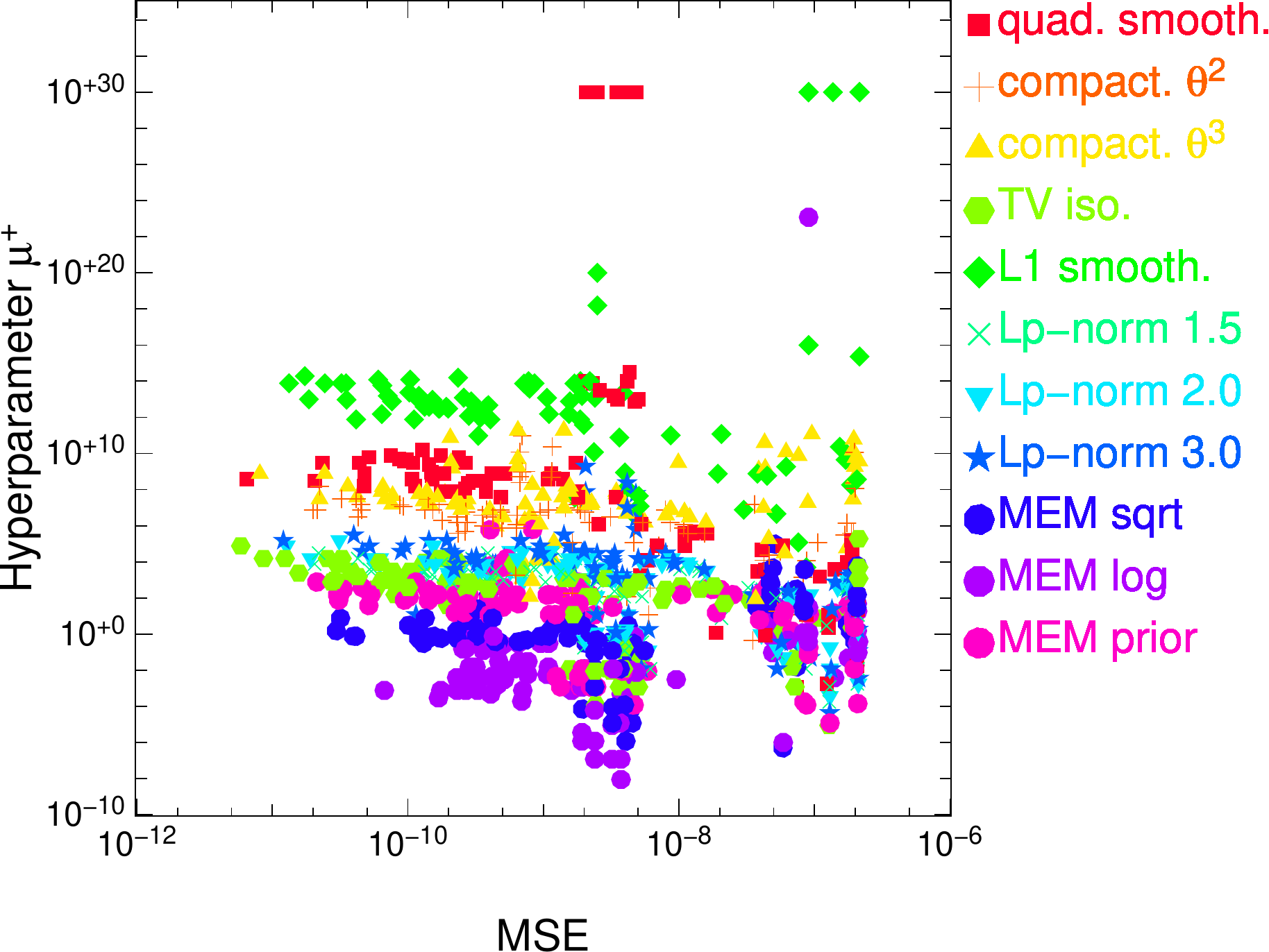}\\
    \includegraphics[width=8.5cm]{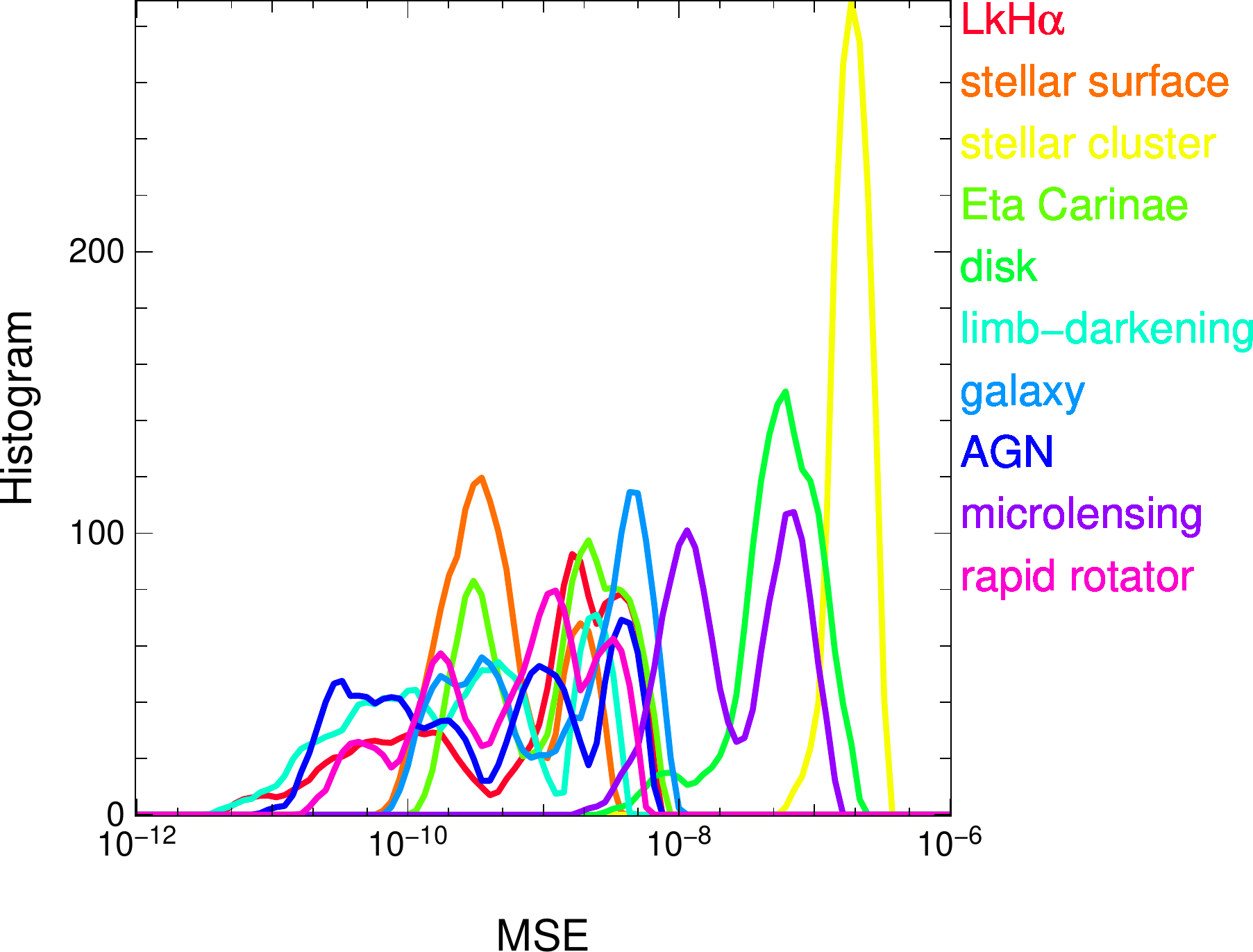}&
    \includegraphics[width=8.5cm]{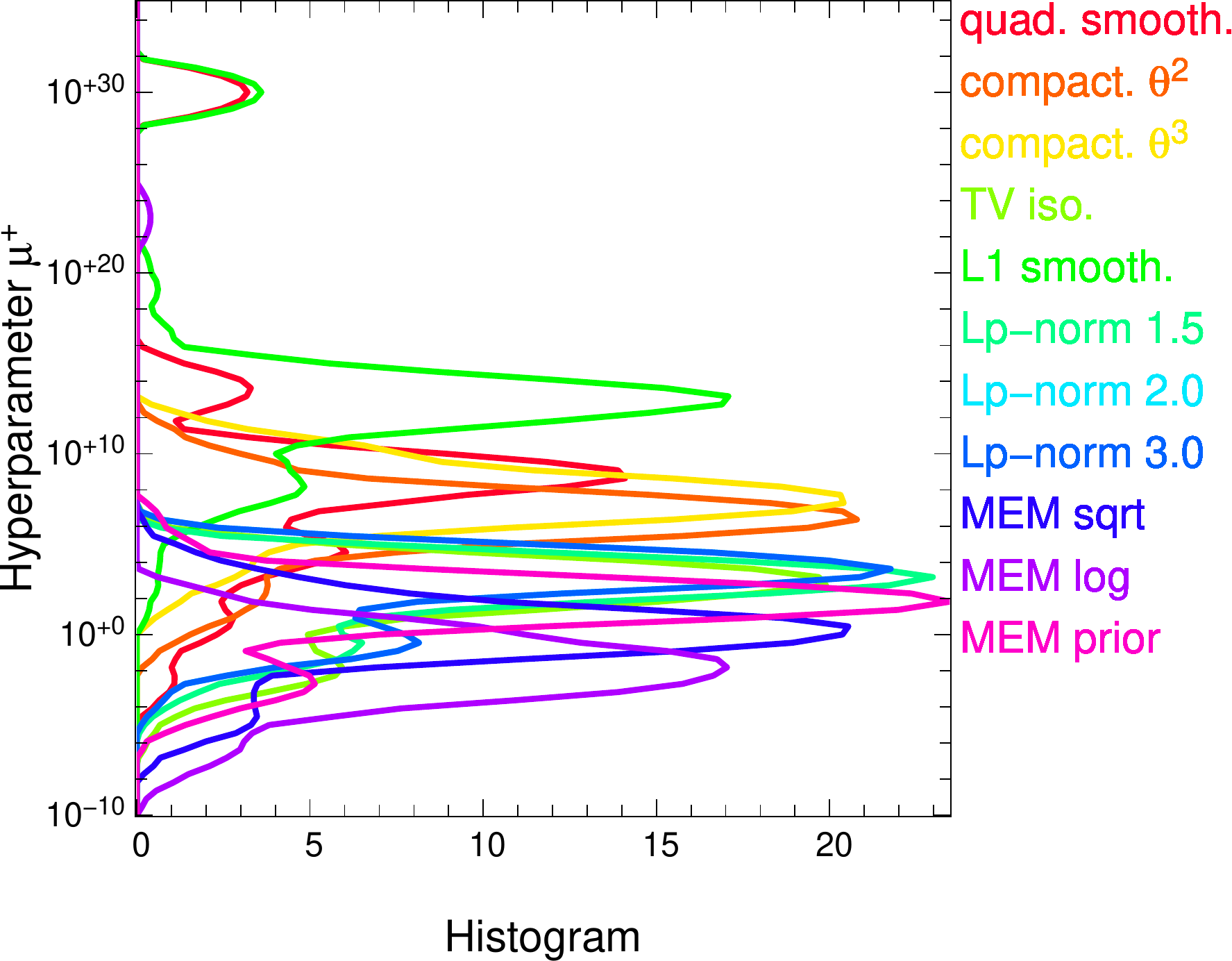}\\
  \end{tabular}
  \caption{ \emph{Top row}: Scatter plots representing the optimal value of the hyperparameter $\mu^+$ as a function of the MSE of the images.  \emph{Bottom row}: The corresponding histograms of MSE and $\mu^+$.  \emph{Left column}: The colors and symbols indicate the different classes of objects.  \emph{Right column}: The colors and symbols indicate the different regularization classes. }
  \label{fig:MSEvsMU}
\end{figure*}

We wish to determine the type of relationship that links the optimized regularization factor $\mu^+$ to the MSE in order to detect any trends. The scatter plot in \Fig{fig:MSEvsMU} reports the values of $\mu^+$ and MSE obtained for each simulation. In the left panel, the different colors and symbols correspond to different objects. In the right panel, the different colors and symbols correspond to different regularizations. In the bottom row, we  present our computed histograms of MSE (left) and $\mu^+$ (right). As in the rest of the paper,  the plotted histograms are approximations of the probability density functions (PDF) of our  results. These curves were computed from our samples following the optimal method described in  \citet{Scott1992}.

In the left part of \Fig{fig:MSEvsMU}, the different colors representing the objects seem to be aligned vertically. We therefore conclude that the MSE depends mostly on the structure of the observed object.

\begin{figure}
  \resizebox{\hsize}{!}{
    \includegraphics{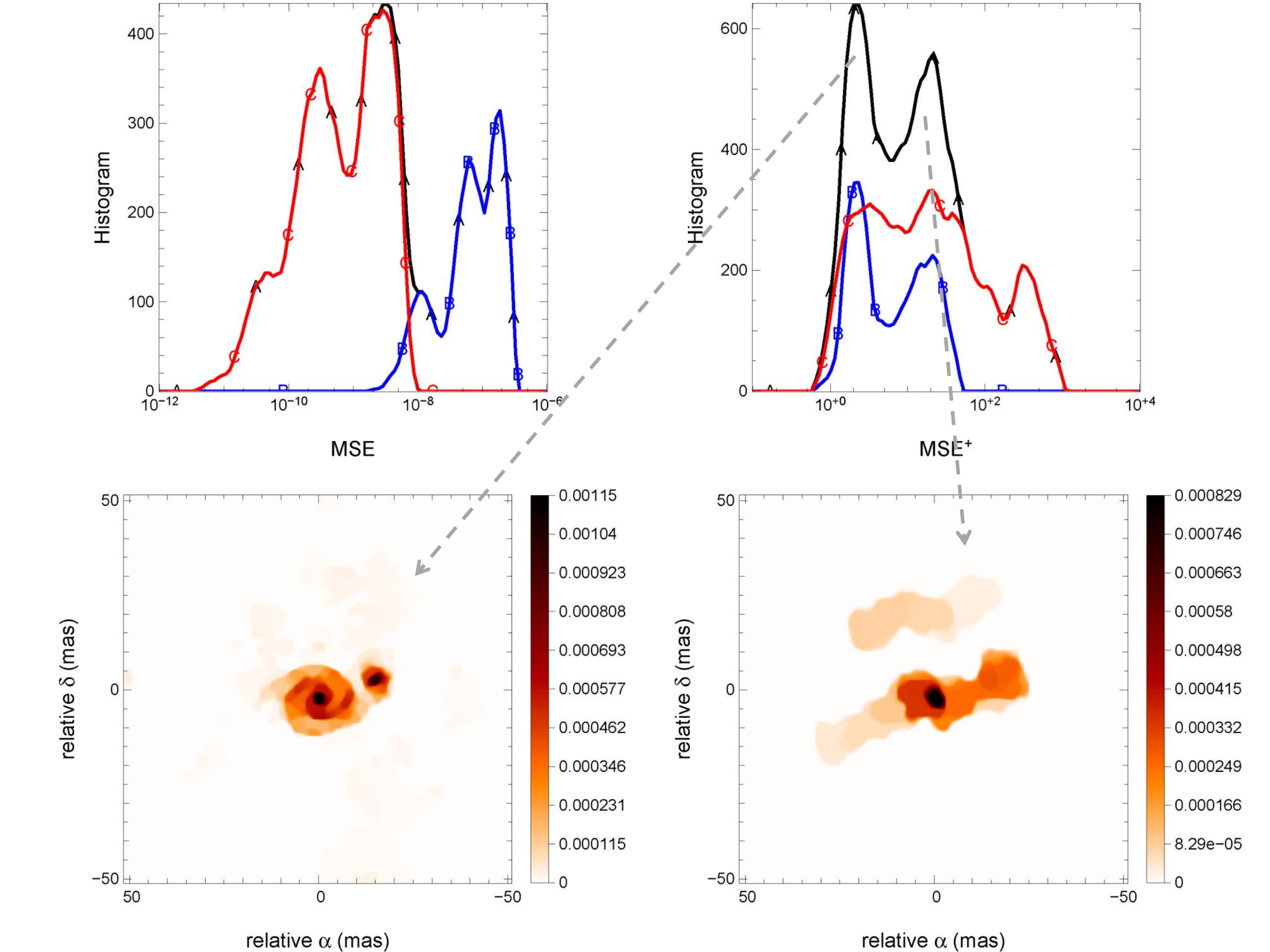}}
  \caption{\emph{Upper row:} Distribution of the MSE (\emph{left}) and the MSE$^+$ (\emph{right}). The colors and letters represent the two classes of objects: \emph{blue/B} for the objects with very compact structures, \emph{red/C} for the others. The total distribution is shown in the \emph{black/A} curve. \emph{Bottom row:} example of reconstructed images for the good (\emph{left}) and bad (\emph{right}) MSE$^+$ peak.}
  \label{fig:MSEpdf}
\end{figure}

More precisely, two classes of objects can be distinguished in the distribution of MSE in the left panel of \Fig{fig:MSEvsMU}. We therefore grouped together the objects with similar behaviors in the top panels of \Fig{fig:MSEpdf}: (i) the objects with very compact sources, \ie the star cluster, the protoplanetary disk and the microlensing (curve in blue labeled B), and (ii) the other objects with extended structures (curve in red labeled C). The MSE is systematically higher for objects of the first class.

\begin{figure}
  \centering
  \resizebox{\hsize}{!}{
    \includegraphics{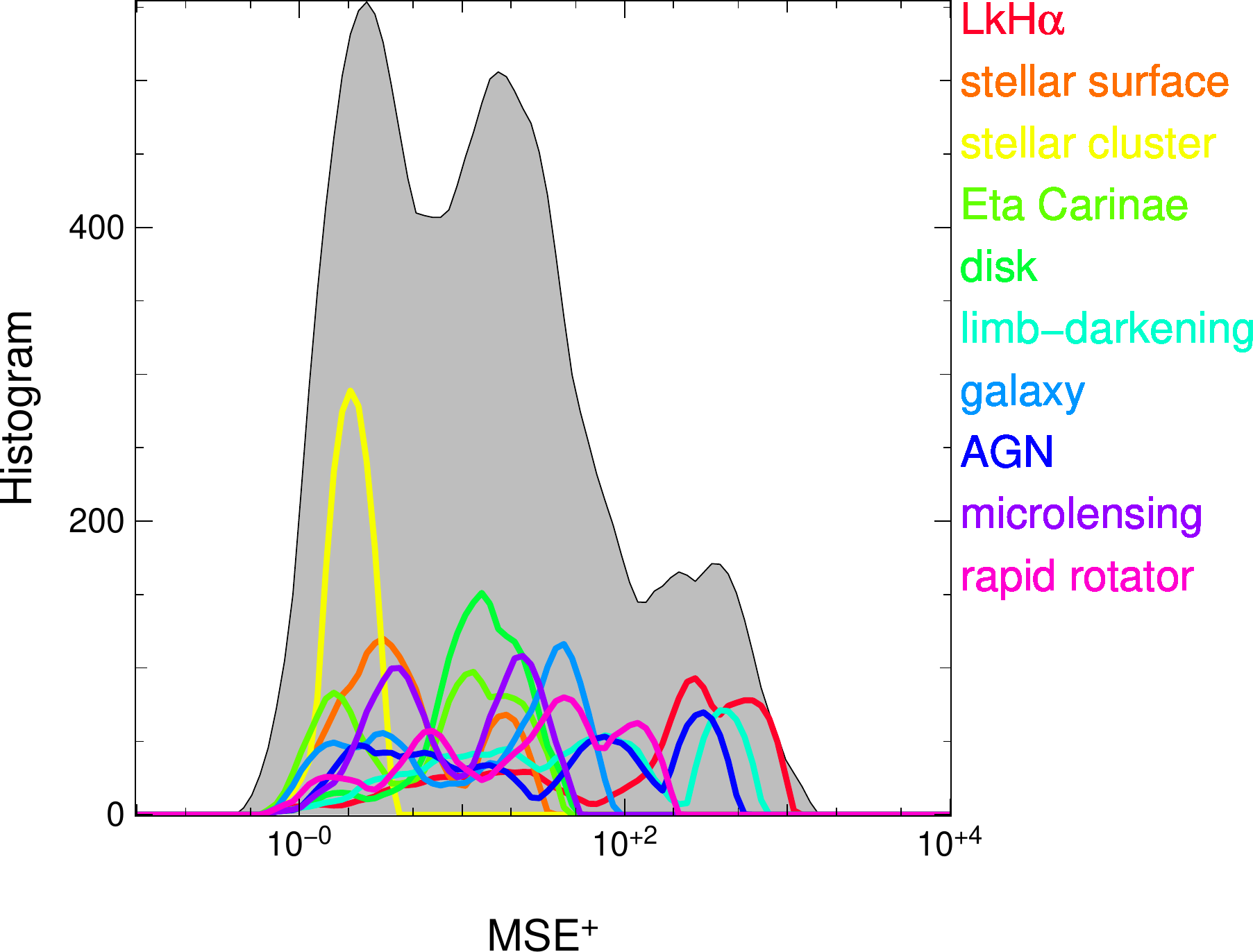}
    \includegraphics{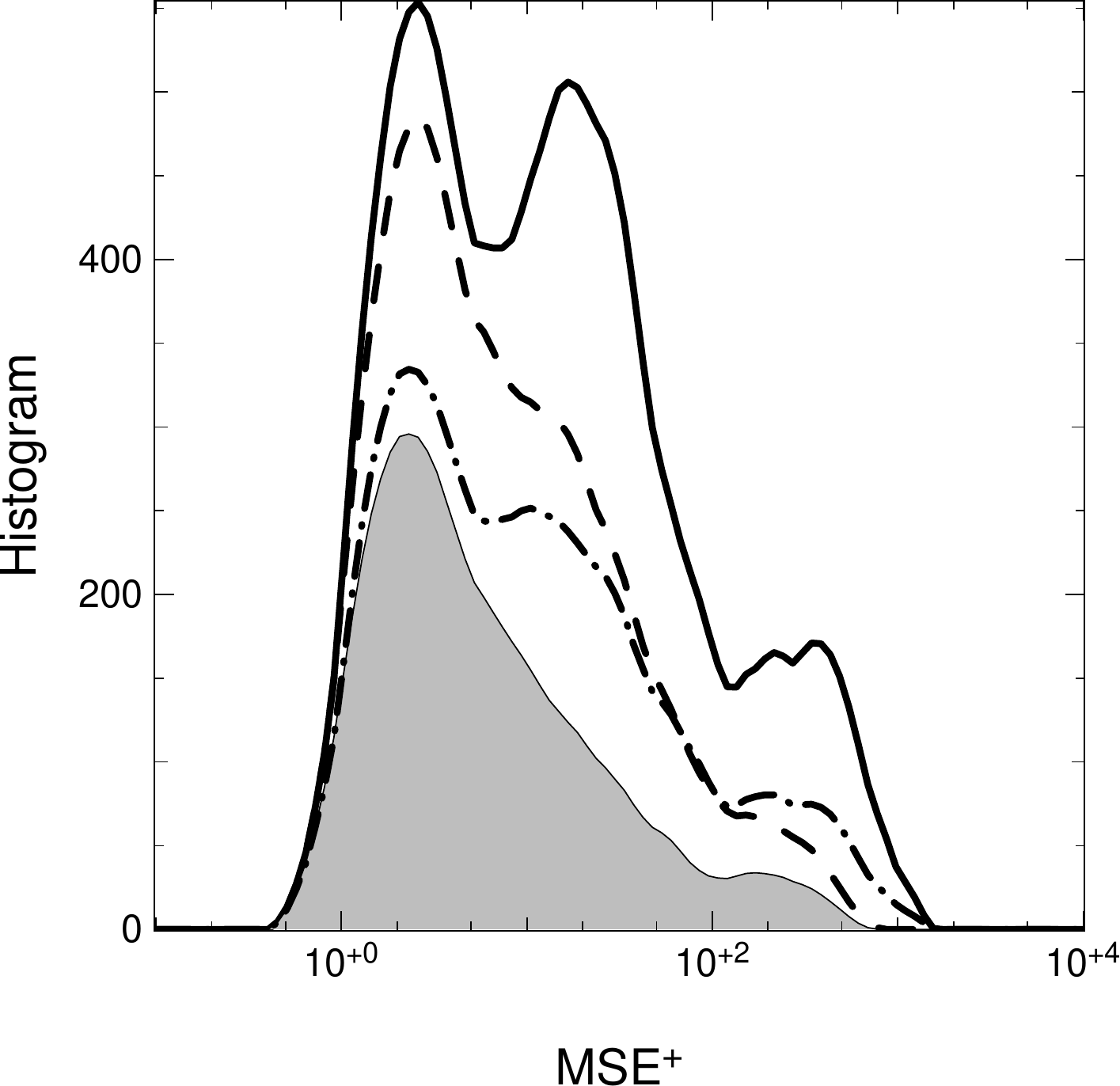}}
  \caption{Distribution of MSE$^+$.  \emph{Left}: histograms of MSE$^+$ for different objects in different colors; the \emph{gray zone} corresponds to the total distribution, all objects confounded. \emph{Right}: \emph{solid line}, all the configurations and regularizations are kept; \emph{dashed line}, with the sparsest \uv coverage removed; \emph{dot-dashed line}, with the bad regularizations removed; \emph{in gray zone}, with the sparsest \uv coverage and bad regularizations removed. }
  \label{fig:data-cleaning}
\end{figure}

Following these observations, we try to find a way of renormalizing the MSE as a function of the object. To join the curves together, we define a new MSE criterion, called MSE$^+$, by dividing each MSE by the smallest MSE for each object separately. As expected, this normalization cancels the different object classes, as seen in the right part of \Fig{fig:MSEpdf}. Two peaks clearly appear on the graph, distinguishing the good reconstructions (left peak) from the bad ones (right peak). An example of each case is shown on the bottom part of \Fig{fig:MSEpdf}. The low quality reconstructions are caused by bad configurations (not enough \uv points, low SNR) and/or bad regularizations as it can be seen in the right part of Fig.~\ref{fig:data-cleaning}  and will be described in the next sections. To be sure that the peaks represent the good and bad  reconstructions and do not come from different object's types, we verify that each object is  represented in each peak, as shown in the left part of Fig.~\ref{fig:data-cleaning}.

By visual inspection, we assess that the value of the MSE$^+$ leads to a correct ordering of the images reconstructed for a given object when the other settings change (data quality, type of regularization, \etc): the lower the MSE$^+$, the higher the quality of the image is. Other attempts at the renormalization of the MSE are explained in Appendix~\ref{app:mse-norm}.

Now that a good quality criterion has been defined and that the optimal value of $\mu$ have been obtained, we can study the other parameters.


\subsection{Limits due to the \uv coverage and the SNR}
\label{sec:limits_uv-snr}

\begin{figure*}
  \centering
    \begin{tabular}{cc}
      \includegraphics[width=8.5cm]{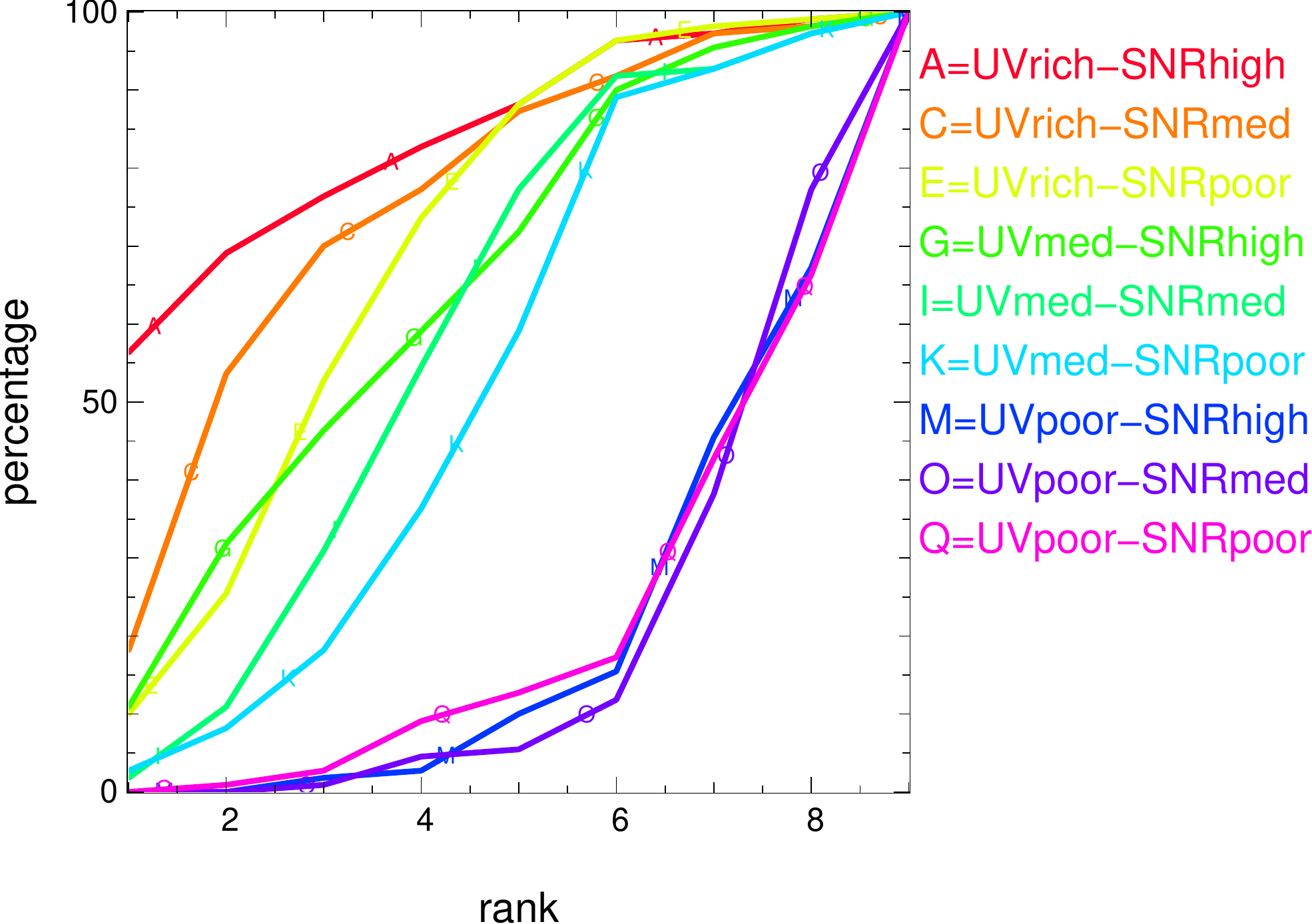}
      \includegraphics[width=8.5cm]{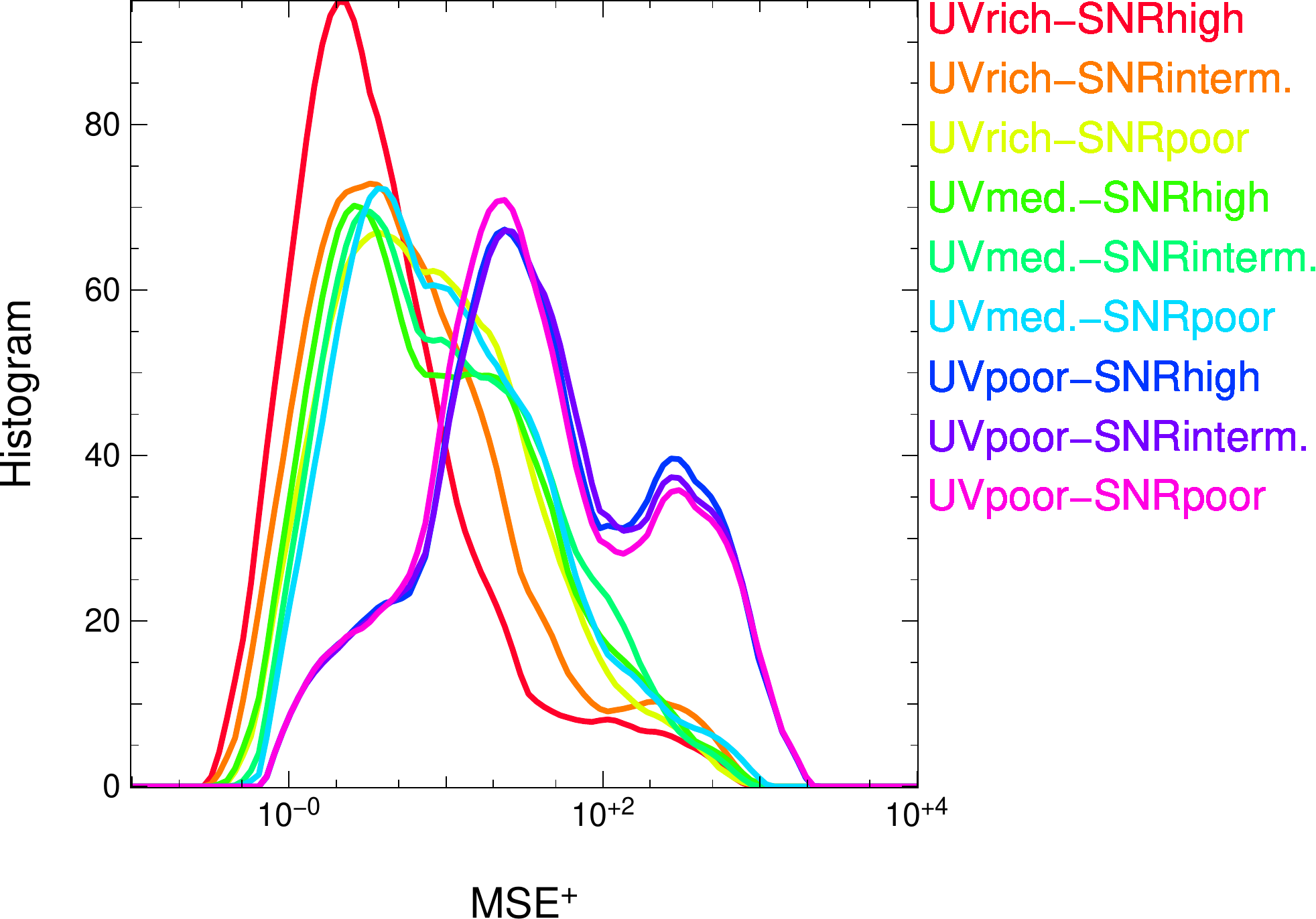}
   \end{tabular}
   \caption{\emph{Left:} cumulative distributions of the ranks reached by the different configurations of \uv coverage and SNR. \emph{Right:} the histograms of the MSE$^+$ for different configurations of \uv coverage and SNR represented by different colors. }
   \label{fig:UV-SNR_classif}
\end{figure*}

In this section, we classify the observational configurations (\uv coverage and SNR) on the basis of the MSE$^+$. For each data set (unique combination of object and regularization), we order the pair [\uv coverage, SNR] according to the value of MSE$^+$ they reach, giving them a rank from 1 for the best configuration (lowest MSE$^+$) to 9 for the worst (highest MSE$^+$). In the left panel of Fig.~\ref{fig:UV-SNR_classif}, we display the cumulative distributions of the ranks reached by every configuration, determining how many times a given configuration reaches at least the first rank, the second rank, \etc The highest quality configurations are the ones in the upper-left part of the plot.

The poor \uv coverage combined with any value of SNR is clearly too sparse to reconstruct good images.  While acceptable reconstruction is possible for any considered SNR when there are enough samples in the \uv coverage.  We deduce that there is a minimal \uv coverage needed to perform image reconstruction, whereas there is no such limit set by the tested SNR.  However, when the \uv coverage is sufficient, the higher the SNR or the more filled the \uv coverage, the higher the quality of the reconstructed image is. The bottom row of the right panel of Fig.~\ref{fig:MSEexpl} shows how the visual quality of the optimal image depends on the SNR. As expected, the higher the SNR, the better the reconstructed image is.

Figure~\ref{fig:UV-SNR_classif} (right) shows the histogram of the MSE$^+$ for different  configurations of \uv coverage and SNR. It indicates that the success image reconstruction is more influenced by the amount of data than by the SNR: the MSE$^+$ is lower for a rich \uv coverage with a poor SNR than for a medium \uv coverage with an intermediate SNR. We note that all the tested \uv coverage are uniform, but we expect that the amount of data has to be sufficient and also \emph{homogeneously spread} in the \uv plan.

After the removal of the sparsest coverage, the MSE$^+$ distribution is shown in  \Fig{fig:data-cleaning} in dashed line. There is still a little bump of bad MSE$^+$ caused by bad regularizations, as discussed in the next section.


\subsection{Quality of the regularizations}
\label{sec:MSE_rgl}

\begin{figure*}
  \centering
  \begin{tabular}{ccc}
    \includegraphics[width=5.5cm]{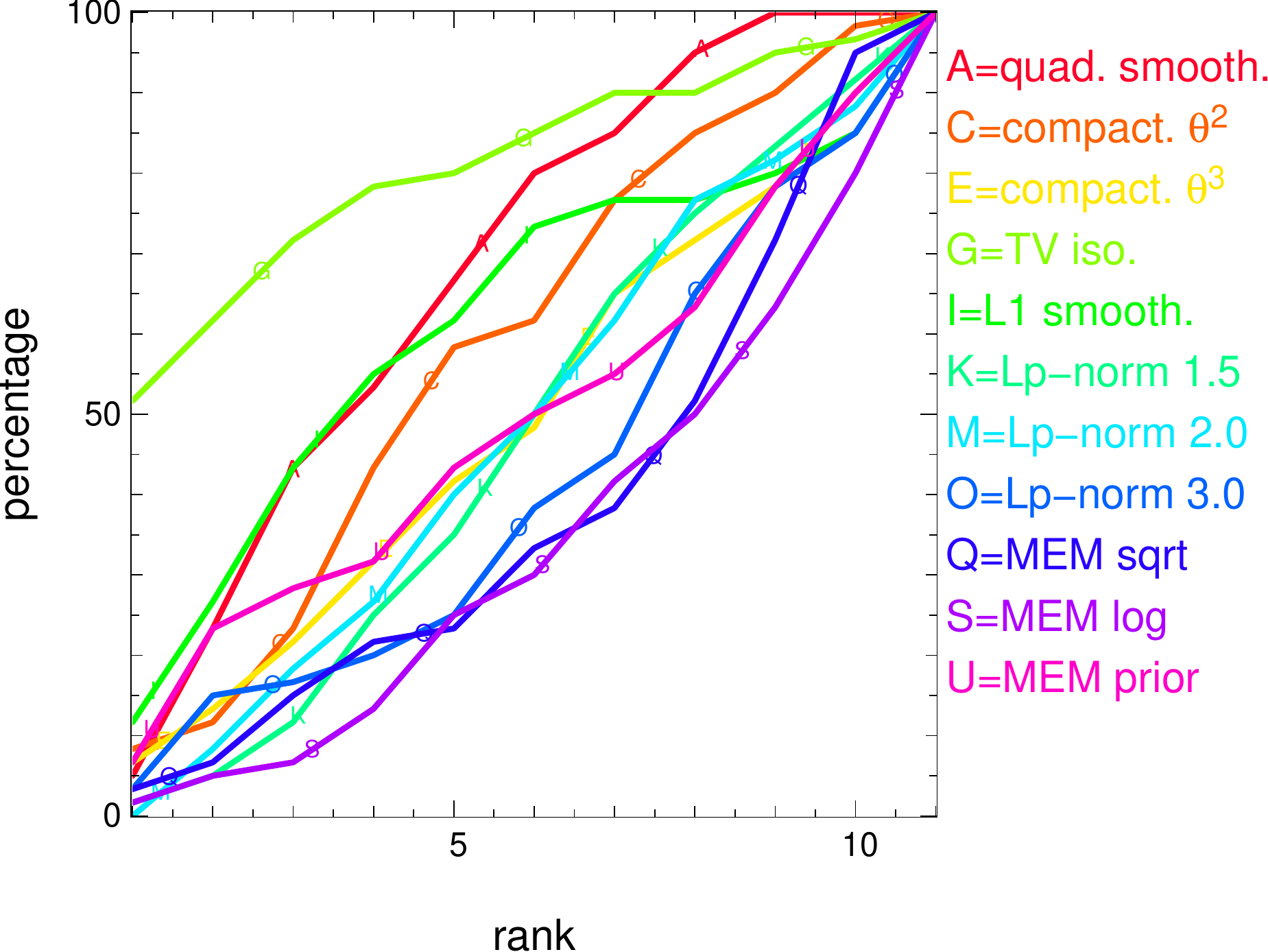}&
    \includegraphics[width=5.5cm]{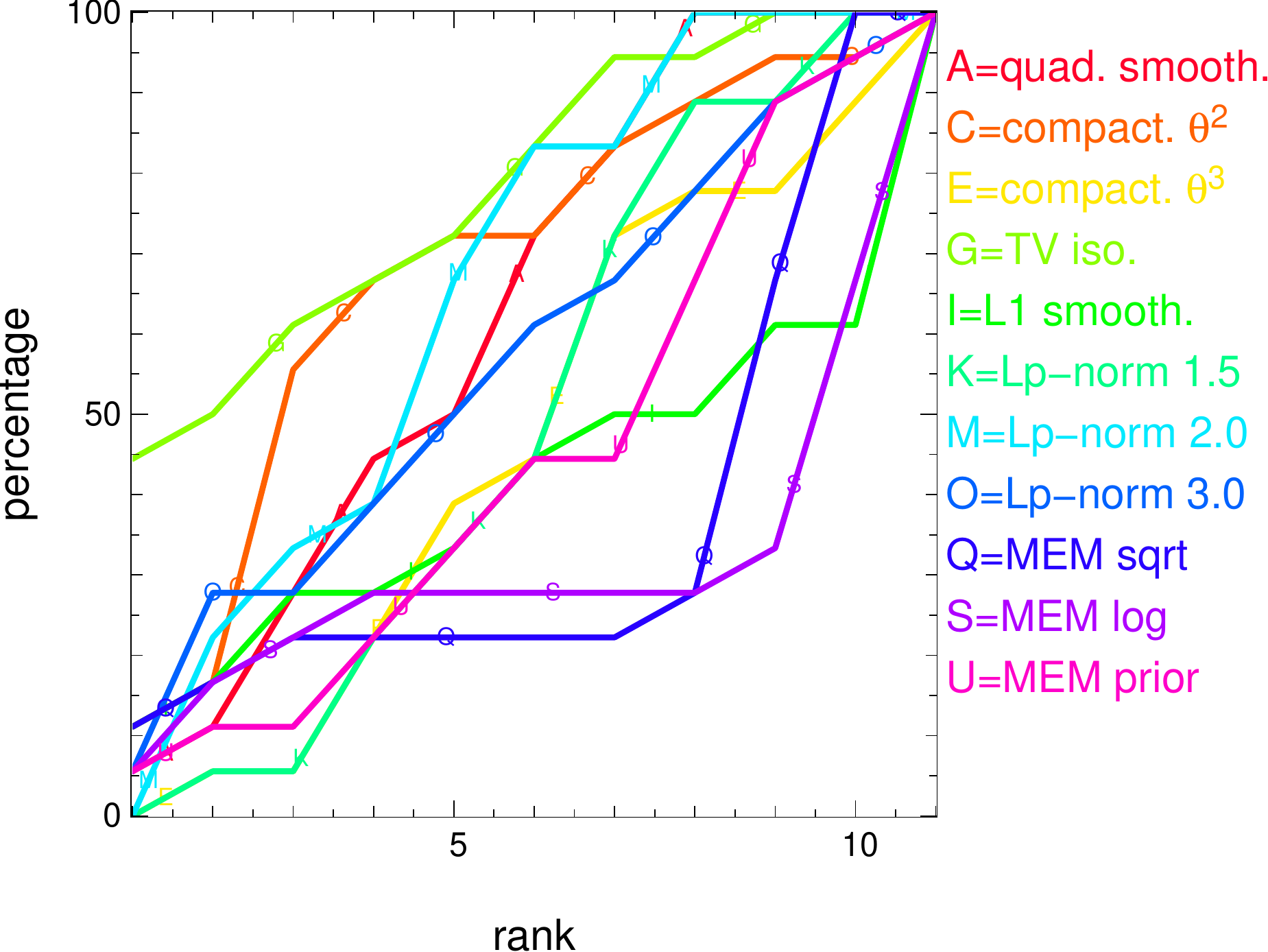}&
    \includegraphics[width=5.5cm]{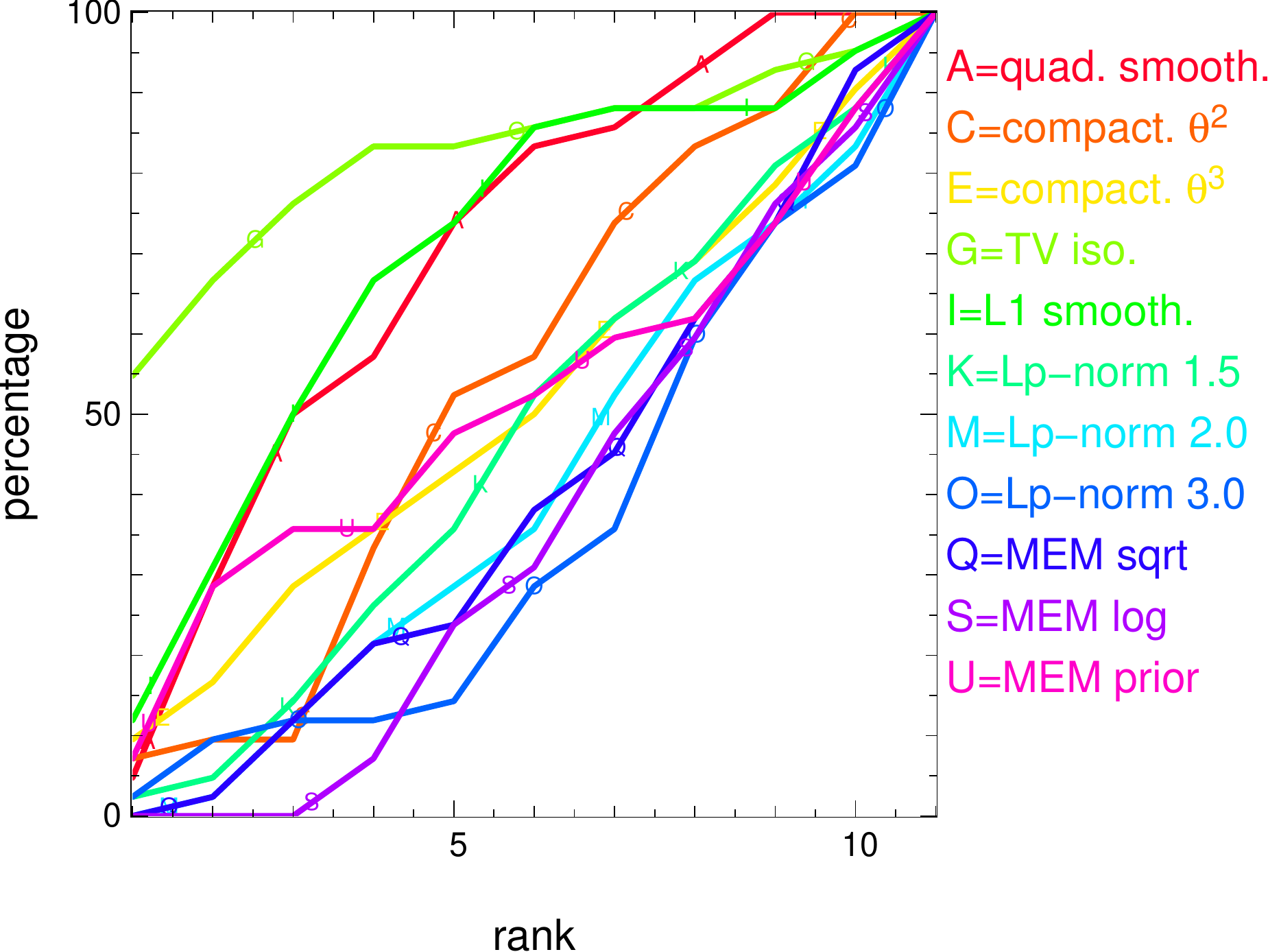}\\
  \end{tabular}
  \caption{Cumulative distributions of the ranks reached by the regularizations. \emph{Left:} all objects; \emph{Middle:} objects with very small structures; \emph{Right:} other objects. }
  \label{fig:Rgl_cum}
\end{figure*}

\begin{figure*}
  \begin{tabular}{cccc}
    \includegraphics[width=4.25cm]{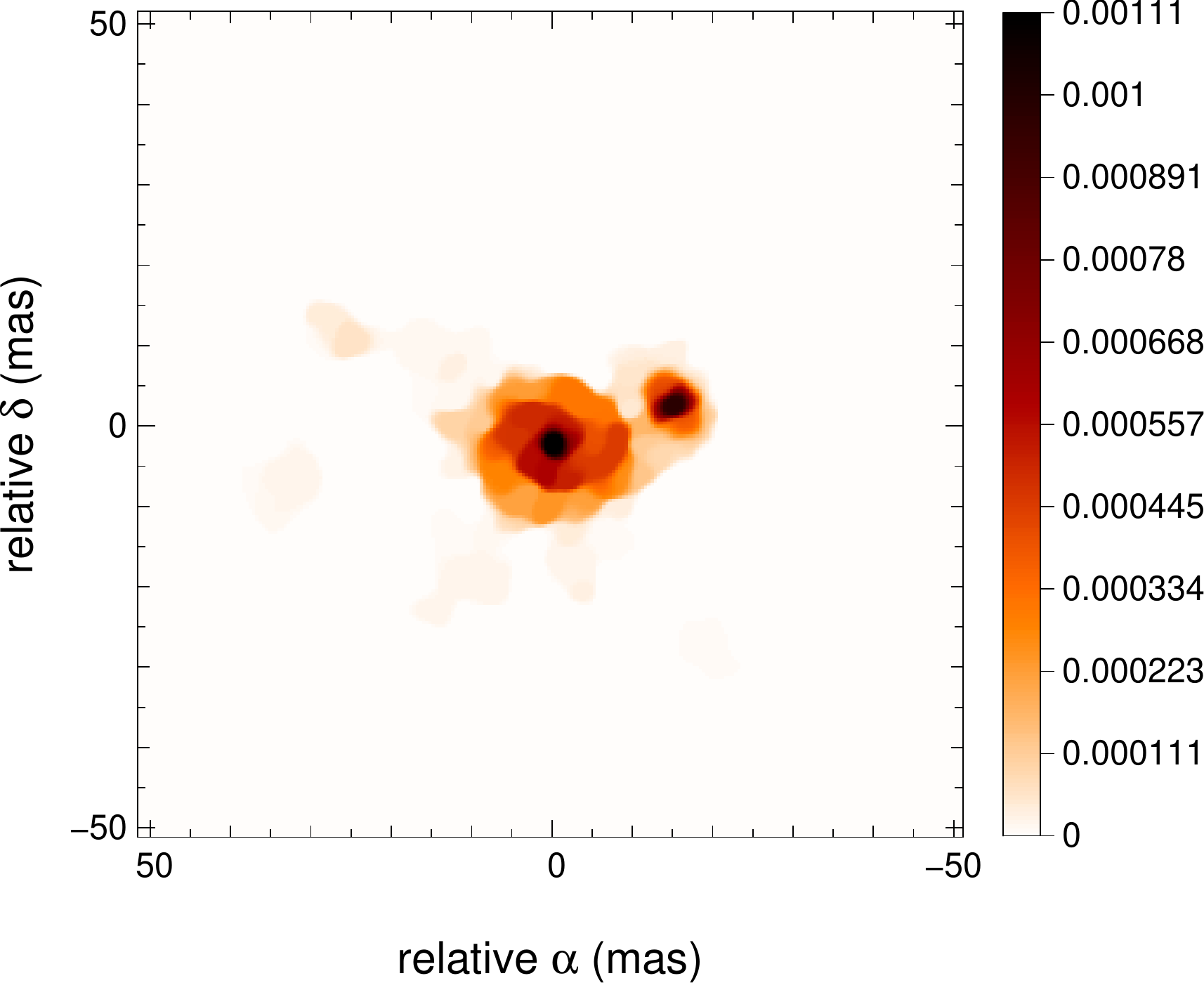}&
    \includegraphics[width=4.25cm]{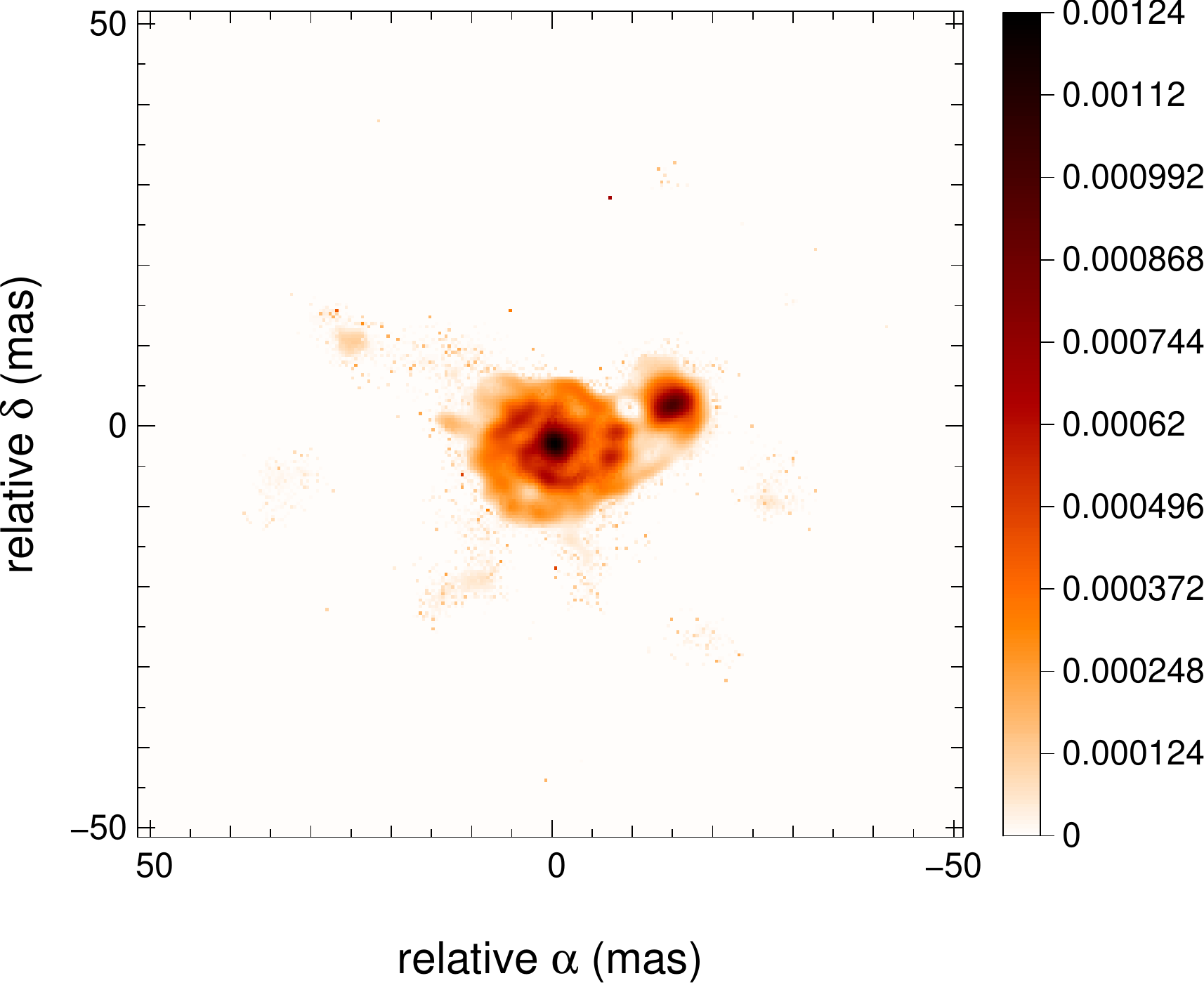}&
    \includegraphics[width=4.25cm]{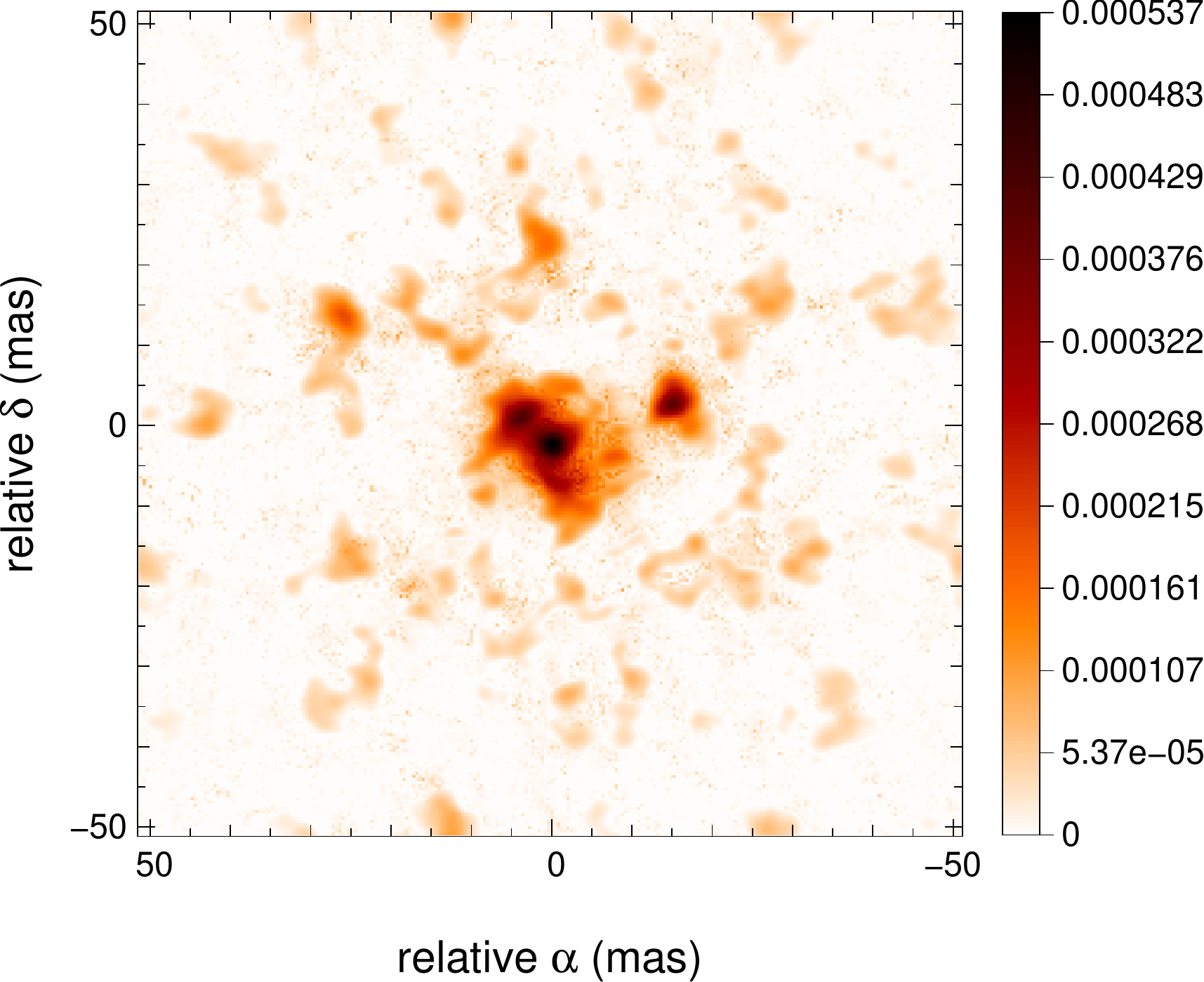}&
    \includegraphics[width=4.25cm]{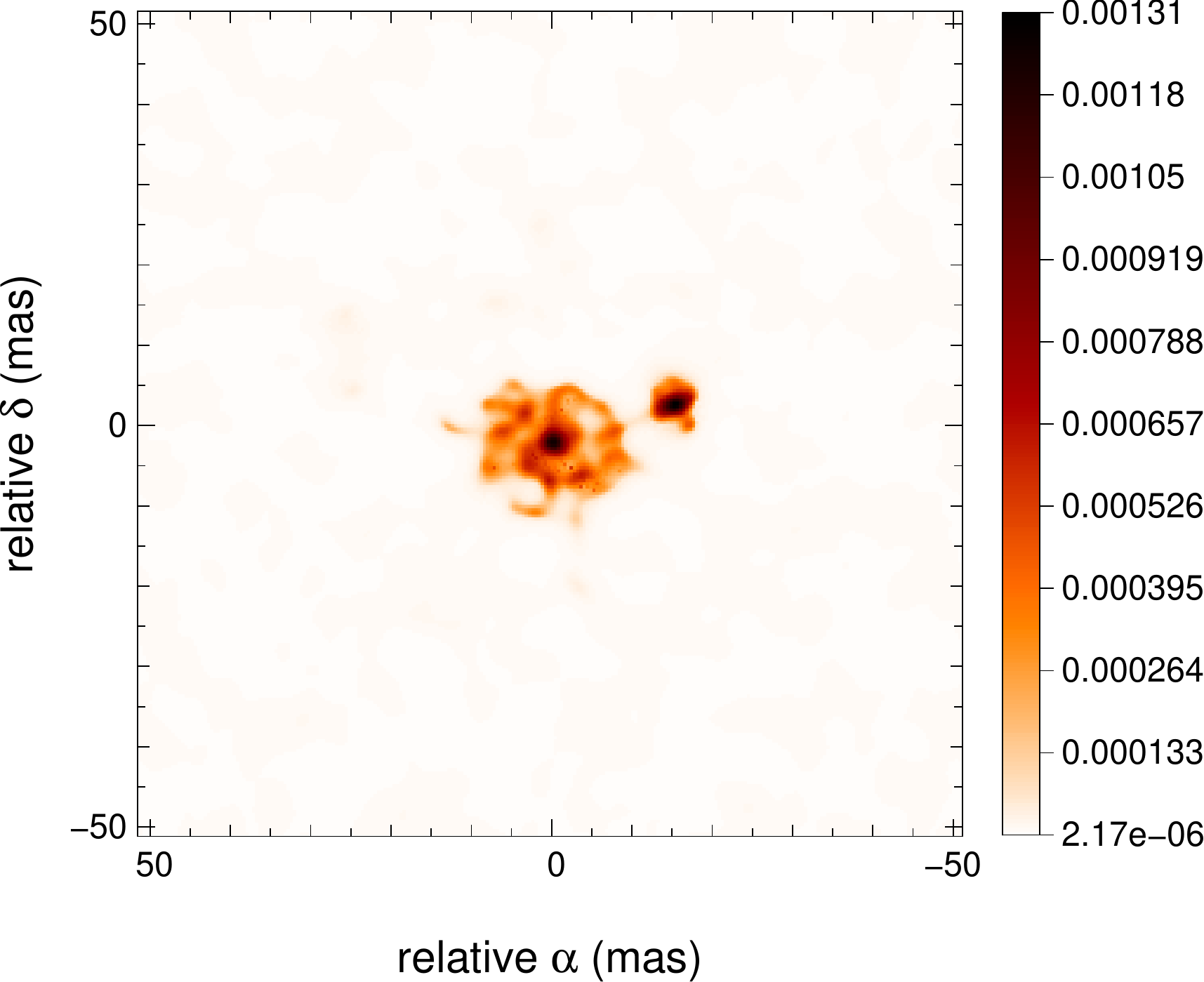}\\
  \end{tabular}
  \caption{Example of reconstructed images of the galaxy image with the medium \uv coverage and the intermediate SNR for different regularizations. \emph{From left to right}, TV, compactness $\theta^2$, $\ell_p$ norm with $p=2$, MEM-log.}
  \label{fig:Rgl-expl}
\end{figure*}

Using the same principles as in the previous section, we classify the regularizations as a function of the MSE$^+$ for different realizations. Figure~\ref{fig:Rgl_cum} shows the corresponding cumulative distributions.  Isotropic TV appears to be the most successful regularization for the two classes of objects. The compactness prior with  $\Weight^\Tag{prior}_n = \abs{\VDirn_n}^2$ is the second highest quality choice for the very compact objects. The worst regularizations are the $\ell_p$-norm, the MEM-sqrt, and the MEM-log, as expected and explained in Sect.~\ref{sec:fprior}. Reconstructions for good and bad regularizations are illustrated in \Fig{fig:Rgl-expl}.

In the MSE$^+$ distribution after the elimination of the bad regularizations (see \Fig{fig:data-cleaning} in dot-dashed line), there is still much evidence of lower quality MSE$^+$. We conclude that the bad MSE$^+$ are mainly caused by the sparsest \uv coverage. However, both have to be eliminated to obtain a cleaned sample of reconstructed images.

\emph{In what follows, we made a selection and only retained the regularizations (quadratic and $\ell_1$ smoothness, compactness, isotropic TV and MEM with a Gaussian prior) and the \uv coverages (rich and medium) that lead to correct reconstructed images.  We kept data for all values of SNR.}


\subsection{Predetermined value of the hyperparameter $\mu^+$?}
\label{sec:mu_analyze}

In a fully Bayesian framework, the data noise and the variability of the object are independent. We therefore expect that $\mu\Fprior$ does not depend on the data \citep{Tarantola-2005-inverse_problem_theory}.  As a result, for a given type of object, the optimal value for $\mu$ should be the same regardless of  the SNR or the \uv coverage.  However, we are not in a truly Bayesian framework because the regularizations are derived from general considerations that must help us to solve for the degeneracies of the problem and are not really derived from the statistics of the brightness distributions of the observed objects. Since the degeneracies of the problem are mostly due to the sparsity of the \uv coverage, this observational parameter may have some influence on the tuning of the regularization weight. Our expectation is thus that, for a given type of object, a quasi optimal value of $\mu^+$ can be derived from simulations and from simple considerations to scale this parameter if different image reconstruction settings are used.

\begin{figure*}
  \centering
  \begin{tabular}{cc}
    \includegraphics[width=8cm]{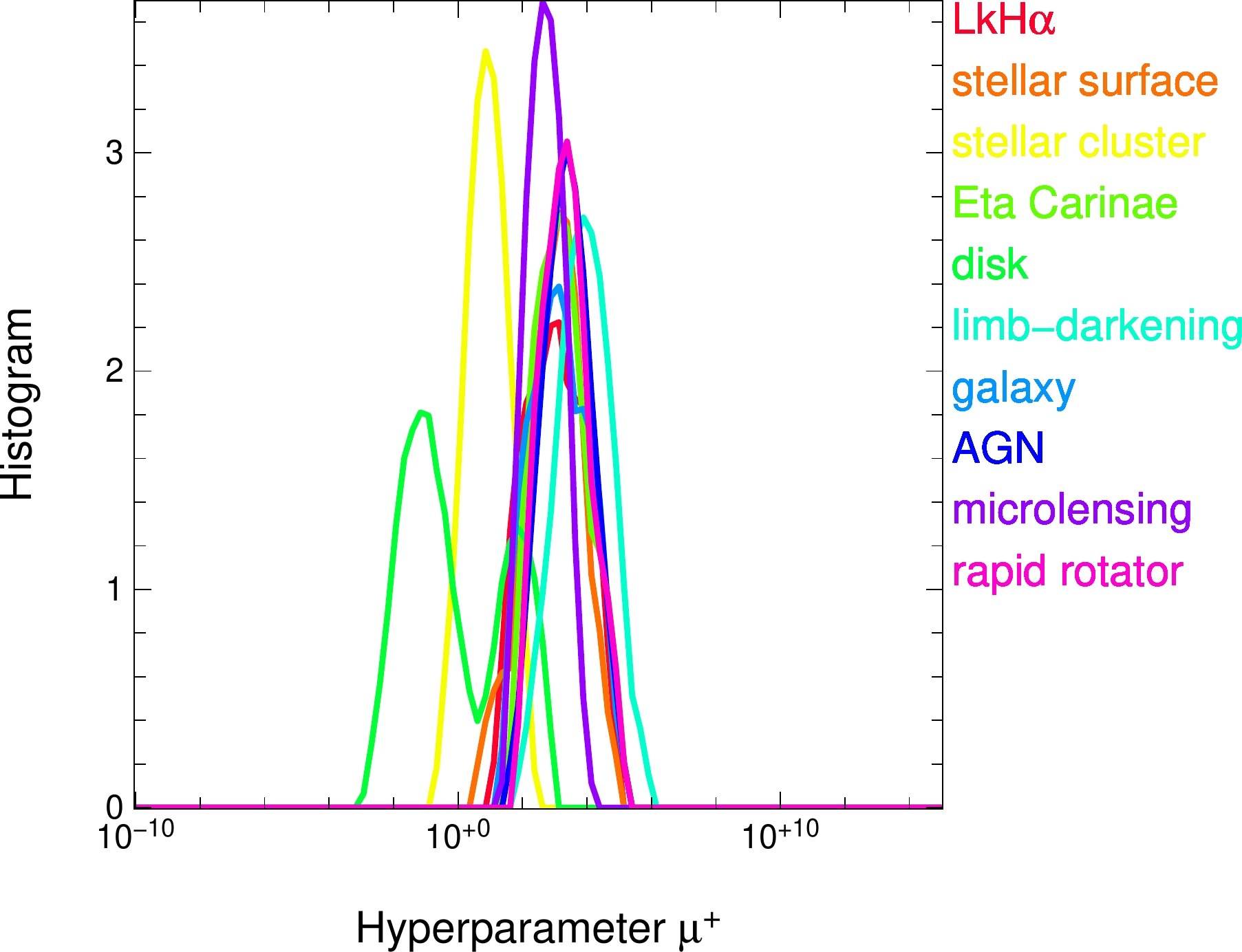} &
    \includegraphics[width=9cm]{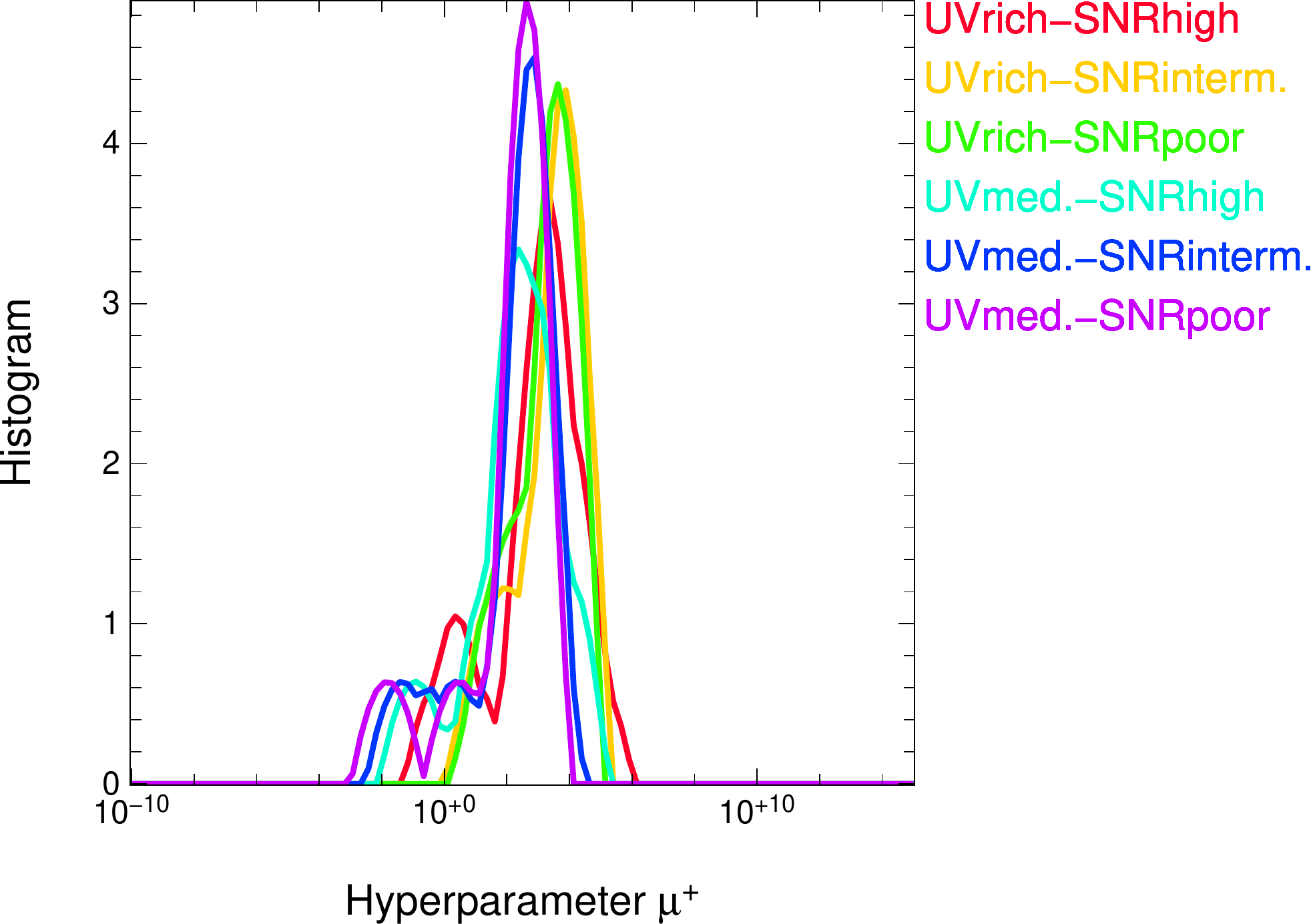}\\
  \end{tabular}
  \caption{Histogram of $\mu^+$ for the TV regularization. \emph{Left:} the colors correspond to different objects. \emph{Right:} the colors correspond to different pair [\uv coverage--SNR].}
  \label{fig:MUpdf}
\end{figure*}

The hyperparameter $\mu^+$ depends mostly on the regularizations, as seen in the right part of \Fig{fig:MSEvsMU}, where the colors, representing the regularizations, appear aligned horizontally and the pre-eminent peaks of the histograms are separated. Moreover, the hyperparameter $\mu$ appears to be quasi independent of the SNR and the \uv coverage, as expected. As seen in  \Fig{fig:MUpdf} for the TV regularization, if there is still a variation in the value of $\mu$, it depends on the object morphology but not on the \uv plane and the SNR value.

This finding allows us to link each regularization to a mean value of the hyperparameter $\mu$. This mean value gives a useful start point for the user (see \Tab{tab:mu_value}). The equations to rescale this value in the case of different image settings are given in Appendix~\ref{app:scaling-hyperparameter}.

\begin{table}
  \centering
  \caption{Table of the mean values of $\mu$ for each regularization.}\label{tab:mu_value}
  \begin{tabular}{ll}
    \hline
    Regularizations & $\mu^+$ \\
    \hline
    \hline
    Quadratic smoothness & $10^9$ \\
    Compactness $r^2$ & $10^7$ \\
    Compactness $r^3$ & $10^7$ \\
    TV isotropic & $10^3$ \\
    $\ell_1$ smoothness & $10^{13}$ \\
    MEM with prior & $10^2$ \\
    \hline
 \end{tabular}
\end{table}
%


\subsection{How different noise realizations affect the MSE and the optimal $\mu$?}
\label{sec:noise_analyze}

\begin{figure}
  \resizebox{\hsize}{!}{
    \includegraphics{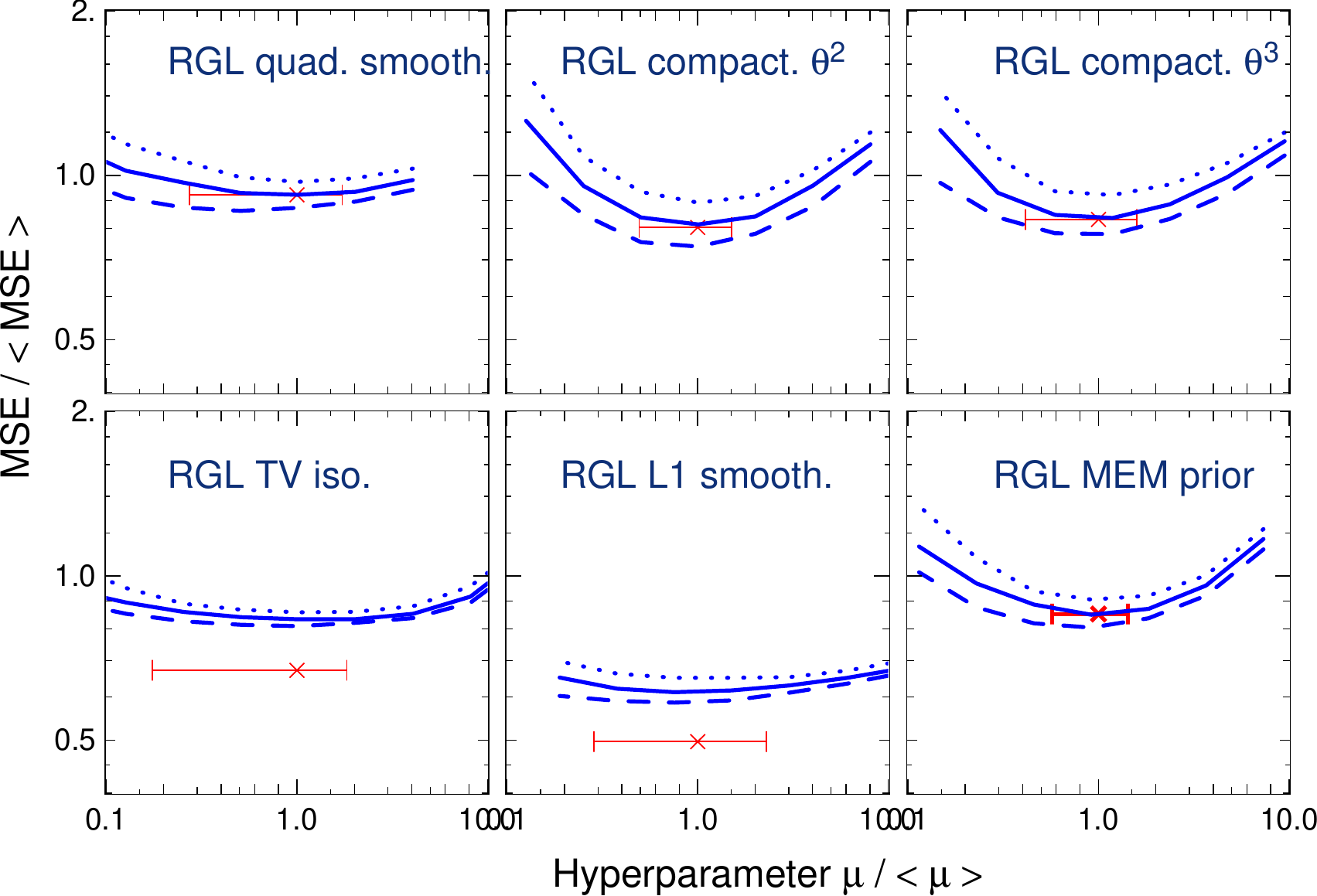}}
  \caption{Variation in MSE and $\mu$ with different noise realizations. \emph{Blue}, the quartile curve of the realizations (25\% in \emph{dash line}, 50\% in \emph{solid line}, and 75\% in \emph{dot line}).  \emph{Red}, the mean optimal $\mu$ (cross) and its variance.}
  \label{fig:noise}
\end{figure}

\begin{figure}
  \resizebox{\hsize}{!}{
    \includegraphics{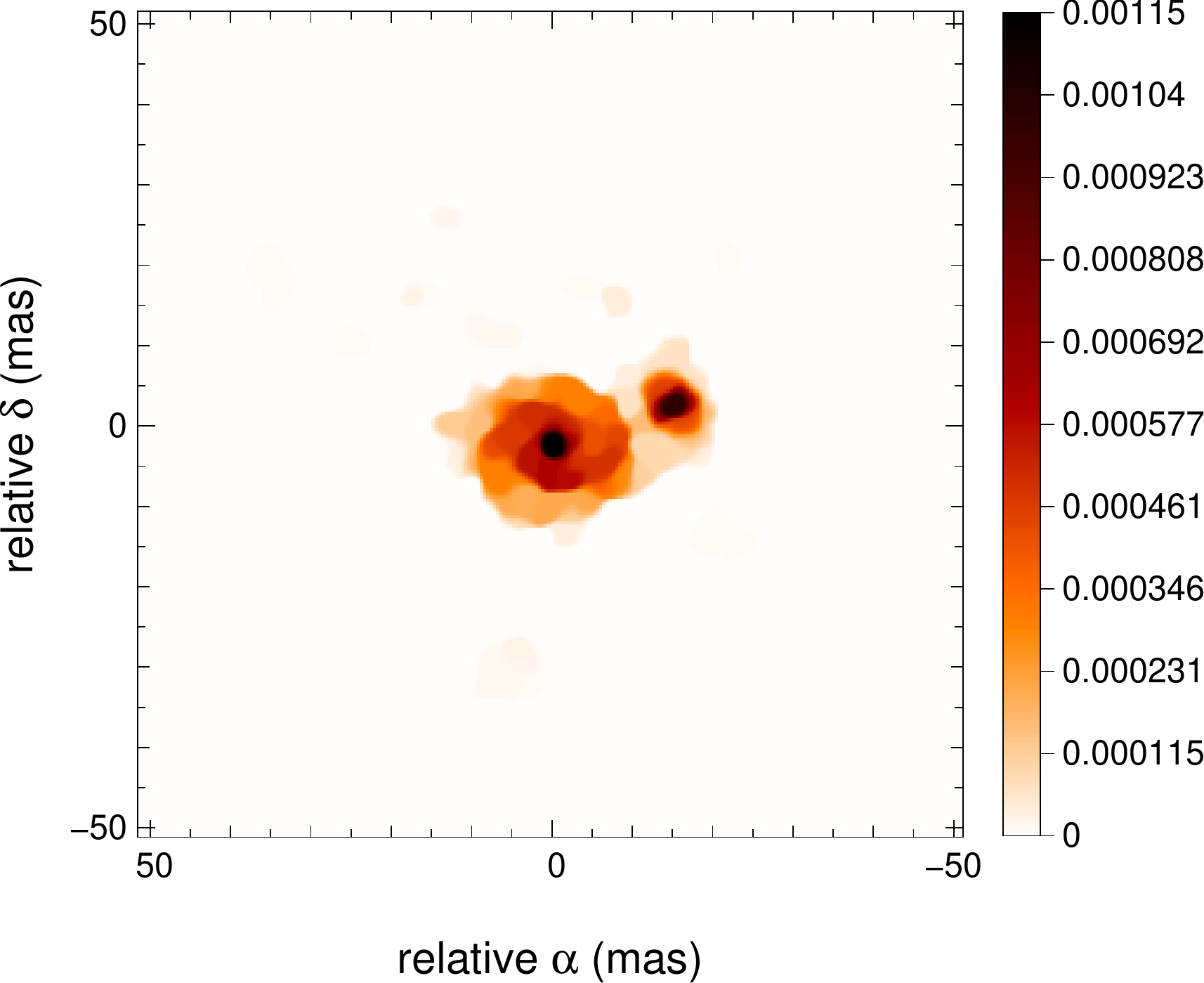}
    \includegraphics{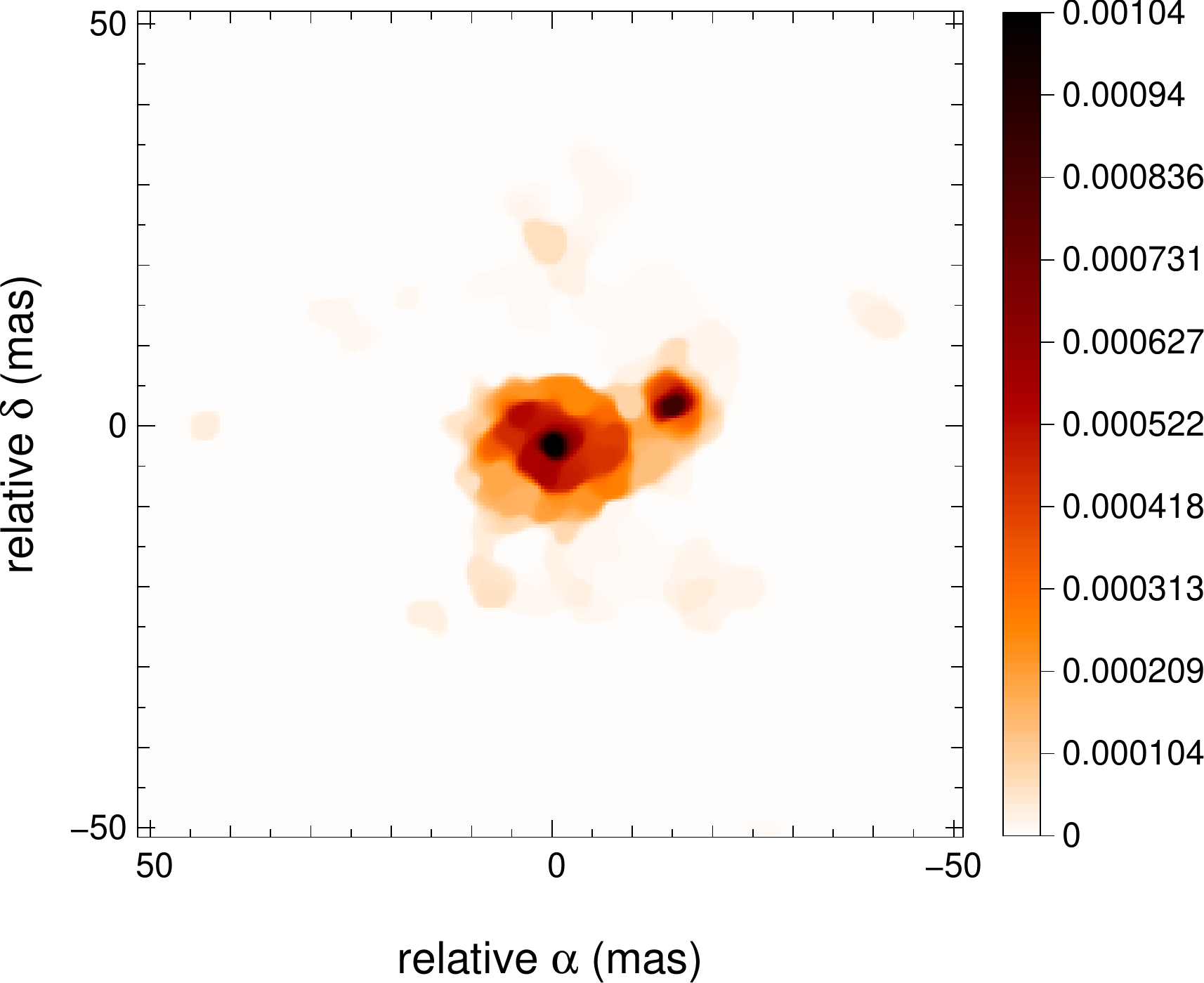}}
  \caption{Reconstructed images for the best (\emph{left}) and the worst (\emph{right}) noise realization in the TV case.}
  \label{fig:noise-expl}
\end{figure}

In all simulations, in order not to influence the classification of the results, the same random seed is used to compute the noisy data. Therefore, we test 100 noise realizations for each regularization in the case of the galaxy with a medium \uv coverage and an intermediate SNR to study its impact on the curves computed from a single realization. From \Fig{fig:noise}, we conclude that the MSE is not very different, thus the image reconstruction does not depend on the particular noise realization (as shown for example in \Fig{fig:noise-expl}). Moreover, the optimal value of $\mu$ varies by less than an order of magnitude.

%


\subsection{The effective spatial resolution}
\label{sec:Fwhm_analyze}

\begin{figure*}
  \begin{minipage}[c]{.5\linewidth}
  \centering
    \includegraphics[width=10cm]{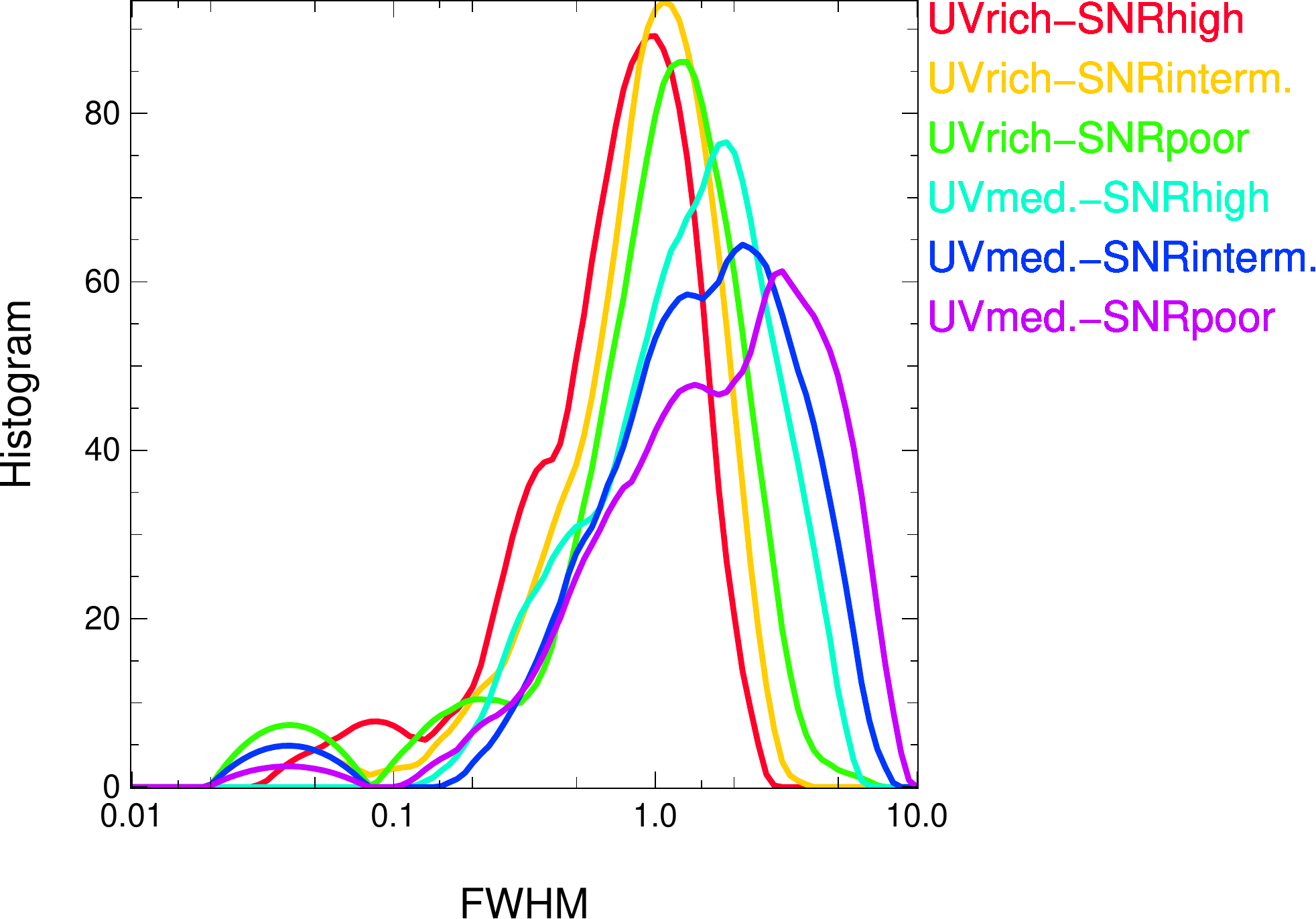}
  \end{minipage}
  \hfill
  \begin{minipage}[c]{.5\linewidth}
  \centering
    \begin{tabular}{c}
      \includegraphics[width=5cm]{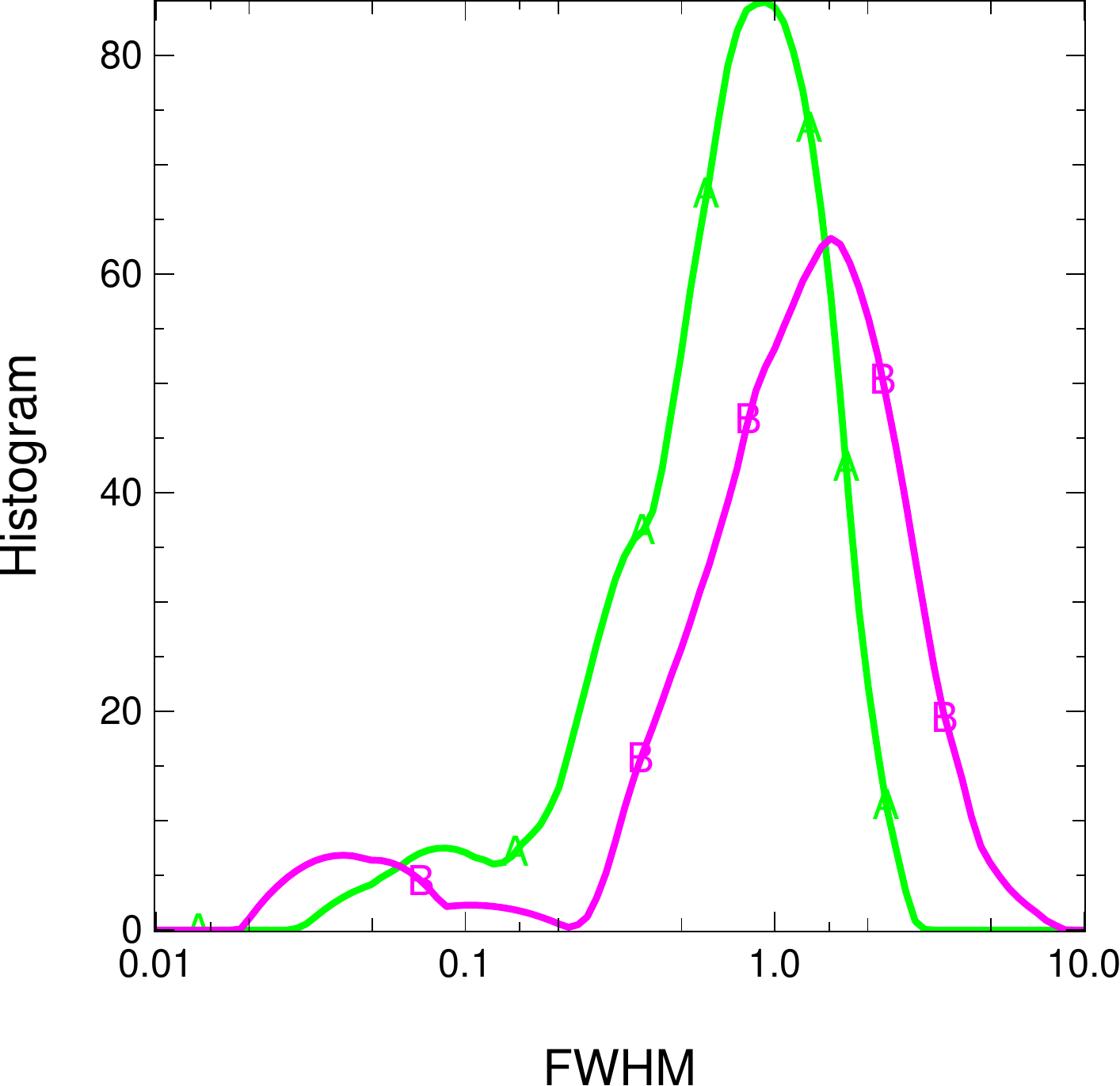} \\
      \hspace{-0.5cm} \includegraphics[width=5cm]{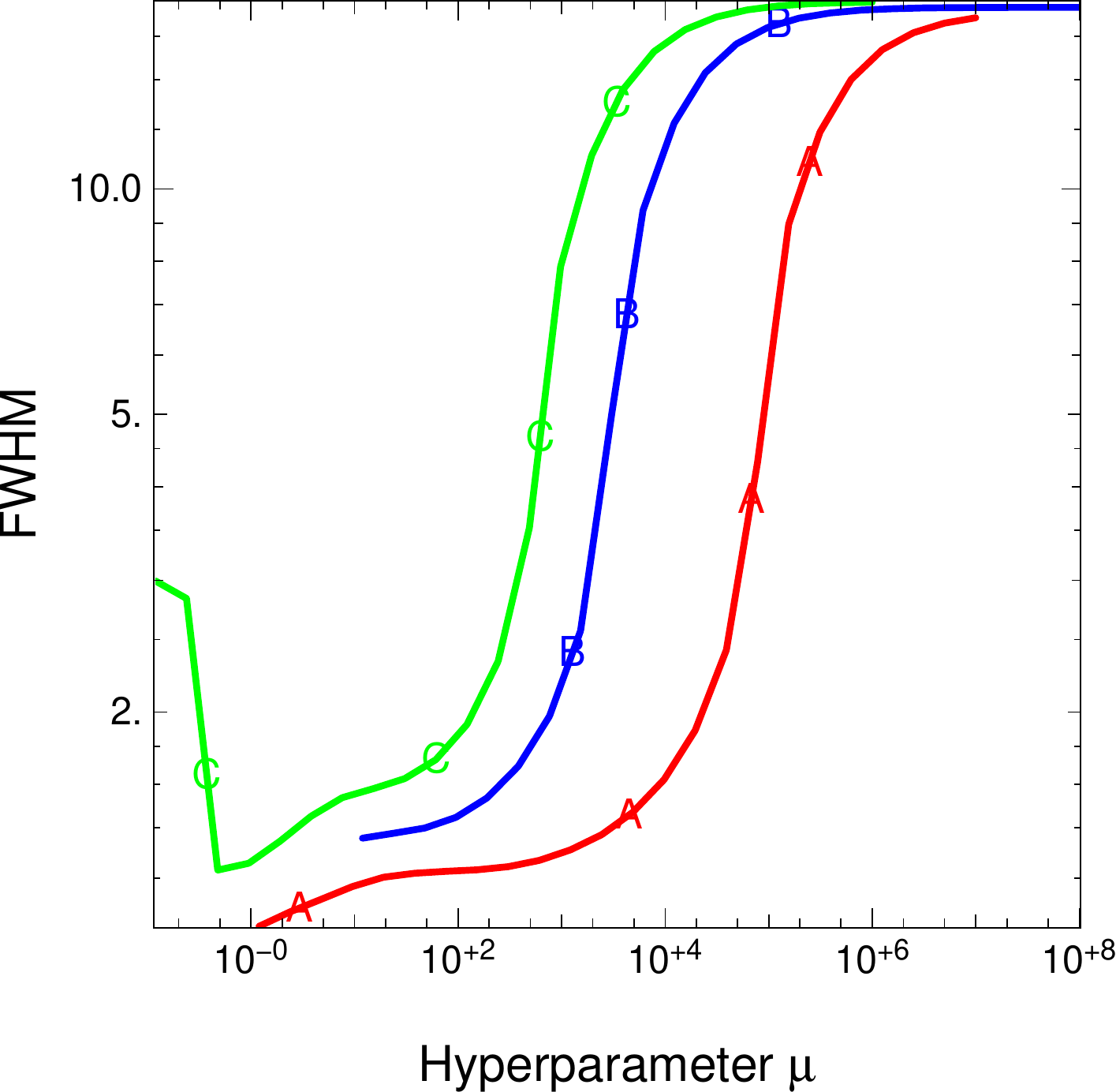} \\
    \end{tabular}
  \end{minipage}
  \caption{\emph{Left:} FWHM of the Gaussian computed for the effective resolution in units of the interferometric resolution of the data. \emph{Up-right:} comparison between the FWHM with (\emph{green}) and without (\emph{magenta}) the constraint of positivity. \emph{Bottom-right:} variation in the effective resolution as a function of the hyperparameter $\mu$ for three different SNR (\emph{red} high, \emph{blue} intermediate, \emph{green} poor). The regularization used is the TV one.}
  \label{fig:fwhm}
\end{figure*}

To investigate whether super-resolution is achievable and to quantify the amount of trustable super-resolution, we estimate the effective resolution of the reconstructed images. We define the effective resolution as the full width at half maximum (hereafter FWHM) of the Gaussian that yields the smallest value of MSE between the reconstructed image and the true image smoothed by that Gaussian
\begin{equation}
  \Tag{FWHM} = \argmin_{w} \Norm{\VParam^\Tag{rec} - \M{G}(w)\cdot\VParam^\Tag{ref}}^2 \, ,
\end{equation}
where $\M{G}(w)$ is a linear smoothing filter that convolves its argument by a Gaussian of FWHM equal to $w$.

As shown in the left part of \Fig{fig:fwhm}, the effective resolution varies with both the amount of data and the SNR: the more extensively the \uv parameter space is  filled and the higher the SNR, the higher the effective resolution is. As for the MSE (\cf Sect.~\ref{sec:limits_uv-snr}), the effective resolution is more influenced by the \uv coverage than by the SNR. Super-resolution can be achieved in only the best  cases.

The bottom-right part of \Fig{fig:fwhm} shows the expected behavior of the effective resolution: the higher the hyperparameter $\mu$, the lower the effective resolution is. In order words, the more the image is regularized, the fewer the details that are visible.

An interesting question is why this super-resolution exists. Although we do not have a definitive answer, we can speculate that it is due to the role of the positivity in the image reconstruction \citep{Biraud1969}. This is  confirmed by our simulations (\cf the upper right part of \Fig{fig:fwhm}): without the positivity constraints, the distribution of the FWHMs has a peak higher than when using the positivity constraints.


\subsection{Other methods for tuning the regularization?}
\label{sec:fdata_analyze}

\begin{figure*}
  \centering
  \begin{tabular}{cc}
    \includegraphics[width=8.5cm]{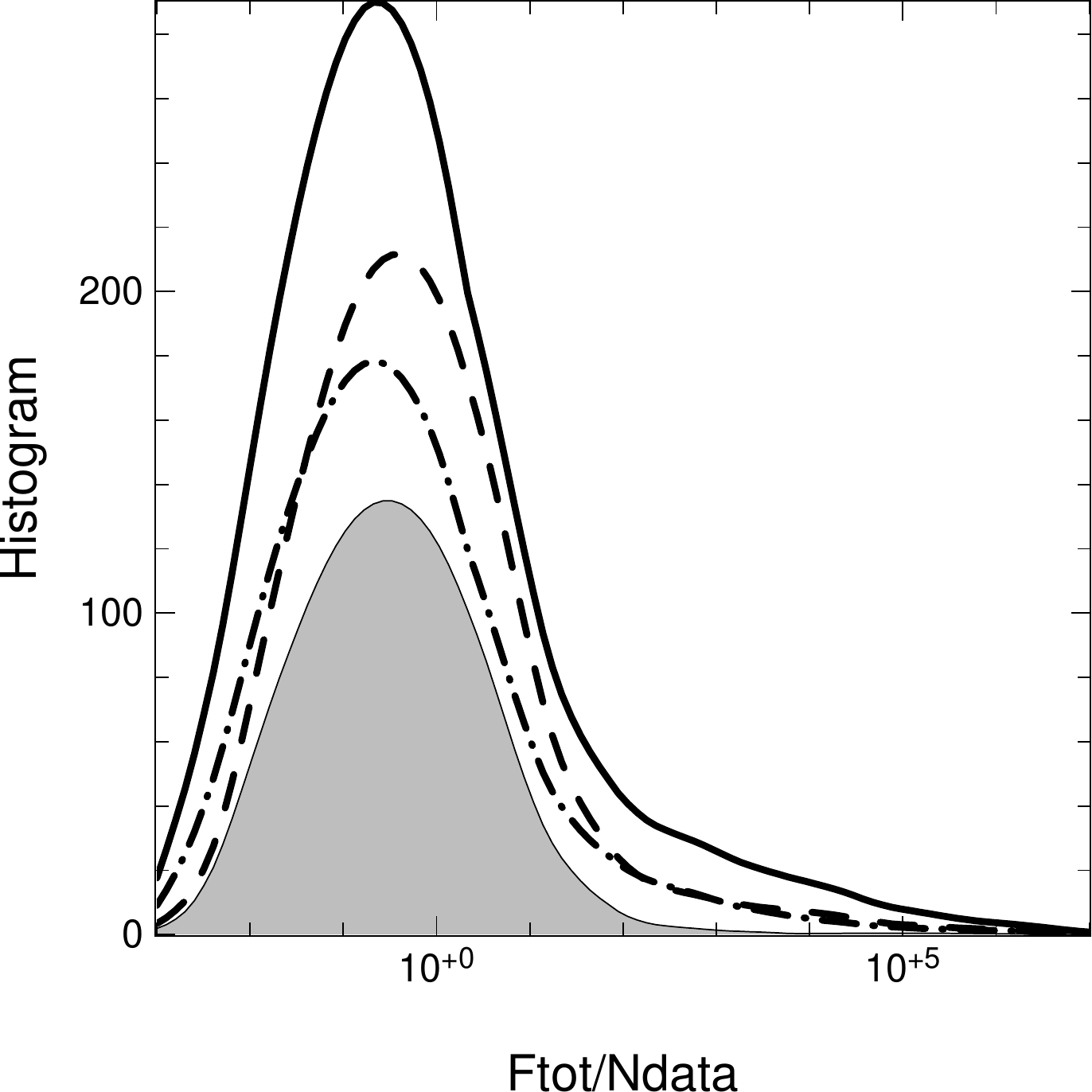}&
    \includegraphics[width=8.5cm]{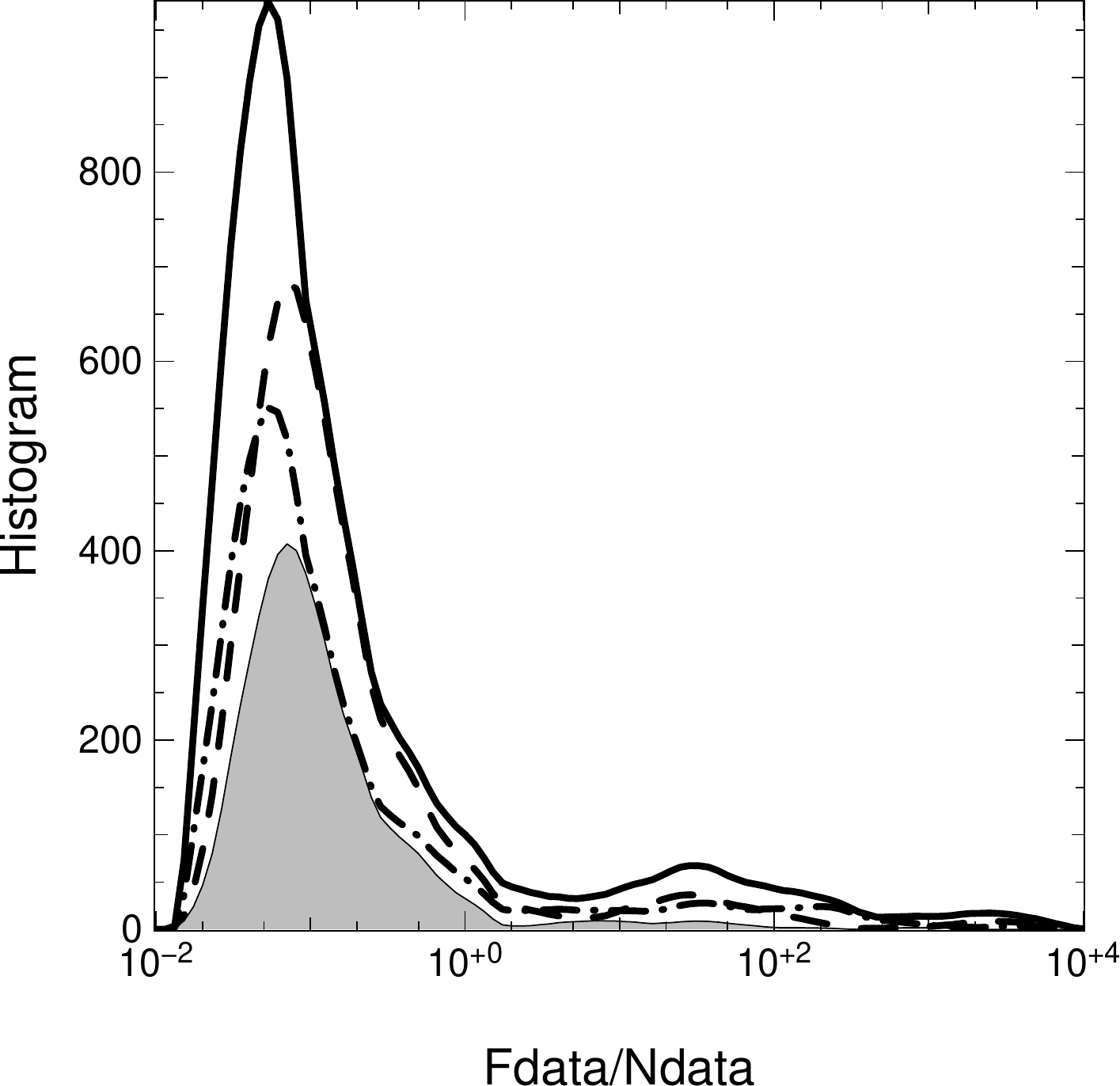} \\
  \end{tabular}
  \caption{Histograms of $\Fcost/\Ndata$ and $\Fdata/\Ndata$. \emph{Solid line:} all configurations and regularizations are kept. \emph{Dashed line:} with the sparsest \uv coverage removed. \emph{Dot-dashed line:} with the bad regularizations removed. \emph{In gray zone:} with the sparsest \uv coverage and bad regularizations removed.}
  \label{fig:Fdata-Ftot}
\end{figure*}

For fully Gaussian statistics (\ie all the likelihood and prior penalties are quadratic and there are no constraints such as the normalization and the positivity), the expected values of the total penalty function $\Fcost(\VParam^{+})$ and the likelihood term $\Fdata(\VParam^{+})$ at the optimal solution $\VParam^{+}$ is given by \citep{Tarantola-2005-inverse_problem_theory}
\begin{align}
  \Expected{\Fcost(\VParam^{+})} &= \Ndata \, , \\
  \Expected{\Fdata(\VParam^{+})} &= \Ndata - \Nedf \, ,
  \label{eq:fdata}
\end{align}
where $\Ndata$ is the number of data points (for both visibility amplitudes and phases), and $\Nedf$ is the number of equivalent degrees of freedom.  Despite our being unable to use fully  Gaussian statistics, the left panel of \Fig{fig:Fdata-Ftot} shows that the distribution of $\Fcost(\VParam^{+})$ peaks approximatively at the value of $\Ndata$.  The spread of this distribution however prevents us to be able to tune the regularization level according to the criterion that $\Fcost(\VParam^{+}) \approx \Ndata$.

The value of $\Fdata(\VParam^{+})/\Ndata$ is around $0.1$, which means that $\Nedf/\Ndata \approx 90\%$ of the data information is resolved. The image reconstruction is able to estimate almost as many parameters as data points. Since the $\Fdata(\VParam^{+})/\Ndata$ ratio has a smaller range (from $\sim$0.3 to $\sim$0.003) than the possible value of the hyperparameter $\mu$ (from $100$ to $10^{13}$), it may be easier to tune the balance between the terms in the penalty function thanks to the $\Fdata(\VParam^{+})/\Ndata$ value instead of $\mu$. However, this criterion may be really variable with the noise statistics and a study of the variation in $\Fdata(\VParam^{+})/\Ndata$ with different noise statistics should be done before using it as a tune factor \citep{Gull1988,Pichon1998}.


%
\begin{figure*}
  \centering
  \includegraphics[width=8.5cm]{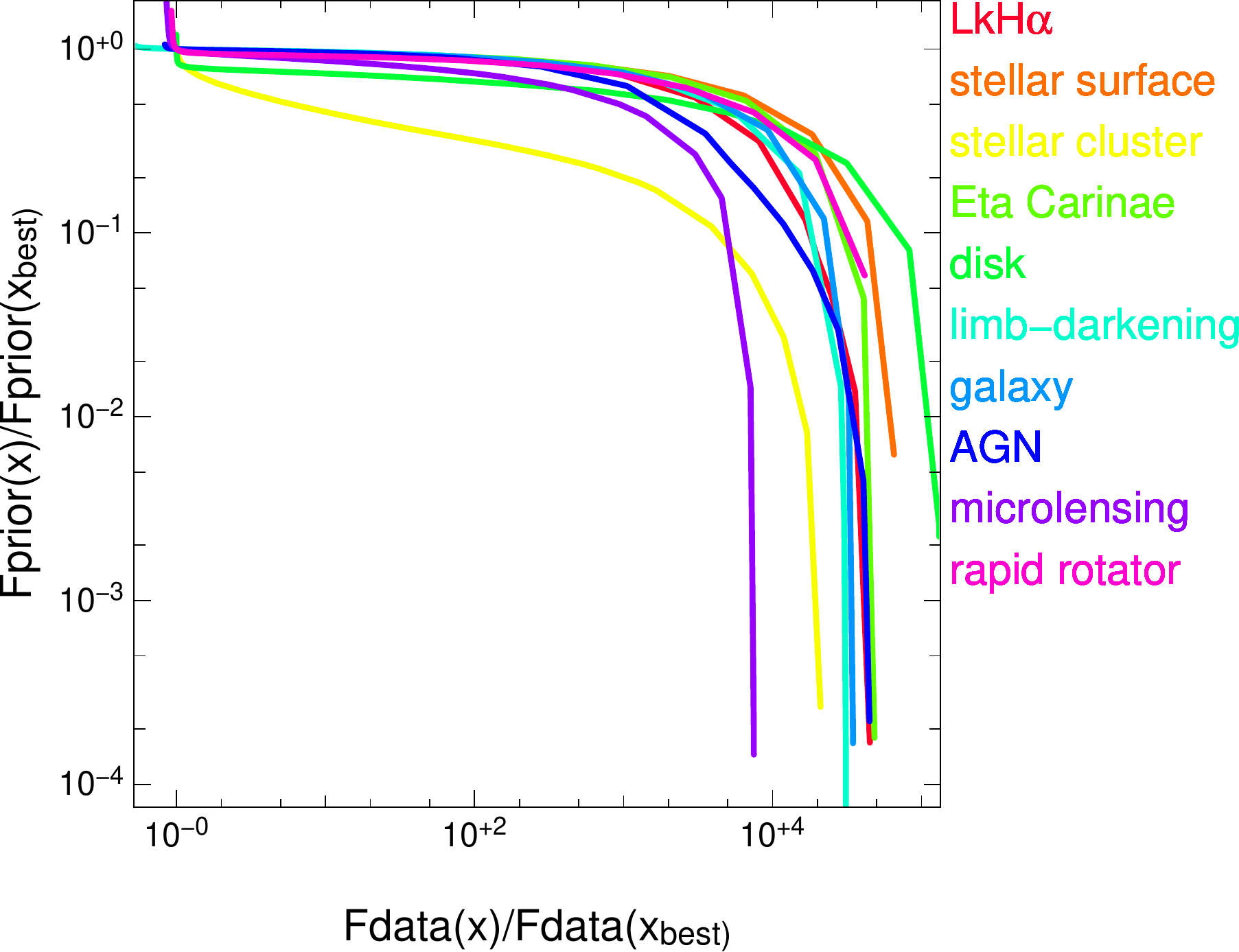}
  \includegraphics[width=8.5cm]{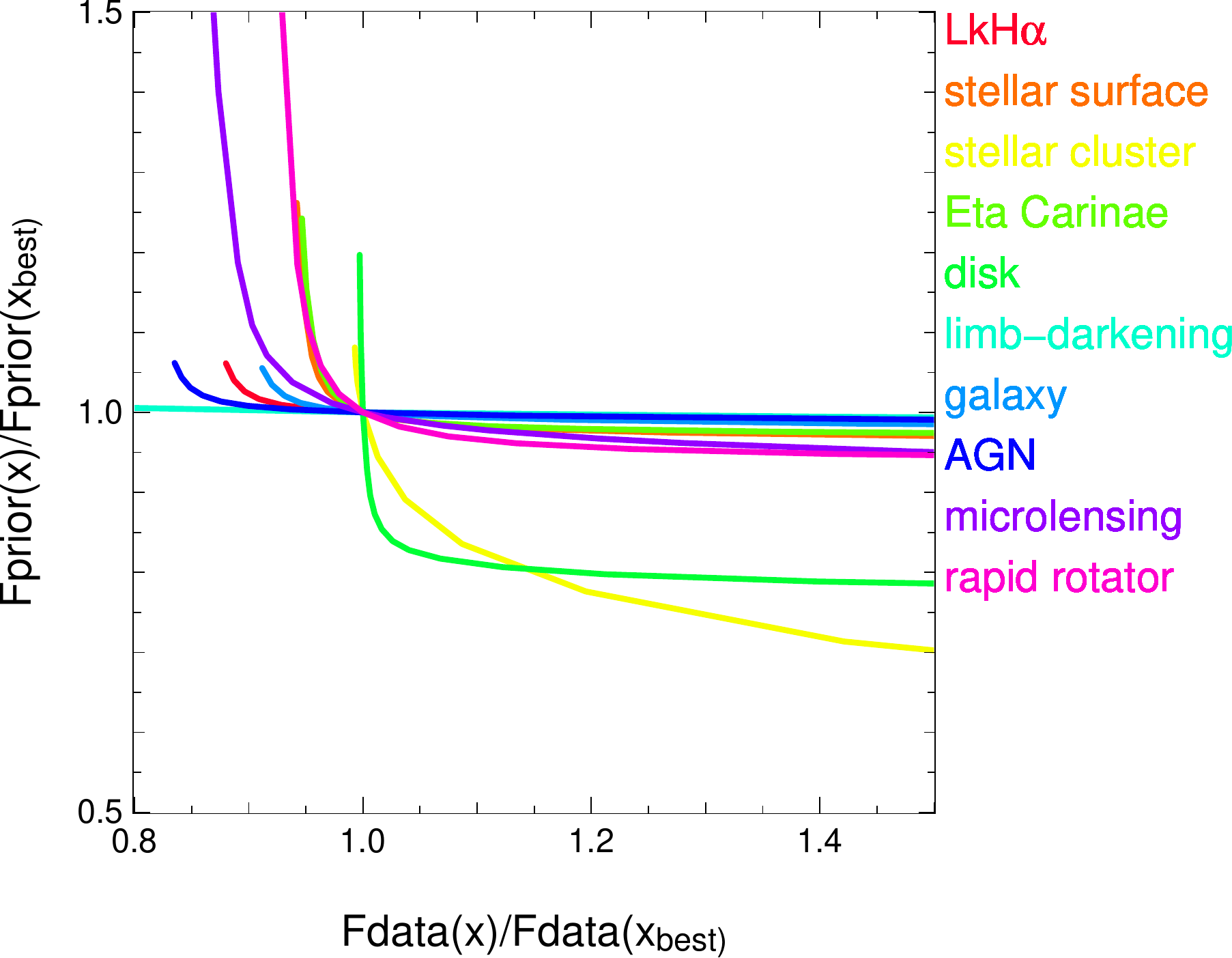}
  \caption{An example of the L-curve in the simulations for the TV regularization with a zoom on the right part.}
  \label{fig:LcurveExpl}
\end{figure*}

Another way of determining the value of $\mu$ is the so-called L-curve. The L-curve is a log-log plot of the regularization term $\Fprior$ versus the likelihood term $\Fdata$ for a range of the hyperparameter $\mu$. In the L-curve criterion, the regularization parameter $\mu$ is such that the corresponding point on the L-curve lies in the corner. This choice is motivated by the corner being the separation between the flat part where the solution is dominated by regularization errors and the vertical part where it is dominated by the perturbation errors \citep{Hansen2000}. The correct behavior of the L-curve is confirmed by the simulations, as seen in \Fig{fig:LcurveExpl} (right): since the curves are plotted as a function of the highest  quality image, they should cross in their corner, which is globally the case. We note that the  shape of the L-curve seems more complicated as there are at least two corners and not only  one (see \Fig{fig:LcurveExpl} left). The L-curve appears to be an appropriate tool for finding the optimal value of $\mu$ but a more general study has to be done to confirm its suitability (see comparison with GCV) and its practical implementation in an image reconstruction algorithm.


\section{Conclusions}
\label{sec:conclusions}

Thanks to the use of a flexible algorithm devoted to image reconstruction in optical  interferometry, we have performed a detailed study of the regularization terms. This study is the  first one to compare such a number of regularizations on an equal footing, \ie with the same  algorithm and using the same data type. Performing  these systematic tests has allowed us to discuss the different parameters and terms in the image reconstruction and to extract some practical rules, which are summarized in the following:
\begin{enumerate}
\item A \textbf{minimal \uv coverage} is necessary to reconstruct an image. Even if the image  quality improves with the SNR, such a limit does not exist for the SNR. In other words, in the image reconstruction technique, increasing the number of telescopes is more interesting than constructing larger ones. The homogeneity of the \uv coverage is probably also critical but has not been tested.
\item Some regularizations are suitable for the optical image reconstruction and others not, regardless of the object being targeted. The holes in the \uv plane are the major issue in optical interferometry and the main role of the regularization is the correct interpolation of the missing data, regardless of the object. The highest quality  regularization among the tested ones is the \textbf{isotropic total variation}.
\item The \textbf{hyperparameter $\mu$} does not depend on the \uv coverage and the SNR as theoretically expected and depends mostly on the type of regularization. An optimal value  for each regularization tested in this paper is given in \Tab{tab:mu_value}. This optimal value may vary by a factor of 2-3 without there being any major changes in the images. A slight dependence on both the object structures and pixel size is also discernible and the equation to rescale the optimal values are computed. It should be interesting to implement regularizations independent of the pixel size.
\item \textbf{Super-resolution} can be achieved in the image reconstruction process and its level rises with the \uv coverage filling.
\item There are various possible ways of tuning the regularization level:
  \begin{itemize}
  \item The visual tuning is enough as the $\mu$ value can slightly vary without causing  any large changes in the reconstructed image.
  \item Setting the likelihood term $\Fdata$ seems to be a more effective way of fixing  the balance between the regularization and the likelihood terms. However, the variation in  the likelihood term with noise statistics needs to be investigated.
  \item The L-curve criterion could give correct results.
  \end{itemize}
\end{enumerate}


\begin{acknowledgements}
  This research has made use of Yorick, a free data processing language written by D.\ Munro (\texttt{http://yorick.sourceforge.net}) and the Service de Calcul Intensif de l'Observatoire de Grenoble.
\end{acknowledgements}


\bibliographystyle{aa} 
\bibliography{srenard-test-mira_ok}


\appendix


\section{The regularization expressions}
\label{app:fprior}

In our simulations, we test 11 different regularization terms commonly used in image reconstruction methods and implemented in \Mira:

\begin{list}{XXXX}{\setlength{\itemsep}{1em}}
\item[1.] \textbf{Quadratic smoothness}:
  \begin{equation}
    \label{eq:regul-quadratic-smoothness}
    \Fprior(\VParam) = \norm{\VParam - \M{S}\cdot\VParam}^2 \, ,
  \end{equation}
  where $\M{S}$ is a smoothing operator implemented via finite differences.
\item[2-3.] \textbf{Compactness} \citep{LeBesnerais2008}:
  \begin{equation}
    \label{eq:regul-compactness}
    \Fprior(\VParam) = \sum_n \Weight^\Tag{prior}_n \VParam_n^2 \, ,
  \end{equation}
  which is a separable quadratic regularization. To enforce compactness, the weights $\Weight^\Tag{prior}_n > 0$ have to increase with the distance from the center of the image. Two different cases were studied in the simulations: $\Weight^\Tag{prior}_n=\norm{\VDirn_n/\Delta\Dirn}^2$ and $\Weight^\Tag{prior}_n = \norm{\VDirn_n/\Delta\Dirn}^3$, where $\norm{\VDirn_n/\Delta\Dirn}$ is the distance in pixels of the $n$-th pixel from the center of the field of view.
\item[4.] \textbf{Total variation} \citep{Strong_Chan-2003-total_variation}:
  \begin{equation}
    \Fprior(\VParam) = \sum_{n_1,n_2}
    \sqrt{\norm{\nabla\Param_{n_1,n_2}}^2 + \epsilon^2} \, ,
    \label{eq:tv-prior}
  \end{equation}
  where
  \begin{displaymath}
    \norm{\nabla\Param_{n_1,n_2}}^2 = (\Param_{n_1+1,n2} - \Param_{n_1,n_2})^2 +
    (\Param_{n_1,n_2+1} - \Param_{n_1,n_2})^2
  \end{displaymath}
  is the squared magnitude of the spatial gradient in the image, $\epsilon > 0$ is a small number inserted to avoid the discontinuity in zero, and $(n_1,n_2) \sim n$ are the two dimensional indices of the $n$-th pixel. In our reconstructions, $\epsilon$ has always been chosen so as to be  negligible compared to the gradient of significant structures.
\item[5.] \textbf{$\ell_2-\ell_1$ smoothness} \citep{Charbonnier_et_al-1997-edge_preserving}:
  \begin{equation}
    \Fprior(\VParam) = \tau^2 \sum_{n}
    \psi\bigl(\norm{(\M{D}\cdot\VParam)_{n}}/\tau\bigr) \, ,
    \label{eq:l1-l2-prior}
  \end{equation}
  where $\psi(z) = z - \log(1 + z)$ is a $\ell_2$-$\ell_1$ norm, $\tau>0$ is a threshold level, and $\M{D}$ is a finite difference operator approximating the $q$-th spatial derivatives of its argument.  When $\norm{(\M{D}\cdot\VParam)_{n}}$ is much smaller than the threshold, $\tau^2\,\psi\bigl(\norm{(\M{D}\cdot\VParam)_{n}}/\tau\bigr) \approx 1/2\,\norm{(\M{D}\cdot\VParam)_{n}}^2$, while, for $\norm{(\M{D}\cdot\VParam)_{n}}$ it is much larger than the threshold, $\tau^2\, \psi\bigl(\norm{(\M{D}\cdot\VParam)_{n}}/\tau\bigr) \approx \tau\,\norm{(\M{D}\cdot\VParam)_{n}}$. This regularization attempts to strongly smooth the weak spatial gradients and slightly smooth the strong gradients.  In our simulations, we take $\M{D}$ to approximate the spatial Laplacian and choose a threshold small enough such that the regularization behaved mostly like a linear smoothness.
\item[6-8.] \textbf{$\ell_p$-norm:}
  \begin{equation}
    \Fprior(\VParam)
    = \sum\nolimits_n \left(\Param_n^2 + \epsilon^2\right)^{p/2}
    \approx \sum\nolimits_n \abs{\Param_n}^p \, ,
    \label{eq:lp-prior}
  \end{equation}
  where $\epsilon>0$ is a small value introduced to avoid the singularity in zero when $p \le 1$. For $p>1$, the $\ell_p$-norm regularization tends to produce a smooth image as it reduces the variance of the pixels.  For $p<1$, the $\ell_p$-norm regularization tends to promote sparsity in the image. This is interesting mostly for objects consisting of point sources. True sparsity constraints would be obtained for $p=0$, although when $p<1$ the regularization is no longer convex and the optimization problem becomes extremely difficult to solve as $p$ gets closer to 0. Results in compress sensing \citep{Candes_Romberg_Tao-2006-robust_uncertainty} have proven that choosing $p=1$ yields the most sparse solution, like $p=0$ for a large class of problems. However, taking $p=1$ yields a non-smooth but convex problem that is much easier to solve than the combinatorial problem resulting from the choice $p=0$.  With positivity and normalization constraints, the $\ell_1$-norm of $\VParam$ is constant. Hence, taking $p=1$ is meaningless in our framework and we consider only $p=1.5$,  $p=2$ and $p=3$.
\item[9-11.] \textbf{Maximum entropy methods} \citep{Narayan1986}:
  \begin{equation}
    \Fprior(\VParam) = - \sum\nolimits_n h(\Param_n;\bar{\Param}_n) \, .
    \label{eq:mem-prior}
  \end{equation}
  Here the prior is to assume that the image is drawn towards a prior model $\bar{\VParam}$ according to a non-quadratic potential $h$, called the \emph{entropy}. We try three entropies:
  \begin{align}
    \text{MEM-sqrt:}\hspace*{4.1ex}
    &h(\Param;\bar{\Param}) = \sqrt{\Param} \, ; \\
    \text{MEM-log:}\hspace*{4.5ex}
    &h(\Param;\bar{\Param}) = \log(\Param) \, ; \\
    \text{MEM-prior:}\hspace*{3ex}
    & h(\Param;\bar{\Param}) = \Param - \bar{\Param} - \Param\,
    \log\left(\Param/\bar{\Param}\right) \, .
    \label{eq:MEM-prior}
  \end{align}
  MEM was first introduced in radioastronomy and is useful for images made of bright point-like sources on a smooth background.  In our simulations, we took the prior image $\bar{\VParam}$ to be the isotropic 2-D Gaussian that most accurately fits the amplitude visibility data.
\end{list}


\section{Renormalization of MSE}
\label{app:mse-norm}

\begin{figure}
  \resizebox{\hsize}{!}{
    \begin{tabular}{cc}
      \includegraphics{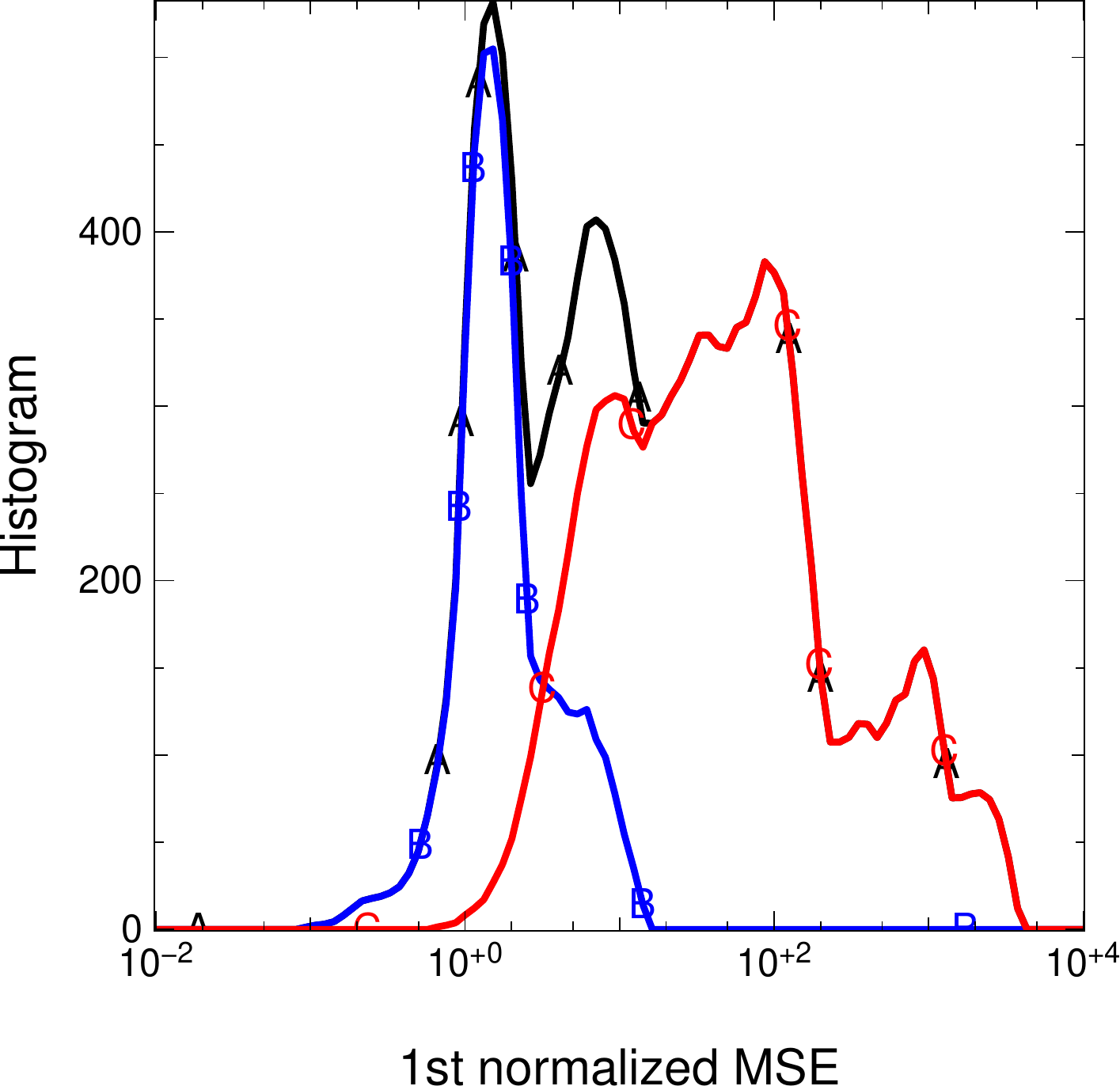}&
      \includegraphics{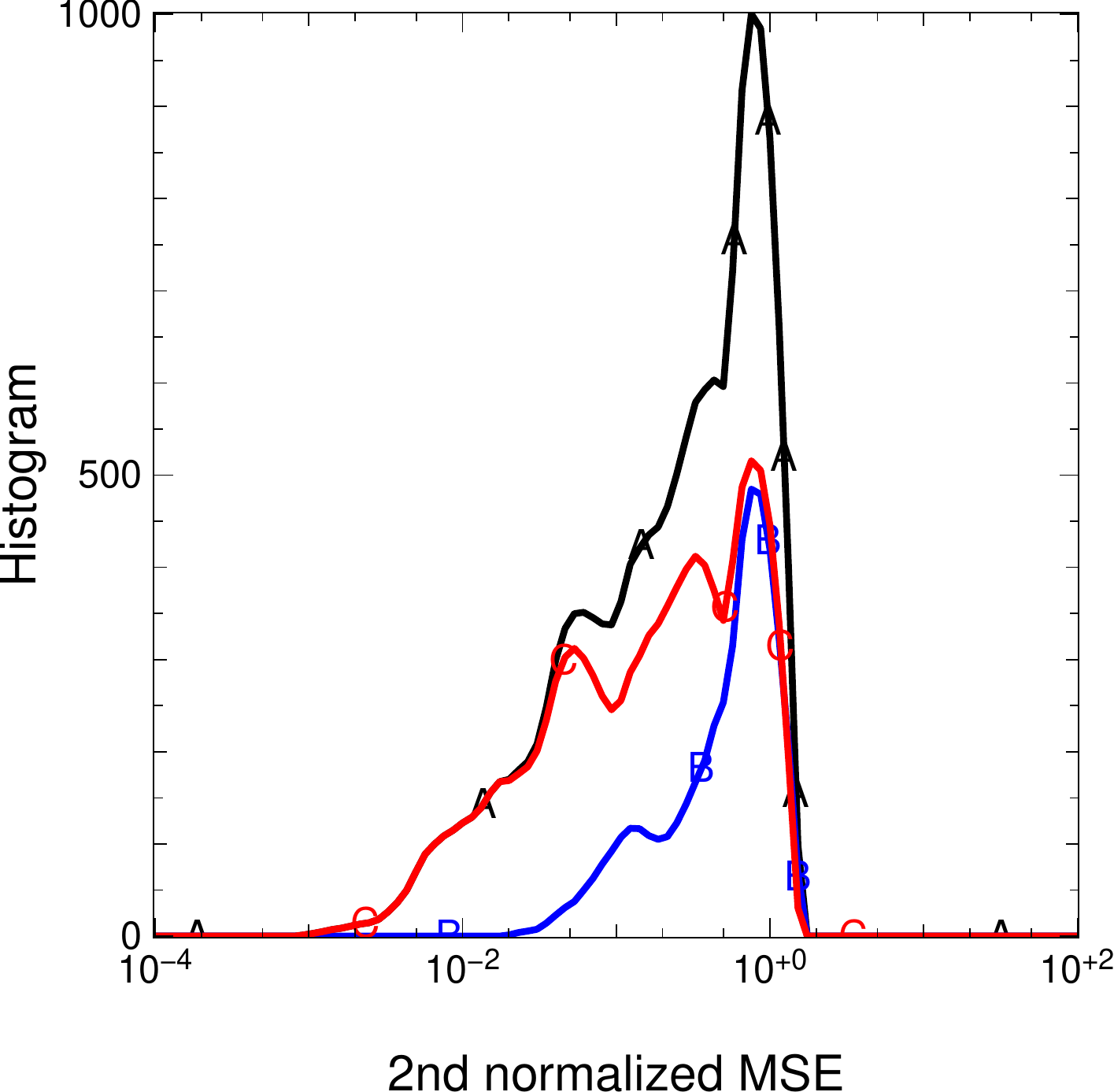}\\
    \end{tabular}}
  \caption{Attempt to renormalize the MSE. Distribution of the first normalized MSE (\emph{left})
  and of the second normalized MSE (\emph{right}).
  The colors and letters represent the two classes of objects:
  \emph{blue/B} for the objects with very compact structures, \emph{red/C} for the others.
  The total distribution is shown in \emph{black/A} curve.}
  \label{fig:MSEnorm}
\end{figure}

We attempted to normalize the MSE with two additional methods:
\begin{enumerate}
\item Since the squared difference between the real image and a smoothed version of the real image is higher for images with sharp or point-like structures, we computed a first \emph{normalized} MSE as
\begin{equation}
\Tag{MSE_\mathrm{norm.,1}} = \frac{\sum_n \left(\Param^\Tag{rec}_n -
  \Param^\Tag{ref}_n\right)^2}{\sum_n \left((\M{S}\cdot\VParam^\Tag{ref})_n
  - \VParam^\Tag{ref}_n\right)^2} \, ,
\end{equation}
where $\M{S}$ is a smoothing operator.  The distribution of this normalized MSE is shown in  the right panel of \Fig{fig:MSEnorm}: the distribution is narrower but the two classes remain,  despite the normalization.
\item In the second normalization, the MSE is compared to the norm of the reference image
\begin{equation}
\Tag{MSE_\mathrm{norm.,2}} = \frac{\sum_n \left(\Param^\Tag{rec}_n -
  \Param^\Tag{ref}_n\right)^2}{\sum_n \left(\VParam^\Tag{ref}_n\right)^2} \, .
\end{equation}
 This normalization is more effective than the previous one, as it joins the curves together. However, the distribution is unimodal and does not enable us to distinguish the good and the bad reconstructions. It is thus not a useful normalization.
\end{enumerate}


\section{Scaling the hyperparameter $\mu$}
\label{app:scaling-hyperparameter}

In a Bayesian framework, the prior penalty $\mu\,\Fprior(\VParam)$ should only depend on both the sought after brightness distribution $\Image(\VDirn)$ and the image parametrization.  In this appendix and from these simple principles, we derive a method to adapt the value of the hyperparameter $\mu$ when the image parameters such as the pixel size are modified.

Using the sampled image model in Eq.~(\ref{eq:sampled-image-model}), the normalization of $\VParam$ implies that
\begin{displaymath}
  1 = \sum\nolimits_n \Param_n = \alpha\,\sum\nolimits_n \Image(\VDirn_n)
  \approx \frac{\alpha}{(\Delta\Dirn)^L}\,\int \Image(\VDirn) \, \mathd^L\VDirn \, ,
\end{displaymath}
where we have used the Riemann approximation of integrals, $\Delta\Dirn$ is the pixel size, and $L=2$ is the number of dimensions in $\VDirn$. Without any loss of generality, we can assume that the sought after brightness distribution is normalized such that $\int \Image(\VDirn) \, \mathd^L\VDirn = 1$; hence
\begin{equation}
  \label{eq:scaling-of-alpha}
  \alpha \approx (\Delta\Dirn)^L \, ,
\end{equation}
and Eq.~(\ref{eq:sampled-image-model}) becomes
\begin{equation}
  \label{eq:sampled-normalized-image-model}
  x_n \approx (\Delta\Dirn)^L \, \Image(\VDirn_n) \, .
\end{equation}

\subsection{Separable $\ell_p$ norm}

Using Eq.~(\ref{eq:sampled-image-model}) and then, the Riemann approximation, the prior penalty for a separable $\ell_p$ norm regularization, after substituting in  Eq.~(\ref{eq:lp-prior}), becomes
\begin{align}
  \mu\,\Fprior(\VParam) &= \mu\,\sum_n \abs{\Param_n}^p
  \approx \mu\,\alpha^p \, \sum_n \abs{\Image(\VDirn_n)}^p \notag \\
  &\approx \mu\,\frac{\alpha^p}{\Delta\Dirn^L}
  \, \int \abs{\Image(\VDirn)}^p \, \mathd\VDirn \, ,
\end{align}
which shows that for a regularization by a separable $\ell_p$ norm
\begin{equation}
  \mu \propto \begin{cases}
    \Delta\Dirn^{L} / \alpha^p & \text{in general;} \\
    \Delta\Dirn^{L\,(1 - p)} & \text{with the normalization constraint.}
  \end{cases}
\end{equation}
Hence, with the normalization constraint, the optimal value of $\mu$ should be the same regardless of the pixel size for a regularization given by a separable $\ell_1$ norm.

\subsection{$\ell_p$ norm on the gradient}

Using 1-D notation to simplify the equations, the prior penalty for a regularization by the $\ell_p$ norm on the gradient is given by
\begin{align}
  \mu\,\Fprior(\VParam)
  &= \mu\,\sum_n \abs{\Param_{n+1} - \Param_{n}}^p  \notag \\
  &\approx \mu\,\alpha^p \, \sum_n
  \abs{\Image(\Dirn_{n} + \Delta\Dirn) - \Image(\Dirn_n)}^p \notag \\
  & \approx \mu\,\alpha^p\,\Delta\Dirn^p \, \sum_n
  \Abs{\partial_{\Dirn}\Image(\Dirn_{n})}^p \, , \notag
\end{align}
where $\partial_{\Dirn}\Image(\Dirn)$ denotes the partial derivative of the brightness distribution along the angular direction.  In $L$ dimensions and using the Riemann approximation, this gives
\begin{align}
  \mu\,\Fprior(\VParam)
  & \approx \mu\,\alpha^p\,\Delta\Dirn^{p - L}
  \, \int \Abs{\partial_{\VDirn}\Image(\VDirn)}^p
  \, \mathd\VDirn \, , \notag
\end{align}
which shows that
\begin{equation}
  \mu \propto \begin{cases}
    \Delta\Dirn^{L - p} / \alpha^p & \text{in general;} \\
    \Delta\Dirn^{L - p\,(L + 1)} & \text{with the normalization constraint.}
  \end{cases}
\end{equation}
Applying this result for a regularization by quadratic smoothness in 2D, \eg Eq.~(\ref{eq:regul-quadratic-smoothness}), we found that, with a normalization constraint, $\mu\propto \Delta\Dirn^{-4}$.

\subsection{Total variation}

The preceding result, with $p=1$, readily applies to regularization by the total variation, that is
\begin{equation}
  \mu \propto \begin{cases}
    \Delta\Dirn^{L - 1} / \alpha & \text{in general;} \\
    \Delta\Dirn^{-1} & \text{with the normalization constraint.}
  \end{cases}
\end{equation}
We can also deduce that, if a relaxed version of TV is used, as in Eq.~(\ref{eq:tv-prior}), the relaxation parameter $\epsilon$ must scale as the pixel size $\Delta\Dirn$ to have the prior penalty approximately insensitive to the pixel size.

We note that, with our particular choice of the threshold $\tau$ for the $\ell_2-\ell_1$ smoothness regularization defined in Eq.~(\ref{eq:l1-l2-prior}), we expect this regularization to behave mostly like TV.

\subsection{Quadratic compactness}

The quadratic compactness we used in \Mira is given by Eq.~(\ref{eq:regul-compactness})
\begin{align}
  \mu\,\Fprior(\VParam)
  &= \mu\,\sum_n \Norm{\frac{\VDirn_n}{\Delta\Dirn}}^q \, \Param_n^2
  \approx \mu \, \frac{\alpha^2}{\Delta\Dirn^{q}} \,
  \sum_n \norm{\VDirn_n}^q \, \Image(\VDirn_n)^2 \notag \\
  &\approx \frac{\mu \, \alpha^2}{\Delta\Dirn^{q + L}} \,
  \int \norm{\VDirn}^q \, \Image(\VDirn)^2 \, \mathd\VDirn \, , \notag
\end{align}
with $q=2$ or $3$.  From this last approximation, we derive the scaling of $\mu$ with the pixel size
\begin{align}
  \mu \propto \begin{cases}
    \Delta\Dirn^{q + L} / \alpha^2 & \text{in general;} \\
    \Delta\Dirn^{q - L} & \text{with the normalization constraint.}
  \end{cases}
\end{align}
Hence, in 2D ($L=2$) and for a normalized image, $\mu$ does not depend on the pixel size for $q=2$, while it scales as $\Delta\Dirn$ for $q=3$.

\subsection{Maximum entropy}

For maximum entropy methods, we have
\begin{displaymath}
  \mu \, \sum_n \sqrt{\Param_n}
  \approx \mu \, \frac{\alpha^{1/2}}{\Delta\Dirn^L}
\int \Image(\VDirn)^{1/2} \, \mathd\VDirn \, ,
\end{displaymath}
and
\begin{displaymath}
  \mu \, \sum_n h\!\left( \Param_n ; \bar{\Param}_n\right)
  \approx
  \mu \, \frac{\alpha}{\Delta\Dirn^L} \,
  \int h\!\left( \Image(\VDirn); \bar{\Image}(\VDirn)\right)
  \, \mathd\VDirn \, ,
\end{displaymath}
with $h(x;\bar{x}) = x - \bar{x} - x\,\log(x/\bar{x})$ as in
Eq.~(\ref{eq:MEM-prior}).  from which we deduce that
\begin{align}
  \mu \propto \begin{cases}
    \Delta\Dirn^{L} / \alpha^{1/2} & \text{for MEM-sqrt;} \\
    \Delta\Dirn^{L} / \alpha & \text{for MEM-prior.}
  \end{cases}
\end{align}
Hence, for a normalized image, $\mu$ does not depend on the pixel size in a MEM-prior regularization.

\end{document}